\documentclass[aps,prd,amsmath,floats,floatfix,superscriptaddress,nofootinbib,noshowpacs,10pt]{revtex4-2}
\usepackage{amssymb,epsfig}

\newcommand{\Real}{\mathbb{R}}

\begin{document}

\title{Accretion of a Vlasov gas by a Kerr black hole}
\author{Patryk Mach}
\email{patryk.mach@uj.edu.pl}
\affiliation{Instytut Fizyki Teoretycznej, Uniwersytet Jagiello\'{n}ski, {\L}ojasiewicza 11, 30-348 Krak\'{o}w, Poland}

\author{Mehrab Momennia}
\email{momennia1988@gmail.com}
\affiliation{Instituto de F\'isica y Matem\'aticas, Universidad Michoacana de San
Nicol\'as de Hidalgo, Edificio C-3, Ciudad Universitaria, 58040 Morelia,
Michoac\'an, M\'exico}

\author{Olivier Sarbach}
\email{olivier.sarbach@umich.mx}
\affiliation{Instituto de F\'isica y Matem\'aticas, Universidad Michoacana de San
Nicol\'as de Hidalgo, Edificio C-3, Ciudad Universitaria, 58040 Morelia,
Michoac\'an, M\'exico}

\date{\today}

\begin{abstract}
We investigate the accretion of a collisionless, relativistic kinetic gas by a rotating Kerr black hole, assuming that at infinity the state of the gas is described by a distribution function depending only on the energy of the particles. Neglecting the self-gravity of the gas, we show that relevant physical observables, including the particle current density and the accretion rates associated with the mass, the energy, and the angular momentum, can be expressed in the form of closed integrals that can be evaluated numerically or approximated analytically in the slow-rotation limit. The accretion rates are computed in this manner for both monoenergetic particles and the Maxwell-J\"uttner distribution and compared with the corresponding results in the non-rotating case. We show that the angular momentum accretion rate decreases the absolute value of the black hole spin parameter. It is also found that the rotation of the black hole has a small but non-vanishing effect on the mass and the energy accretion rates, which is remarkably well described by an analytic calculation in the slow-rotation approximation to cubic order in the rotation parameter. The effects of rotation on the morphology of the accretion flow are also analyzed.
\end{abstract}

\maketitle

\section{Introduction}

The problem of relativistic Bondi-type accretion of matter on a rotating (Kerr) black hole has a long history, dating back to about 40 years ago. In 1988 Petrich, Shapiro, and Teukolsky obtained an analytic solution in the hydrodynamical regime, assuming a zero-vorticity potential flow and the ultra-hard equation of state \cite{lPsSsT1988}. It was derived by exploiting a correspondence between potential flows with the ultra-hard equation of state and solutions of the massless Klein-Gordon equation, which is known to be separable on the Kerr background spacetime. The solution obtained in~\cite{lPsSsT1988} is valid for moving black holes, and it naturally reduces to the standard Bondi problem if the black hole velocity is set to zero. In this case, it has a particular property: in Boyer-Lindquist coordinates, both polar and covariant azimuthal components of the four-velocity vanish. Unfortunately, generalizations of this solution to other (more physically relevant) equations of state have only been achieved at the approximate level~\cite{vByP95,vP96}. Another approximate solution was derived in~\cite{eTpTjM13}, based on the ballistic approximation. For a systematic numerical study of the spherical Bondi-type accretion of an ideal fluid into a Kerr black hole, see~\cite{aAeToSdL21}.

In this paper we present an exact stationary solution representing the relativistic Bondi-type accretion of the collisionless (Vlasov) gas in a rotating subextremal Kerr black hole background. By ``Bondi-type'' accretion we mean an idealized, asymptotically spherical stationary model, in which the accretion flow is sustained by a homogeneous reservoir of gas at infinity.

Within the Newtonian approximation, the accretion of a Vlasov gas was investigated by Novikov and Zel'dovich~\cite{ZelNovik-Book}. First steps towards generalization to the relativistic setting were undertaken by Shapiro and Teukolsky~\cite{Shapiro-Book} (see also Appendix~A in Ref.~\cite{sSsT85a}), who computed the accretion rate of a Vlasov gas of monoenergetic particles spherically infalling onto a Schwarzschild black hole. A modern analysis of spherically symmetric steady relativistic accretion of the collisionless Vlasov gas onto a Schwarzschild black hole was presented in~\cite{pRoS16,pRoS17} by Rioseco and Sarbach. Their construction was later repeated for the Reissner-Nordstr\"{o}m spacetime in~\cite{aCpM20} (see also~\cite{pLjYsX25}), with the aim of mimicking the effects associated with the black hole spin (replaced by the electric charge) in a spherically symmetric spacetime. Stationary spherical accretion flows of the Vlasov gas in various exotic black hole spacetimes were studied in~\cite{BardeenBH,SchwLikeBH,KalbRamondBH}. A generalization of spherical accretion models to asymptotically flat and generic spherically symmetric black hole spacetimes was recently shown in~\cite{mMoS25}. For a generalization of the Bondi-type model on the Schwarzschild spacetime, where the boundary conditions are specified at a sphere of finite radius instead of infinity, see~\cite{aGetal21}.

At the same time, non-spherical flows in the Schwarzschild or Kerr spacetime were investigated within the same framework in~\cite{cGoS22a,cGoS22c,sS23,mLyYaC25,cGrR2025} (axially symmetric disks), or~\cite{pMoA21a,pMoA21b,pMaO22b,aCpMaO24} (accretion onto moving black holes). A Bondi-type model confined to the equatorial plane of the Kerr spacetime was derived in \cite{aCpMaO22} and \cite{gKpM25}. Analytic kinetic models describing the relaxation (due to phase-space mixing) of macroscopic variables to a stationary state around a Kerr black hole have been described in~\cite{pRoS18,pRoS24}. For self-gravitating shells or disks made of Vlasov matter around black holes, see~\cite{gR93,fJ21,hA21,fJ22}.

A preliminary analysis of the Bondi-type accretion model of the Vlasov gas in the Kerr spacetime (without restrictions to the equatorial plane) was presented in \cite{pLyLxZ23}. Unfortunately, this paper does not discuss the corresponding phase-space limits, nor does it make an attempt to compute total accretion rates.

Bondi-type models are specified by asymptotic conditions. For simplicity, in this paper we focus on a simple gas~\cite{wI63}, that is a gas consisting of spinless and uncharged classical point particles of the same mass and consider asymptotic distributions depending only on the energy of the gas particles. In particular examples, we concentrate on monoenergetic particles and the Maxwell-J\"{u}ttner distribution. In all cases, we only take into account unbound particle trajectories (extending to infinity). 
Particles moving along bound trajectories can still be included in our model, but since we neglect the collisions between the particles of the gas, as well as its self-gravity, they would not affect results concerning unbound trajectories.

We show that the particle number and energy accretion rates decrease with the increasing black hole spin parameter. This stays in agreement with previous results, in particular the model of accretion onto Reissner-Nordstr\"{o}m black holes \cite{aCpM20,pLjYsX25} and the planar model discussed in \cite{aCpMaO22}. This is also a general feature present in hydrodynamical models, e.g., in \cite{aAeToSdL21}. In contrast to the results of \cite{aCpMaO22}, both accretion rates decrease with increasing asymptotic temperature of the gas. This fact emphasizes the difference between planar and non-planar models, mostly due to their asymptotic properties. Further, we show that the angular momentum accretion rate is roughly proportional to the spin parameter, slowing the black hole rotation down. This stays in agreement with the results of \cite{aCpMaO22}, however it is not obvious---the hydrodynamical model of \cite{lPsSsT1988} is characterized by a vanishing angular momentum accretion rate.

Our solution is exact in the sense that the relevant observable quantities (such as the components of the particle current density or the components of the energy-momentum-stress tensor) can be expressed in the form of closed integrals. These integrals are then evaluated numerically. As an example, we compute the components of the particle current density, illustrating the morphology of the flow in the vicinity of the black hole.

This paper is organized as follows. In Section \ref{sec:preliminaries} we discuss relevant properties of timelike unbound geodesics in the Kerr spacetime, including the properties of the polar and radial geodesic motion. This section also allows us to establish the notation and conventions of this paper.
In Section~\ref{sec:observables} we define observable quantities associated with the kinetic description of the gas and focus on the characterization of phase-space regions available for particles moving along unbound trajectories. An essential element described in this chapter is a novel system of phase-space coordinates. We provide explicit expressions for the particle current density and introduce related quantities. Section~\ref{sec:accretionrates} discusses accretion rates that characterize the flow. In Section~\ref{sec:slowrotapprox} we investigate an analytic approximation of our model, valid for slowly rotating black holes. Numerical examples are discussed in Section~\ref{sec:numerics}, along with a comparison of exact results with the slow-rotation approximation. Section~\ref{sec:conclusions} contains a brief summary of our results. For clarity, many important but technical aspects of our analysis are described in the appendices. Appendix~\ref{App:Qc} demonstrates the properties of our new phase-space coordinates. In Appendices~\ref{App:QcAnalytic} and \ref{App:Qmax} we provide an exact parametrization of the phase-space regions associated with scattered and plunging orbits, including the Kerr separatrix expressed in our coordinate system. The smoothness properties of the observables are analyzed in Appendix~\ref{App:smoothness}, and some integrals used in our numerical calculations are collected in Appendix~\ref{App:Qintegrals}.

Throughout this paper we use geometrized units with $c = G = 1$, where $c$ denotes the speed of light, and $G$ is the gravitational constant. We assume the metric signature $(-,+,+,+)$.

\section{Preliminaries}
\label{sec:preliminaries}

In this section we briefly review (mostly) known facts about future-directed timelike geodesics in the Kerr spacetime that are relevant to this work. For more details on this topic, see for instance Refs.~\cite{Chandrasekhar-Book,ONeill-Book,pRoS24} and references therein. In terms of Kerr-like horizon penetrating coordinates\footnote{These coordinates are related to the Boyer-Lindquist coordinates $(t_\mathrm{BL},r_\mathrm{BL},\vartheta_\mathrm{BL},\varphi_\mathrm{BL})$ according to $r_\mathrm{BL} = r$, $\vartheta_\mathrm{BL} = \vartheta$ and
$$
t_\mathrm{BL} = t - 2M\int^r \frac{r dr}{\Delta},\qquad
\varphi_\mathrm{BL} = \varphi - a\int^r\frac{dr}{\Delta}.
$$
Note that
$$
\frac{\partial}{\partial t} = \frac{\partial}{\partial t_\mathrm{BL}},\quad
\frac{\partial}{\partial r} = \frac{\partial}{\partial r_\mathrm{BL}} - \frac{2M r}{\Delta}\frac{\partial}{\partial t_\mathrm{BL}} - \frac{a}{\Delta}\frac{\partial}{\partial \varphi_\mathrm{BL}},\quad
\frac{\partial}{\partial\vartheta} = \frac{\partial}{\partial \vartheta_\mathrm{BL}},\quad
\frac{\partial}{\partial\varphi} = \frac{\partial}{\partial \varphi_\mathrm{BL}}.
$$
} $(t,r,\vartheta,\varphi)$ the Kerr metric is given by
\begin{equation}
g = -dt^2 +dr^2 -2a\sin^2\vartheta dr d\varphi 
 + \left( r^2 + a^2 \right)\sin^2 \vartheta d\varphi ^2
 + \rho^2 d\vartheta ^2 
 + \frac{2M r}{\rho^2}  \left(dt + dr - a\sin ^2 \vartheta d\varphi \right)^2,
\label{Eq:Kerr}
\end{equation}
where $\rho^2 := r^2 + a^2\cos^2 \vartheta$, $M$ is the black hole mass, and $a$ is the rotation parameter, related to the black hole angular momentum $J$ by $a=J/M$. The inverse metric has components
\begin{equation}
(g^{\mu\nu}) = \frac{1}{\rho^2}\left( \begin{array}{cccc}
 -(\rho^2 + 2M r) & 2Mr & 0 & 0 \\
 2Mr & \Delta & 0 & a \\
 0 & 0 & 1 & 0 \\
 0 & a & 0 & \sin^{-2}\vartheta
\end{array} \right),
\label{Eq:KerrInverse}
\end{equation}
where $\Delta := r^2 -2M r + a^2$. We restrict ourselves to the subextremal family of black hole solutions, for which the rotational and mass parameters $a$ and $M$ satisfy $|a| < M$. Without loss of generality, we assume that $a$ is positive or zero, such that 
\begin{equation}
0\leq a < M,
\end{equation}
where the limit $a=0$ describes the Schwarzschild black hole. For the purpose of this work, we shall restrict ourselves to the region $r > r_-$ outside the Cauchy horizon, where $r_\pm := M \pm \sqrt{M^2 - a^2}$ denote the roots of $\Delta$ and $r_+$ refers to the event horizon radius. Note that the $t = \mathrm{const}$ surfaces are spacelike. We will also make use of the Killing vector fields
\begin{equation}
k := \frac{\partial}{\partial t},\qquad
\eta := \frac{\partial}{\partial\varphi}
\label{Eq:KVF}
\end{equation}
for the physical interpretation of our results.

Geodesic equations can be written in the Hamiltonian form
\begin{equation}
\label{HamiltonsEqs}
\frac{d x^\mu}{d \tilde \tau} = \frac{\partial H}{\partial p_\mu}, \qquad \frac{d p_\nu}{d \tilde \tau} = - \frac{\partial H}{\partial x^\nu},
\end{equation}
where $(x^\mu,p_\mu)$ are phase-space coordinates and the Hamiltonian $H$ reads $H(x,p) = \frac{1}{2}g^{\mu \nu}(x) p_\mu p_\nu$. The first equation implies that the four-momentum is given by
$p^\mu = g^{\mu\nu} p_\nu = d x^\mu/d \tilde \tau$ along the integral curves. We assume a standard normalization with $H = - \frac{1}{2}m^2$, where $m$ denotes the particle's rest mass. For $m > 0$ this implies that the affine parameter $\tilde \tau$ and the proper time $\tau$ are related by $\tilde \tau = \tau/m$.

The geodesic motion in the Kerr spacetime possesses four integrals of motion, given by
\begin{subequations}
\begin{eqnarray}
g^{\mu\nu} p_\mu p_\nu &=& -m^2,
\label{Eq:Ham}\\
p_t &=& -E,
\label{Eq:Energy}\\
p_\varphi &=& L_z,
\label{Eq:AngMom}\\
p_\vartheta^2 + \left(\frac{p_\varphi}{\sin\vartheta} + a\sin\vartheta p_t\right)^2 
+ m^2 a^2\cos^2\vartheta &=& L^2,
\label{Eq:Carter}
\end{eqnarray}
\end{subequations}
with $(E,L_z,L^2)$ describing the particle's energy, azimuthal angular momentum, and Carter constant, respectively. In the following, we denote by $L \geq 0$ the square root of $L^2$. Note that in the non-rotating limit $a=0$, $L$ reduces to the expression for the total angular momentum in a spherically symmetric spacetime.

The constants of motion $(m,E,L_z,L^2)$ allow one to separate the motion in phase space as follows:
\begin{subequations}
\begin{eqnarray}
(t,p_t) &:& p_t = -E,
\label{Eq:tPlaneRestriction}\\
(\varphi,p_\varphi) &:& p_\varphi = L_z,
\label{Eq:phiPlaneRestriction}\\
(\vartheta,p_\vartheta) &:& p_\vartheta^2 = \Theta(\vartheta),\qquad
\Theta(\vartheta) := L^2 - \left(\frac{L_z}{\sin\vartheta} - a\sin \vartheta E \right)^2 
 - a^2 m^2 \cos^2\vartheta,
\label{Eq:thetaPlaneRestriction}\\
(r,p_r) &:& \left( \Delta p_r - 2M E r + a L_z \right)^2 = R(r),\qquad
R(r) := \left[ E(r^2 + a^2 ) - a L _z  \right]^2 - \Delta\left(L^2 + m^2 r^2 \right).
\label{Eq:rPlaneRestriction}
\end{eqnarray}
\end{subequations}
Furthermore, Hamilton's equations of motion~(\ref{HamiltonsEqs}) imply $\rho^2 \frac{d\vartheta}{d\tilde{\tau}} = p_\vartheta$ and $\rho^2\frac{dr}{d\tilde{\tau}} = \Delta p_r - 2M E r + a L_z = \Delta p_{r_\mathrm{BL}}$. In what follows, we will frequently replace $L_z$ and $L$ with the new constants
\begin{equation}
\label{defhatlz}
\hat L_z := L_z - a E, \qquad
\beta := \frac{\hat L_z}{L},
\end{equation}
since this leads to a simple form of the effective potential describing the radial motion~\cite{Tejeda-PhD,eTpTjM13}. 

A recent analysis of Kerr geodesic equations in Kerr-like horizon penetrating coordinates (\ref{Eq:Kerr}) can be found in \cite{zBetal25}. For a version using Boyer-Lindquist coordinates see, e.g., \cite{aCeHpM23}. In the following subsections, we review the qualitative behaviors of the functions $\Theta(\vartheta)$ and $R(r)$ which are related to the effective potentials for the polar and radial motion, respectively.

\subsection{Properties of the polar motion}

The polar motion is characterized by the effective potential
\begin{equation}
K(\vartheta) := \left(\frac{L_z}{\sin\vartheta} - a\sin \vartheta E \right)^2  + a^2 m^2 \cos^2\vartheta,
\qquad
0 < \vartheta < \pi,
\end{equation}
and according to Eq.~(\ref{Eq:thetaPlaneRestriction}), the polar coordinate $\vartheta$ is restricted by the condition $K(\vartheta)\leq L^2$. As discussed, for example in Appendix~A of Ref.~\cite{pRoS24} and illustrated in Fig.~\ref{Fig:PolarPot}, the function $K$ possesses two types of behavior, depending on the values of $a$, $m$, $E$, and $L_z$. When $a^2(E^2-m^2)\leq L_z^2$ (case A), it describes an infinite potential well with a global minimum at $\vartheta=\pi/2$, where $K(\vartheta) = \hat{L}_z^2$. In contrast, when $a^2(E^2-m^2) > L_z^2$ (case B), it describes a double-well potential, with global minima at $\vartheta=\vartheta^*$ and $\vartheta=\pi-\vartheta^*$ where
\begin{equation}
K(\vartheta^*) = K(\pi - \vartheta^*) = \hat{L}_z^2 - \left( a\sqrt{E^2-m^2} - |L_z| \right)^2,
\label{Eq:SpecialCaseCond1}
\end{equation}
and a local maximum at $\vartheta=\pi/2$, where $K(\pi/2) = \hat{L}_z^2$.

\begin{figure}[ht]
\begin{center}
\includegraphics[width=0.49\textwidth]{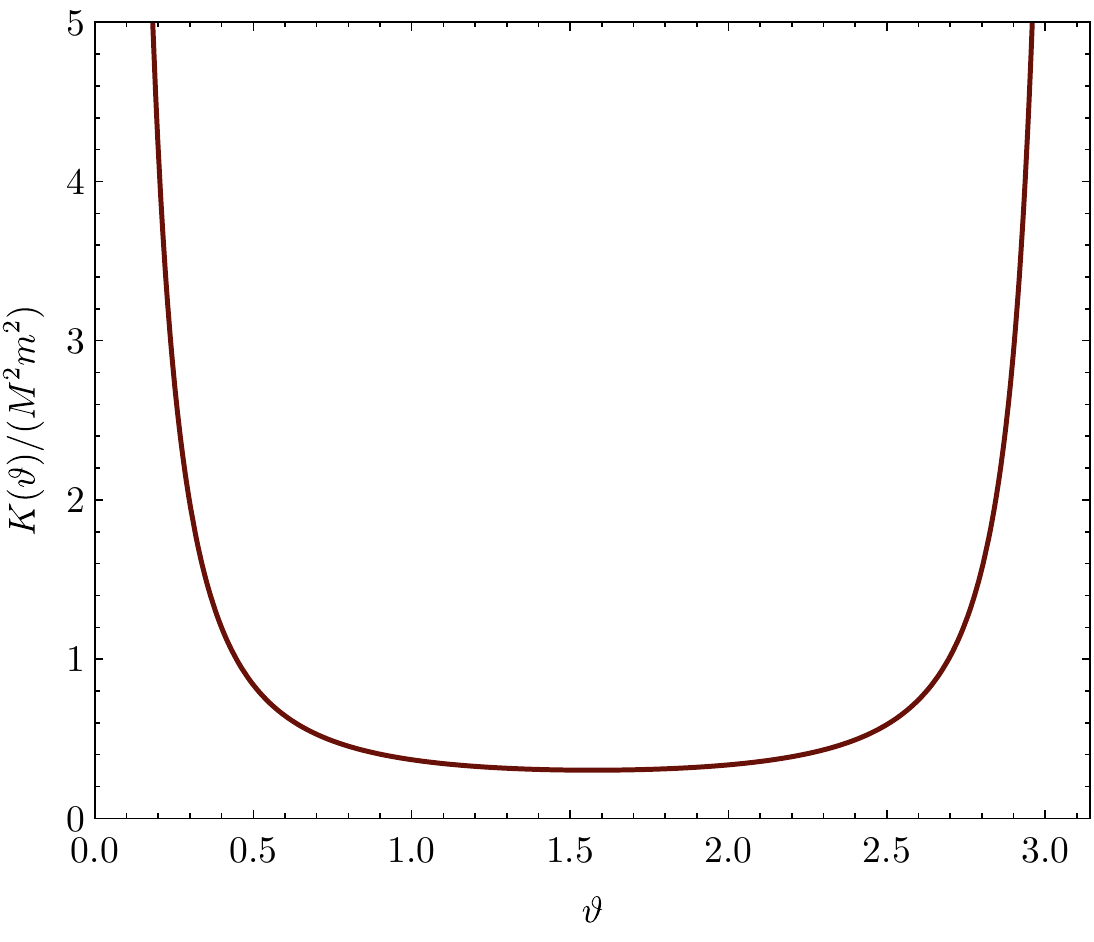}
\includegraphics[width=0.49\textwidth]{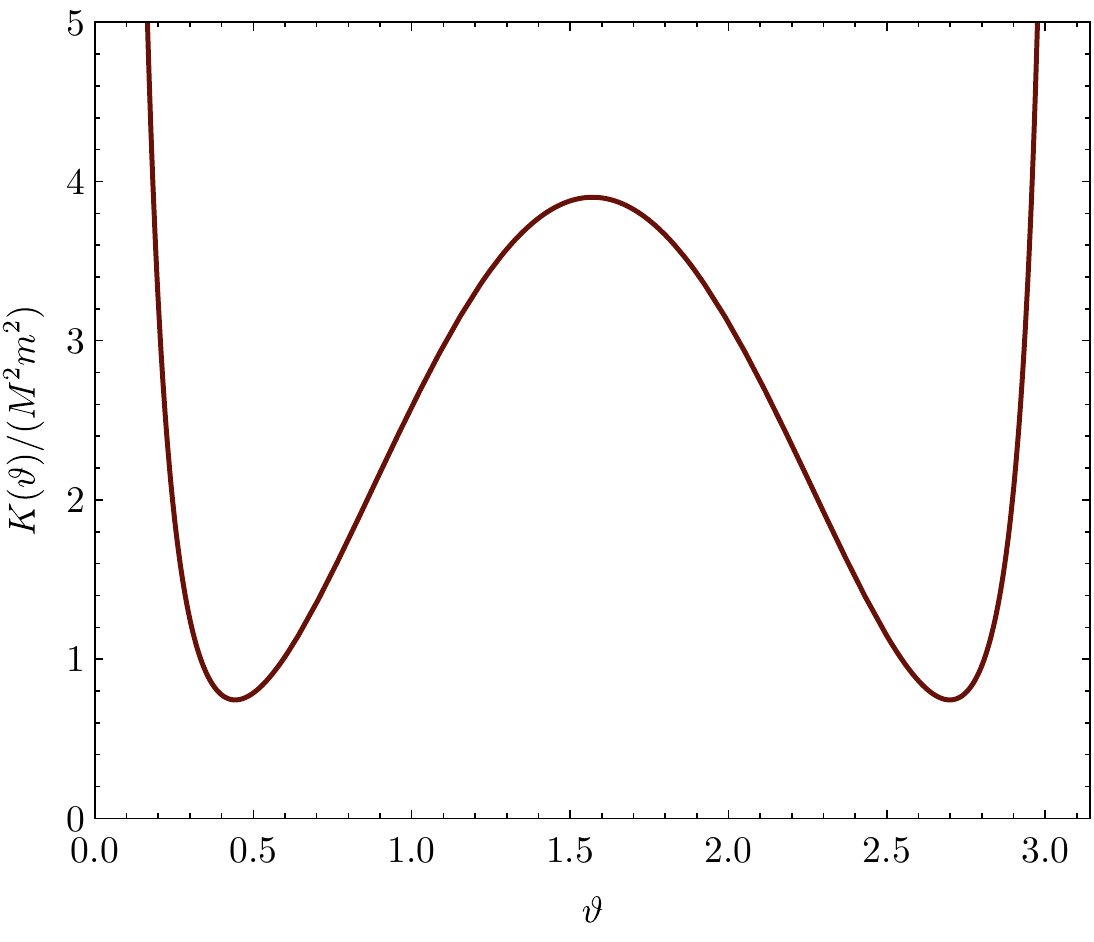}
\end{center}
\caption{Left plot (case A): Function $K(\vartheta)$ for the parameter values $a = 0.95M$, $L_z = 0.4Mm$, and $E = m$. There is a global minimum at $\vartheta = \pi/2$. Right plot (case B): Same as in the previous case except that the energy is $E = 2.5m$. The global minima are located at $\vartheta=\vartheta^*$ and $\vartheta = \pi - \vartheta^*$ with $\vartheta^* \approx 0.443$.}
\label{Fig:PolarPot}
\end{figure}

For bound orbits, $0 < E < m$, and thus only case A can occur, such that $L^2 \geq \hat{L}_z^2$. Consequently, in this case, the parameter $\beta$ defined in Eq.~(\ref{defhatlz}) is restricted to the interval $[-1,1]$,  the limits $\beta = 1$ and $\beta=-1$ describing prograde and retrograde equatorial orbits, respectively. In contrast, for the case of unbound orbits which is the relevant one in the present article, $E > m$, and hence both cases A and B can occur. In case B, the minimum value for $L^2$ can be smaller than $\hat{L}_z^2$ and thus $\beta$ is not necessarily restricted to the interval $[-1,1]$ anymore. Expressed in terms of $\hat{L}_z$, the inequality $L_z^2 < a^2(E^2 - m^2)$ is equivalent to
\begin{equation}
-a(E + \sqrt{E^2-m^2}) < \hat{L}_z < -a(E - \sqrt{E^2-m^2}),
\label{Eq:SpecialCaseCond2}
\end{equation}
and hence it follows that $\hat{L}_z$ is always negative in case B, and thus the value of $\beta$ can be smaller than $-1$. In the next subsection, we shall prove that such orbits, having $\beta < -1$, correspond to plunging particles.

\subsection{Properties of the radial motion}

\begin{figure}[ht]
\begin{center}
\includegraphics[width=0.49\textwidth]{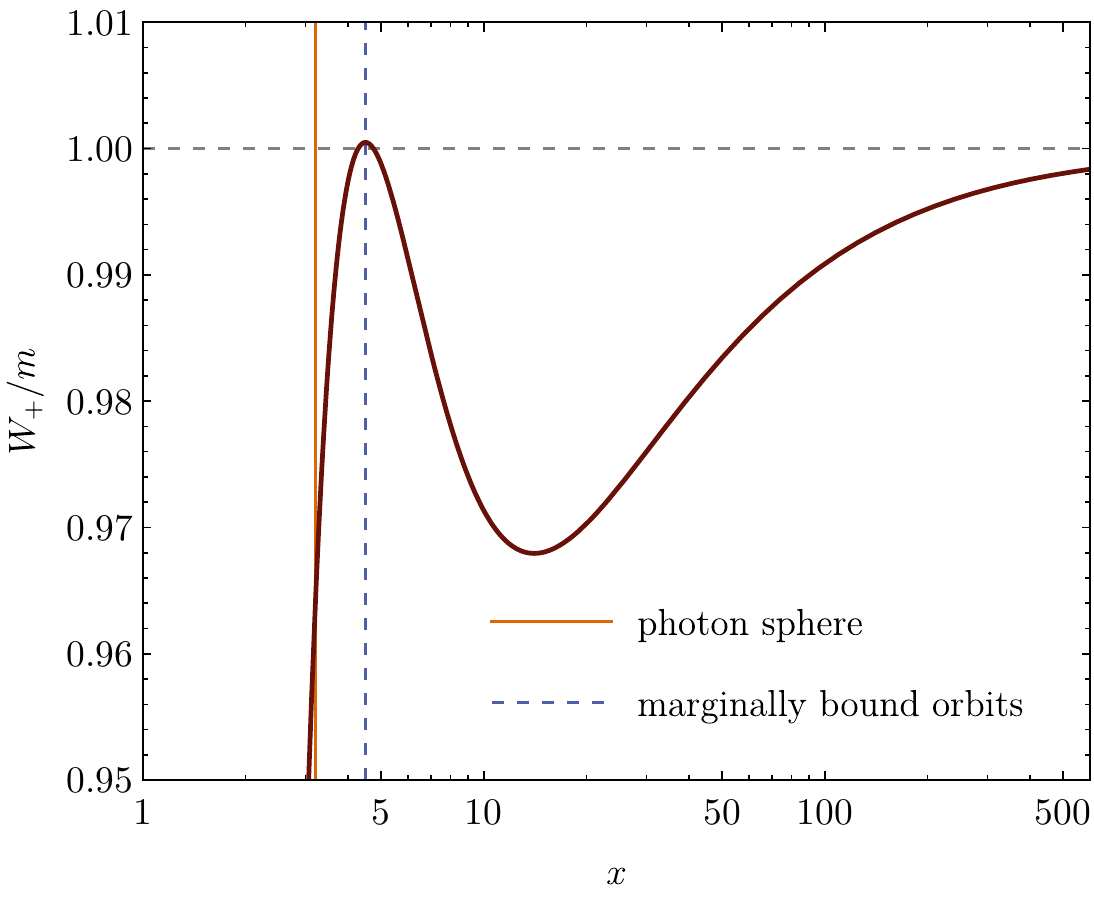}
\includegraphics[width=0.49\textwidth]{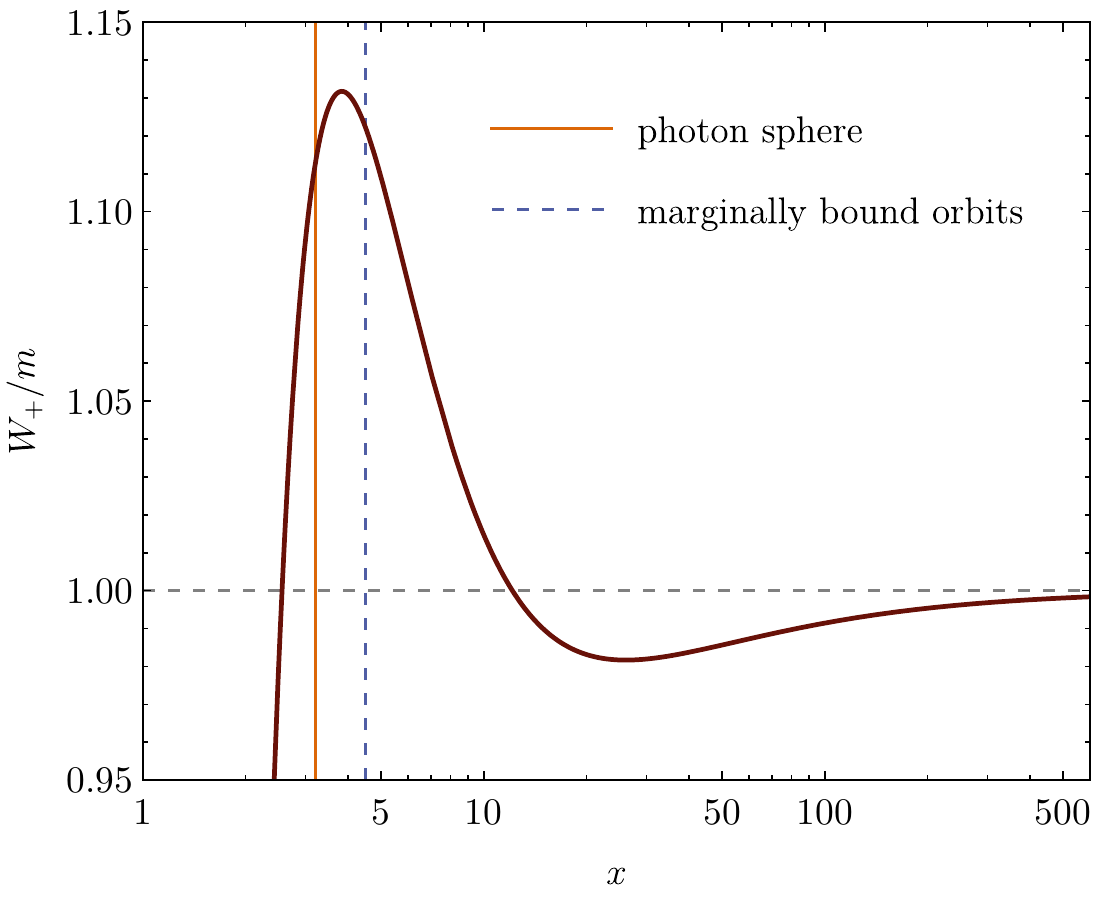}
\end{center}
\caption{\label{Fig:RadialPot} Left plot: Function $W_+(r)$ for the parameter values $a = 0.95M$, $\beta = \hat{L}_z/L = -0.6$, and $L = 4.83 Mm$. The maximum of the potential well approximately coincides with the asymptotic value $m$ and is located at $r\approx r_\mathrm{mb} \approx 4.505 M$. Right plot: Same as in the previous case except that $L = 6Mm$.}
\end{figure}

The function $R(r)$ characterizing the radial motion (see Eq.~(\ref{Eq:rPlaneRestriction})) can be factorized in the form
\begin{equation}
R(r) = r^4[ E - W_+(r)][ E - W_-(r) ],
\end{equation}
with
\begin{equation}
W_\pm(r) := \frac{a\hat{L}_z}{r^2} \pm \frac{\sqrt{\Delta(m^2 r^2 + L^2)}}{r^2},
\end{equation}
and thus the motion is restricted to the sets $W_+(r)\leq E$ or $W_-(r)\geq E$. As shown in~\cite{pRoS24}, only the former possibility corresponds to future-directed trajectories outside the horizon $r > r_+$, and thus
\begin{equation}
W_+(r) \leq E,\qquad
W_+(r) := \frac{a\hat{L}_z}{r^2} + \frac{\sqrt{\Delta(m^2 r^2 + L^2)}}{r^2}.
\end{equation}
The qualitative properties of the effective potential $W_+(r)$ as a function of the parameters $\beta := \hat{L}_z/L$ and $L$ have been thoroughly analyzed in Appendix~B in~\cite{pRoS24} for the case for which $|\beta|\leq 1$. For the following, it is useful to introduce the dimensionless quantities
\begin{equation}
\alpha := \frac{a}{M},\qquad
x := \frac{r}{M},\qquad
\varepsilon := \frac{E}{m},\qquad
\lambda := \frac{L}{Mm}.
\end{equation}
It follows from the analysis in~\cite{pRoS24} that for each $\beta\in [-1,1]$ there is a critical value $L_\mathrm{mb} = L_\mathrm{mb}(\alpha,\beta)$ such that for $L > L_\mathrm{mb}(\alpha,\beta)$ the effective potential $W_+(r)$ has a unique potential barrier with a maximum greater than $m$ and a unique well with a minimum smaller than $m$ (see Fig.~\ref{Fig:RadialPot} for an illustration). The location of the maximum lies in the interval $(r_\mathrm{ph},r_\mathrm{mb})$, with $r_\mathrm{ph} = r_\mathrm{ph}(\alpha,\beta)$ and $r_\mathrm{mb} = r_\mathrm{mb}(\alpha,\beta)$ denoting the radii of the photon and marginally bound orbits, respectively (see also the comments after Eq.\ (\ref{Eq:hDef}) below). As $L$ increases from $L_\mathrm{mb}$ to infinity, the location of the maximum moves continuously from $r_\mathrm{mb}$ to $r_\mathrm{ph}$ and the maximum value increases monotonically from $m$ to $\infty$. For convenience, we will also denote $x_\pm := r_\pm/M$, $x_\mathrm{ph} := r_\mathrm{ph}/M$, $x_\mathrm{mb} := r_\mathrm{mb}/M$, etc.

The extrema of the effective potential describe spherical orbits (i.e. orbits with constant value of $r$, which however are in general not confined to a plane unless $\beta=\pm 1$ or $\alpha=0$), and they can be parametrized in terms of their dimensionless radii $x > x_\mathrm{ph}$ as follows (see Eqs.\ (B12) and (B13) in~\cite{pRoS24})\footnote{For orbits confined to the equatorial plane one has $\beta = \pm 1$ and Eqs.\ (\ref{lsph}) and (\ref{esph}) reduce to the following expressions
\begin{subequations}
\begin{eqnarray}
    \lambda_\mathrm{sph}(x) & = & \frac{x^{5/4} - \mathrm{sgn}(\beta) \alpha x^{3/4}}{\sqrt{2 \mathrm{sgn} (\beta) \alpha + \sqrt{x}(x-3)}}, \\
    \varepsilon_\mathrm{sph}(x) & = & \frac{\sqrt{x}(x-2) + \mathrm{sgn}(\beta) \alpha}{x^{3/4} \sqrt{2 \mathrm{sgn} (\beta) \alpha + \sqrt{x}(x-3)}},
\end{eqnarray}
\end{subequations}
which coincide with Eqs.\ (53) in \cite{aCpMaO22} (see also \cite{jBwPsT72}).}
\begin{subequations}
\begin{eqnarray}
\label{lsph}
    \lambda_\mathrm{sph}(x) & = & x \frac{\sqrt{x - \alpha^2 (1 - \beta^2)} - \alpha \beta}{\sqrt{h(x)}}, \\
    \label{esph}
    \varepsilon_\mathrm{sph}(x) & = & \frac{x^2 - 2 x + \alpha^2 (1 - \beta^2) + \alpha \beta \sqrt{x - \alpha^2 (1 - \beta^2)}}{x \sqrt{h(x)}},
\end{eqnarray}
\end{subequations}
where
\begin{equation}
h(x) = x (x - 3) + 2 \alpha^2 (1 - \beta^2) + 2 \alpha \beta \sqrt{x - \alpha^2 (1 - \beta^2)}.
\label{Eq:hDef}
\end{equation}
The unique zero of the function $h$ on the interval $(x_+,\infty)$ corresponds to the dimensionless radius $x_\mathrm{ph}$ of the photon orbit, with $h$ being negative on $(x_+,x_\mathrm{ph})$ and positive on $(x_\mathrm{ph},\infty)$. The dimensionless radius $x_\mathrm{mb}$ of the marginally bound orbit satisfies the relation $\varepsilon_\mathrm{sph}(x_\mathrm{mb}) = 1$.

For the situation analyzed in this article, it is the maximum of $W_+(r)$ which plays a crucial role, since it distinguishes between unbounded trajectories emanating from the asymptotic region which are absorbed by the black hole and those that are scattered at the potential barrier.

The analysis carried out in~\cite{pRoS24} was restricted to the interval $-1\leq\beta\leq 1$ since that work focused on bound orbits. However, as discussed in the previous subsection, for unbound orbits $\beta < -1$ is possible in case B. We now show that $\beta < -1$
can only occur when $R(r) > 0$ for all $r\geq r_+$, which implies that this situation corresponds to orbits that plunge into the black hole. To prove this, we use $L^2 < \hat{L}_z^2$ and $\Delta < r^2$ for $r\geq r_+$ and estimate
\begin{eqnarray}
R(r) &=& (Er^2 - a\hat{L}_z)^2 - \Delta(L^2 + m^2 r^2) \nonumber \\ &>&  (Er^2 - a\hat{L}_z)^2 - r^2(\hat{L}_z^2 + m^2 r^2) = r^2\left[ (E^2-m^2)r^2 - L_z^2 \right] + a^2(\hat{L}_z^2 + E^2 r^2).
\end{eqnarray}
Since $r > a$ and $(E^2 - m^2)a^2 > L_z^2$, the claim follows. Note that $R(r) > 0$ and $E > m$ also imply $W_+(r) < E$ for all $r\geq r_+$, such that orbits characterized by $\beta < -1$ are, in fact, future-directed plunging orbits. They are characterized by the inequalities~(\ref{Eq:SpecialCaseCond2}) and $K(\vartheta^*) < L^2 < \hat{L}_z^2$, with $K(\vartheta^*)$ as in Eq.~(\ref{Eq:SpecialCaseCond1}).

\subsection{Absorbed and scattered trajectories}
\label{SubSec:AbsorbedScattered}

From the previous discussions, we arrive at the following conclusions, which are important for the remainder of this paper. Future-directed timelike trajectories emanating from infinity are characterized by $E > m$. They are scattered at the effective potential if
\begin{equation}
L > L_c(\alpha,\beta,E),\qquad
\beta := \hat{L}_z/L\in [-1,1],
\label{Eq:Lc}
\end{equation}
and otherwise they are plunging into the black hole. Here, $L_c(\alpha,\beta,E)$ refers to the unique value of $L$ corresponding to the case in which the maximum of the effective potential $W_+$ is equal to $E$. For fixed values of $\alpha$ and $\beta$ it can be given in parametric form as follows:
\begin{equation}
r_c = M x_c,\qquad
E_c = m\varepsilon_\mathrm{sph}(x_c),\quad
L_c = M m\lambda_\mathrm{sph}(x_c),
\qquad x_\mathrm{ph} < x_c < x_\mathrm{mb}.
\end{equation}
Since $\varepsilon_\mathrm{sph}$ and $\lambda_\mathrm{sph}$ are monotonically  decreasing functions in the interval $(x_\mathrm{ph},x_\mathrm{mb})$, the specification of $E_c > m$ yields a unique value of $x_c$, and this in turn uniquely determines $L_c$. Furthermore, for fixed values of $\alpha$ and $\beta$, the value of $L_c$ increases monotonically with $E$.

In the Schwarzschild limit $a=0$, expressions for $L_c$ and $x_c$ become independent of $\beta$ and can be written in an explicit form as
\begin{equation}
L_c = M m\frac{x_c}{\sqrt{x_c-3}},
\qquad
x_c = \frac{8}{\sqrt{9\varepsilon_c^4 - 8\varepsilon_c^2} + 4 - 3\varepsilon_c^2},
\label{Eq:LcSchwarzschild}
\end{equation}
where $E_c = m \varepsilon_c$. As $\varepsilon_c$ increases from $1$ to $\infty$, $x_c$ decreases from $4$ to $3$ and $L_c$ increases from $4M m$ to $\infty$.

\section{Spacetime observables for the accretion of a Vlasov gas}
\label{sec:observables}

In this section and the following ones, we analyze a collisionless gas consisting of identical classical point particles of mass $m$. Let $(\mathcal M, g)$ denote the spacetime manifold. The cotangent space at $x \in \mathcal M$ and the cotangent bundle will be denoted by $T^\ast_x \mathcal M$ and $T^\ast \mathcal M$, respectively. The state of the gas is described by the one-particle distribution function $f \colon \Gamma_m^+ \to \Real$, which is defined on the future mass shell
\begin{equation}
\Gamma_m^+ = \{ (x,p) \in T^\ast \mathcal{M} \colon g^{\mu\nu}(x) p_\mu p_\nu = - m^2, \, p \text{ is future-directed} \},
\end{equation}
and satisfies the collisionless Boltzmann (or Vlasov) equation\footnote{See, for instance, Refs.~\cite{oStZ14b,rAcGoS22} for recent reviews on the manifestly covariant formulation of relativistic kinetic theory on the tangent or cotangent bundle, and Ref.~\cite{hA11} for a review on mathematical results on the Einstein-Vlasov system.}
\begin{equation}
g^{\mu \nu} p_\nu \frac{\partial f}{\partial x^{\mu}}
 - \frac{1}{2}\frac{\partial g^{\alpha\beta}}{\partial x^{\mu}}p_{\alpha}p_{\beta}
 \frac{\partial f}{\partial p_{\mu}} = 0.
 \label{Eq:Vlasov}
\end{equation}

In this work, we search for a stationary and axially symmetric solution of Eq.\ (\ref{Eq:Vlasov}), taking into account unbound particle trajectories, emanating from infinity. As discussed in the previous section, such trajectories can either plunge into the black hole (we will usually refer to these trajectories as absorbed ones), or they can be scattered at the centrifugal potential and return to infinity. We neglect a zero-measure set of trajectories that asymptote to bound spherical orbits. For simplicity, we consider a uniform asymptotic distribution of the gas, described by the one-particle distribution function $F_\infty(E)$ depending only on the energy $E$ of the particles; however, most of the formulas derived below are also valid if $F_\infty$ is a function of the conserved quantities $E$, $L$, and $L_z$. Equation~(\ref{Eq:Vlasov}) then implies that the one-particle distribution function $f(x,p)$ must be globally of the form
\begin{equation}
f(x,p) = F_\infty(E),
\end{equation}
when restricted to the region of the phase space corresponding to scattered or absorbed orbits. For simplicity, we assume that $f$ is identically zero in the complementary region (which includes bound orbits and orbits emanating from the white hole).

The main challenge of the analysis presented in this paper consists in computing the relevant spacetime observables: the particle current density and the energy-momentum-stress tensor, defined as
\begin{equation}
J_\mu(x) := \int\limits_{P_x^+(m)} p_\mu f(x,p) \mbox{dvol}_x(p),\qquad
T_{\mu\nu}(x) := \int\limits_{P_x^+(m)} p_\mu p_\nu f(x,p) \mbox{dvol}_x(p).
\label{Eq:Observables}
\end{equation}
Here
\begin{equation}
P_x^+(m) = \{ p \in T_x^\ast \mathcal{M} \colon g^{\mu\nu}(x) p_\mu p_\nu = - m^2, \, p \text{ is future-directed} \},
\end{equation}
denotes the future mass hyperboloid at $x$ and $\mbox{dvol}_x(p)$ is the volume form on $P_x^+(m)$ (see Eq.~(\ref{Eq:dvolx}) below). One of the difficulties faced when computing the integrals~(\ref{Eq:Observables}) is related to the correct identification of the regions in the phase space $\Gamma^+_m$ corresponding to scattered and absorbed particle trajectories, originating at infinity. In practice, a restriction to suitable regions in $\Gamma^+_m$ is achieved by specifying appropriate integration limits. To treat this problem, it is convenient to first re-express the integrals in terms of the constants of motion $(E,L,L_z)$, since they allow for a characterization of the relevant types of orbits. However, it turns out that another change of variables will be necessary to avoid a complication arising from the different types of behavior of the effective potential $K(\vartheta)$ corresponding to the polar motion.

\subsection{Representation in terms of the constants of motion}

Let us first express the volume form $\mbox{dvol}_x(p)$ in terms of the constants of motion $E$, $L$, and $L_z$. For this, we start with the general expression (see, for instance, Eq.~(C3) in Ref.~\cite{rAcGoS22})
\begin{equation}
\mbox{dvol}_x(p) = \sqrt{-\det(g^{\mu\nu})} \frac{dp_r dp_\vartheta dp_\varphi}{p^t}.
\label{Eq:dvolx}
\end{equation}
From Eq.~(\ref{Eq:KerrInverse}) one finds $\sqrt{-\det(g^{\mu\nu})} = 1/(\rho^2\sin\vartheta)$ and
\begin{equation}
p^t = \left( 1 + \frac{2M r}{\rho^2} \right)E + \frac{2M r}{\rho^2} p_r,
\end{equation}
such that
\begin{equation}
\mbox{dvol}_x(p) = \frac{dp_r dp_\vartheta dp_\varphi}{\left[ (\rho^2 + 2M r) E + 2M r p_r \right]\sin\vartheta}.
\end{equation}
Next, using Eqs.~(\ref{Eq:phiPlaneRestriction}), (\ref{Eq:thetaPlaneRestriction}), and (\ref{Eq:rPlaneRestriction}) one finds
\begin{equation}
dp_r dp_\vartheta dp_\varphi = \frac{L}{\sqrt{R(r)}\sqrt{L^2 - K(\vartheta)}}\left[ 2M r p_r + (\rho^2 + 2M r)E \right] dE dL dL_z,
\end{equation}
such that
\begin{equation}
\mbox{dvol}_x(p) = \frac{dE (L dL) dL_z}{\sin\vartheta\sqrt{R(r)}\sqrt{L^2 - K(\vartheta)}}.
\end{equation}
Together with
\begin{equation}
p_t = -E,\quad
p_r = \frac{2M E r - aL_z \pm \sqrt{R(r)}}{\Delta},\quad
p_\vartheta = \pm \sqrt{L^2 - K(\vartheta)},\quad
p_\varphi = L_z,
\label{Eq:FourMomentum}
\end{equation}
and the identification of the correct ranges for $(E,L,L_z)$ this allows one, in principle, to express the observables~(\ref{Eq:Observables}) in terms of explicit integrals. However, as mentioned previously, a practical problem with this approach is the correct identification of the integration limits, which is enhanced by the fact that the effective potential $K(\vartheta)$ exhibits two distinct behaviors. For this reason, in the following we introduce a new parametrization which replaces the conserved quantities $(L,L_z)$ with new variables $(Q,\chi)$, whose integration range is simpler to determine.

\subsection{New parametrization}
\label{SubSec:NewParametrization}

The new parametrization can be introduced by rewriting condition~(\ref{Eq:thetaPlaneRestriction}) in the form
\begin{equation}
p_\vartheta^2 + \left(\frac{L_z}{\sin\vartheta} - a\sin \vartheta E \right)^2 = L^2 - a^2 m^2\cos^2\vartheta.
\end{equation}
This suggests substituting for $L$ and $L_z$ a new variable $Q > 0$ and an angle $\chi$, such that $Q^2 = L^2 - a^2 m^2 \cos^2\vartheta$ and
\begin{subequations}
\begin{eqnarray}
p_\vartheta &=& Q\cos\chi,\\
L_z &=& Q\sin\vartheta\sin\chi + a\sin^2\vartheta E.
\end{eqnarray}
\end{subequations}
Note that in the Schwarzschild limit of $a=0$, the angle $\chi$ coincides with the one introduced in~\cite{pRoS16}. Also, note that for points lying on the equatorial plane ($\vartheta=\pi/2$) one has $L = Q$ and $\hat{L}_z = Q\sin\chi$, such that in this particular case $\beta = \hat{L}_z/L = \sin\chi$.

In terms of the variables $(E,Q,\chi)$, the Lorentz-invariant volume form on the future mass hyperboloid $P_x^+(m)$ has the form
\begin{equation}
\mbox{dvol}_x(p) = \frac{dE (Q dQ) d\chi}{\sqrt{R(r)}},
\end{equation}
where expressed in terms of $Q$ and $\chi$,
\begin{equation}
R = \left( E\rho^2 - aQ\sin\vartheta\sin\chi \right)^2 - \Delta(Q^2 + m^2\rho^2).
\label{Eq:RQchi}
\end{equation}
Furthermore, the components of the momentum $p$ and its associated vector field are\footnote{In terms of Boyer-Lindquist coordinates one obtains exactly the same expressions as in Eq.~(\ref{Eq:FourMomentumBis}) except for $p_{r_\mathrm{BL}} = \pm \sqrt{R}/\Delta$.}
\begin{equation}
p_t = -E,\quad
p_r = -E + \Pi_\pm,\quad
p_\vartheta = Q\cos\chi,\quad
p_\varphi = Q\sin\vartheta\sin\chi + a\sin^2\vartheta E,
\label{Eq:FourMomentumBis}
\end{equation}
and
\begin{equation}
p^t = E + \frac{2Mr}{\rho^2}\Pi_\pm,\quad
p^r = \frac{\pm\sqrt{R}}{\rho^2},\quad
p^\vartheta = \frac{Q\cos\chi}{\rho^2},\quad
p^\varphi = \frac{Q\sin\chi}{\rho^2\sin\vartheta} + \frac{a}{\rho^2}\Pi_\pm,
\label{Eq:FourMomentumUp}
\end{equation}
where we have defined
\begin{equation}
\Pi_\pm := \frac{\pm \sqrt{R} + E\rho^2  - a Q\sin\vartheta\sin\chi}{\Delta}.
\end{equation}
Note that $\Pi_-$ is regular at the horizon; it can be rewritten in a manifestly regular way as follows:
\begin{equation}
\Pi_- = \frac{Q^2 + m^2\rho^2}{\sqrt{R} + E\rho^2  - a Q\sin\vartheta\sin\chi}.
\label{Eq:Pim}
\end{equation}

\begin{figure}
    \begin{center}
    \includegraphics[width=0.49\textwidth]{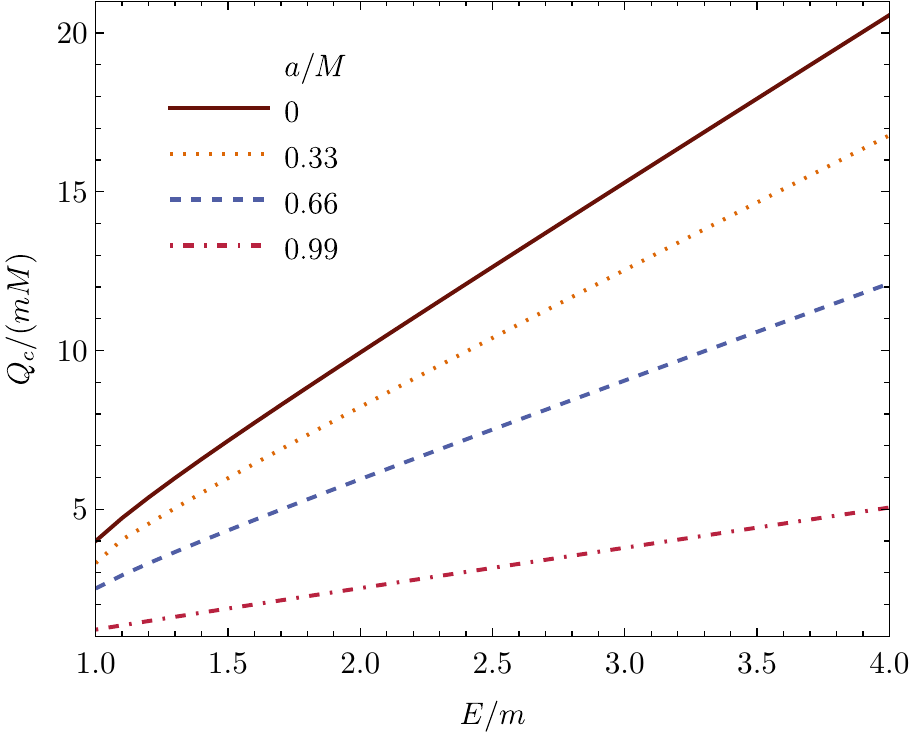}
    \includegraphics[width=0.49\textwidth]{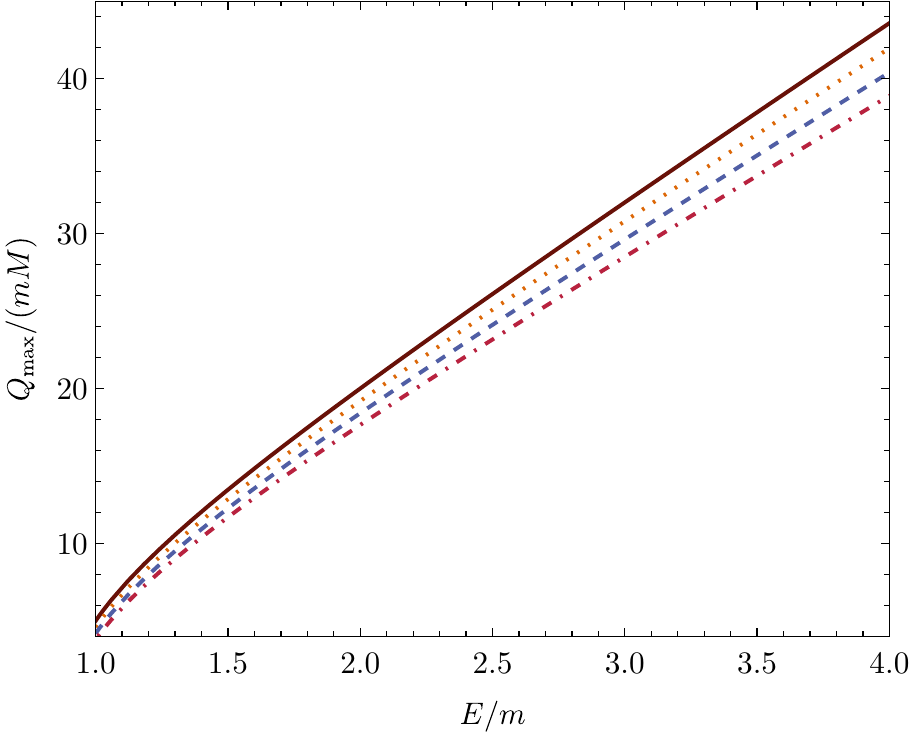}
    \end{center}
    \caption{\label{fig:QcQmax}
    Profiles of critical values of $Q$: $Q_c$ (left panel) and $Q_\mathrm{max}$ (right panel) versus the energy $\varepsilon=E/m$ for various values of the rotation parameter $\alpha=a/M$ and fixed $\vartheta=\pi/2$, $\chi=\pi/2$, and $r=10M$. }
\end{figure}

The integration ranges for $E$, $Q$, and $\chi$ are analyzed in detail in Appendices~\ref{App:Qc}--\ref{App:Qmax}. They can be summarized as follows: $E > m$, $0 < \chi < 2\pi$, and $0 < Q < Q_c$ for plunging orbits, whereas $Q_c < Q < Q_\mathrm{max}$ for scattered orbits. Expressions for $Q_c$ and $Q_\mathrm{max}$ are derived in the appendices. Here, we just mention their most important properties that are necessary for the correct computation of the observables: 
\begin{enumerate}
\item[(i)] For $E > m$, $Q_c$ is a smooth function of $a$, $\vartheta$, $\sin\chi$, and $E$. Simlarly, for $E > m$ and $r > r_c$, $Q_\mathrm{max}$ is a smooth function of $a$, $r$, $\vartheta$, $\sin\chi$, and $E$.

\item[(ii)] For points lying on the equatorial plane one has $L=Q$ and $\beta=\sin\chi$, and thus
\begin{equation}
Q_c = L_c(\alpha,\sin\chi,E),\qquad
Q_\mathrm{max} = L_\mathrm{max}(\alpha,r,\sin\chi,E),
\end{equation}
where $L_c(\alpha,\beta,E)$ is defined below Eq.~(\ref{Eq:Lc}) and $L_\mathrm{max}(\alpha,r,\sin\chi,E)$ is given by
\begin{equation}
\label{Lmaxeqplane}
L_\mathrm{max}(\alpha,r,\sin\chi,E)
 = \frac{r(r^2 E^2 - \Delta m^2)}{r E a\sin\chi + \sqrt{\Delta}\sqrt{r^2 E^2 - \Delta m^2 + m^2 a^2\sin^2\chi}}.
\end{equation}
In particular, for trajectories which are confined to the equatorial plane, such that $p_\vartheta=0$ in addition to $\vartheta=\pi/2$, one has $\chi=\pi/2$ (prograde) or $\chi=3\pi/2$ (retrograde), and Eq.\ (\ref{Lmaxeqplane}) reduces to the known formula (see, e.g., Eq.~(62) in \cite{aCpMaO22}).

\item[(iii)] In the Schwarzschild limit $\alpha=0$, $Q_c = L_c(E)$, and $Q_\mathrm{max} = L_\mathrm{max}(r,E)$ are independent of $\alpha$, $\sin\chi$, and $\vartheta$ and reduce to the corresponding expressions  in~\cite{pRoS16}.

\item[(iv)] The function $Q_c$ can be given analytically in parametric form (see Appendix~\ref{App:QcAnalytic}) and explicitly in the slowly rotating case (see Sec.~\ref{sec:slowrotapprox}). The function $Q_\mathrm{max}$ can be computed analytically (see Appendix~\ref{App:Qmax}):
\begin{equation}
Q_\mathrm{max} = \frac{\rho(\rho^2 E^2 - \Delta m^2)}{\rho E a\sin\vartheta\sin\chi + \sqrt{\Delta}\sqrt{\rho^2 E^2 - \Delta m^2 + m^2 a^2\sin^2\vartheta\sin^2\chi}}.
\label{Eq:Qmax}
\end{equation}
\end{enumerate}
The behaviors of $Q_c$ and $Q_\mathrm{max}$ as functions of $E$ (for fixed values of $\alpha$, $\vartheta$, and $\chi$) are illustrated in Fig.~\ref{fig:QcQmax}.

\subsection{Explicit expressions for the observables}

After these observations we are ready to give explicit expressions for the spacetime observables defined in Eq.~(\ref{Eq:Observables}). For definiteness, we only specify expressions for the particle current density; the procedure for computing the energy-momentum-stress tensor is similar. Hence, in what follows we focus on the particle current density, specifying separate formulas for the contributions arising from the absorbed and scattered particles. Here, one should take into account that in the former case only the minus sign should be considered in the expression for $p_r$ in Eq.~(\ref{Eq:FourMomentumBis}) (which is regular at $r = r_+$), while in the latter case one should take into account the contribution from both sign choices. Finally, when computing the integral over scattered orbits, one has to make sure that only orbits lying to the right of the potential barrier are taken into account. This can be assured by requiring that only orbits with $r_c < r$ contribute to the integrals.

Taking all these observations into account, one finds
\begin{subequations}
\label{Eq:Jmu}
\begin{eqnarray}
J_t^\mathrm{(abs)} &=& -\int\limits_m^\infty dE E F_\infty(E) \int\limits_0^{2\pi} d\chi \int\limits_0^{Q_c} \frac{Q dQ}{\sqrt{R}},
\label{Eq:Jtabs}\\
J_t^\mathrm{(scat)} &=& -2\int\limits_m^\infty dE E F_\infty(E) \int\limits_0^{2\pi} d\chi 1_{r_c < r}
\int\limits_{Q_c}^{Q_\mathrm{max}} \frac{Q dQ}{\sqrt{R}}, 
\label{Eq:Jtscat}\\
J^r_\mathrm{(abs)} &=& - \frac{1}{\rho^2} \int\limits_m^\infty dE F_\infty(E) \int\limits_0^{2\pi} d \chi \int\limits_0^{Q_c} Q dQ,
\label{Eq:Jrabs}\\
J^r_\mathrm{(scat)} &=& 0, 
\label{Eq:Jrscat}\\
J_\vartheta^\mathrm{(abs)} &=& 0,
\label{Eq:Jthetaabs}\\
J_\vartheta^\mathrm{(scat)} &=& 0,
\label{Eq:Jthetascat}\\
J_\varphi^\mathrm{(abs)} &=& - a \sin^2 \vartheta J_t^\mathrm{(abs)} + \sin \vartheta\int\limits_m^\infty dE F_\infty(E) \int\limits_0^{2\pi} d\chi \sin \chi \int\limits_0^{Q_c} \frac{Q^2 dQ}{\sqrt{R}}, 
\label{Eq:Jphiabs}\\
J_\varphi^\mathrm{(scat)} &=& - a \sin^2 \vartheta J_t^\mathrm{(scat)} + 2 \sin \vartheta\int\limits_m^\infty dE F_\infty(E) \int\limits_0^{2\pi} d\chi 1_{r_c < r}\sin\chi \int\limits_{Q_c}^{Q_\mathrm{max}} \frac{Q^2 dQ}{\sqrt{R}}.
\label{Eq:Jphiscat}
\end{eqnarray}
\end{subequations}
In the expressions for $J_t^\mathrm{(scat)}$ and $J_\varphi^\mathrm{(scat)}$, the function $1_{r_c < r}$ is one if $r_c < r$ and zero otherwise, and it ensures that only orbits lying to the right of the potential barrier are taken into account, as explained above. The reason for the vanishing of the polar components originates from the vanishing of the integrals
\begin{equation}
\int\limits_0^{2\pi} d\chi \cos\chi \int\limits_0^{Q_c} \frac{Q^2 dQ}{\sqrt{R}},\qquad
\int\limits_0^{2\pi} d\chi 1_{r_c < r} \cos\chi \int\limits_0^{Q_c} \frac{Q^2 dQ}{\sqrt{R}},
\end{equation}
which in turn is a consequence of the fact that $R$, $r_c$, and $Q_c$ are smooth functions of $\sin\chi$.

For $r > r_+$, the components $J^t$, $J^\varphi$, and $J_r$ can be obtained from Eq.\ (\ref{Eq:Jmu}) using the transformation
\begin{equation}
\begin{pmatrix}
J^t \\ J^\varphi \\ J_r
\end{pmatrix}
 = \frac{1}{\Delta}\begin{pmatrix}
  -\frac{\Sigma}{\rho^2} & -\frac{2aMr}{\rho^2} & 2Mr \\
  -\frac{2 aMr}{\rho^2} & \frac{\Delta-a^2\sin^2\vartheta}{\rho^2\sin^2\vartheta} & a \\
  -2M r & -a & \rho^2
 \end{pmatrix}
\begin{pmatrix}
J_t \\ J_\varphi \\ J^r
\end{pmatrix},
\label{Eq:JTransform}
\end{equation}
where we have defined
\begin{equation}
\Sigma := (r^2 + a^2)^2 - a^2\Delta\sin^2\vartheta.
\label{Eq:DefSigma}
\end{equation}
In spite of the singular factor $1/\Delta$, all the components of $J$ are regular at the horizon, as follows from Eqs.~(\ref{Eq:FourMomentumBis}), (\ref{Eq:FourMomentumUp}), and (\ref{Eq:Pim}).

Equations~(\ref{Eq:Jmu}) hold for any distribution function depending only on the constants of motion, that is, they also hold if $F_\infty(E)$ is replaced with a function depending on $E$, $L$, and $L_z$. However, assuming that $F_\infty$ only depends on the energy allows one to carry out the integrals over $Q$ explicitly. This can be done by making use of the integral formulas
\begin{eqnarray}
\int\frac{Q dQ}{\sqrt{R}} &=& -\frac{\sqrt{R}}{A} - \frac{B}{A}S,
\label{Eq:QInt1}\\
\int\frac{Q^2 dQ}{\sqrt{R}} &=&
 -\left( \frac{Q}{2A} + \frac{3B}{4A^2}\right)\sqrt{R} - \left( \frac{3B^2}{4A^2} + \frac{C}{A} \right) S,
\label{Eq:QInt2}
\end{eqnarray}
where we write $R = -A Q^2 + B Q + C = -A(Q-Q_+)(Q-Q_-)$ with
\begin{equation}
A = \Delta - a^2\sin^2\vartheta\sin^2\chi,\qquad
B = -2\rho^2 E a\sin\vartheta\sin\chi,\qquad
C = \rho^2(\rho^2 E^2 - \Delta m^2),
\label{Eq:RQPoly}
\end{equation}
and where we have defined
\begin{equation}
S := \frac{1}{\sqrt{A}}\arctan\left( \sqrt{\frac{Q_+ - Q}{Q - Q_-}} \right)
 = \frac{1}{\sqrt{-A}}\mbox{arctanh}\left( \sqrt{\frac{Q_+ -Q}{Q_- - Q}} \right).
\end{equation}
Here, the first expression for $S$ should be used when $A > 0$ and $Q_- < Q < Q_+$, whereas the second one should be used when $A < 0$ and $Q < Q_+ < Q_-$ or $Q > Q_+ > Q_-$. Note that $S = 0$ for $Q=Q_+$. Note also that for unbound orbits located outside the Cauchy horizon ($r > r_-$), the constant $C$ is always positive, since $E > m$ and $\rho^2 > \Delta$. This fact follows directly from the following estimates:
\begin{equation}
    \rho^2 E^2 - \Delta m^2 > m^2 (\rho^2 - \Delta) = m^2 (2 M r - a^2 \sin^2 \vartheta) \ge m^2 (2 M r - a^2).
\end{equation}
For $r > r_-$, we have
\begin{equation}
m^2 (2 M r - a^2) > m^2 (2 M^2 - 2 M \sqrt{M^2 - a^2} - a^2) \ge 0,
\end{equation}
since $0 \le a < M$. In contrast to $C$, which is always positive, $A$ admits both signs. Explicit expressions for the roots $Q_\pm$ are given in Appendix~\ref{App:Qmax}. They are always real and distinct from each other outside the event horizon $r > r_+$, and since $-A Q_+ Q_- = C> 0$, it follows that $Q_\pm$ have opposite signs for $A > 0$ and equal signs for $A < 0$. Further details on the integrals, including the case with $A=0$, are given in Appendix~\ref{App:Qintegrals}.

In the Schwarzschild limit $a=0$, one has $A = \Delta$, $B=0$, and $C = r^2(r^2 E -\Delta m^2)$, and the integrals over $Q$ (as well as $r_c$) are independent of $\chi$. The integration over $\chi$ gives a factor of $2 \pi$ except in the expressions for $J_\varphi^\mathrm{(abs)}$ and $J_\varphi^\mathrm{(scat)}$ where the integral over $\chi$ gives zero. In this way, Eq.~(\ref{Eq:Jmu}) leads to  expressions (61) and (64) in Ref.~\cite{pRoS16}.

\subsection{Physically relevant scalar quantities}

In the following, we discuss relevant scalar quantities that can be defined from the current density $J$ and the Killing vector fields $k$ and $\eta$ defined in Eq.~(\ref{Eq:KVF}). First, we will frequently use the particle number density, defined covariantly as
\begin{equation}
n(x) = \sqrt{- J(x) \cdot J(x)},
\label{Eq:nCovariant}
\end{equation}
where $J \cdot J = J_\mu J^\mu$. From this, one can also define the mean particle four-velocity $u^\mu(x) := J^\mu(x)/n(x)$. Contracting with $k$ and $\eta$, one obtains
\begin{equation}
e := -k\cdot u = -\frac{J_t}{n},\qquad
\ell_z := \eta\cdot u = \frac{J_\varphi}{n},
\end{equation}
which describe the specific energy and azimuthal angular momentum associated with an observer co-moving with the mean particle flow. Next, for $r > r_+$ one can decompose
\begin{equation}
J = N(k + \Omega \eta) + J^\perp,
\label{Eq:JDecomposition}
\end{equation}
where $N > 0$ denotes a normalization constant, $J^\perp$ is orthogonal to both $k$ and $\eta$, and $\Omega$ can be interpreted as an angular velocity. By taking scalar products with $k$ and $\eta$ on both sides of Eq.\ (\ref{Eq:JDecomposition}) and eliminating $N$, one arrives at the explicit expression
\begin{equation}
\Omega = \frac{-(k \cdot \eta)(k \cdot J) + (k \cdot k)(\eta \cdot J)}{(\eta \cdot \eta)(k \cdot J) - (k \cdot \eta)(\eta \cdot J)},
\end{equation}
which shows that $\Omega$ is a scalar. In terms of components $J^t$, $J^\varphi$, and $J^r$ associated with the horizon-penetrating coordinates $(t,r,\vartheta,\varphi)$ used in this article, one finds\footnote{In terms of Boyer-Lindquist coordinates it follows directly from Eq.~(\ref{Eq:JDecomposition}) that $\Omega = J_\mathrm{BL}^\varphi/J_\mathrm{BL}^t$.}
\begin{equation}
\label{OmegaHP}
\Omega = \frac{2aMr J_t - (\Delta\sin^{-2}\vartheta - a^2)J_\varphi}{\Sigma J_t + 2aM r J_\varphi} = \frac{\Delta J^\varphi - a J^r}{\Delta J^t - 2M r J^r}.
\end{equation}
If $J_\varphi$ vanished, $\Omega$ would reduce to the angular velocity of a zero angular momentum observer (ZAMO) \cite{jB70},
\begin{equation}
\Omega_\mathrm{ZAMO} = \frac{2a M r}{\Sigma},
\label{Eq:OmegaZAMO}
\end{equation}
where $\Sigma$ is defined as in Eq.~(\ref{Eq:DefSigma}). However, as will become clear in the following subsection, the mean particle flow characterizing the Vlasov gas in our situation has $J_\varphi = n\ell_z\neq 0$ unless $a=0$, and thus $\Omega$ differs from $\Omega_\mathrm{ZAMO}$ and $\Omega-\Omega_\mathrm{ZAMO}$ can be interpreted as the angular velocity of the mean particle flow relative to ZAMOs. Of course, this interpretation breaks down at the event horizon, since ZAMOs only exist in the region $\Delta > 0$. In fact, as the horizon is approached, $r\to r_+$, it follows from Eq.~(\ref{OmegaHP}) that $\Omega\to \Omega_+ := \left. \Omega_\mathrm{ZAMO} \right|_{r=r_+} = a/(2Mr_+)$, since $J^t$, $J^\varphi$, and $J^r$ are regular and $J^r\neq 0$. Therefore, although $\Omega$ is finite at the event horizon, it does not provide any information regarding the properties of the flow at the horizon, since it always coincides with its angular velocity $\Omega_+$. This is a consequence of the fact that the decomposition~(\ref{Eq:JDecomposition}) is ill-defined at $r=r_+$, since the plane spanned by $k$ and $\eta$ contains a null direction.

Note that the exact solution by Petrich, Shapiro, and Teukolsky~\cite{lPsSsT1988} describing the Bondi-type accretion of the ultra-relativistic stiff fluid onto a Kerr black hole (Eqs.\ (32) in~\cite{lPsSsT1988} with $u_\infty = 0$) has both $u_\vartheta = 0$ and $u_\varphi=0$, such that observers which are co-moving with the fluid flow are ZAMOs. In contrast, as mentioned above, our solutions have $u_\vartheta=0$ and $u_\varphi\neq 0$ when $a\neq 0$, such that $\Omega\neq \Omega_\mathrm{ZAMO}$.

\subsection{Asymptotic conditions}
\label{SubSec:Asymptotic}

In examples discussed in the remainder of this work we will consider two particular choices for $F_\infty(E)$. For so-called monoenergetic models, we set
\begin{equation}
\label{Fmono}
F_\infty = A m \delta(E - E_0) = A \delta(\varepsilon - \varepsilon_0),
\end{equation}
where $A$ and $E_0 = m \varepsilon_0$ are constants. In this case, the gas comprises of same-energy particles. Our second choice would be to set
\begin{equation}
\label{FMJ}
F_\infty = A \exp(-z \varepsilon),
\end{equation}
where $z = m/(k_\mathrm{B}T)$, $k_\mathrm{B}$ denotes the Boltzmann constant, and $T$ is the associated temperature. In the asymptotically flat limit $r \to \infty$, this choice leads to an equilibrium Maxwell-J\"{u}ttner distribution~\cite{fJ11a,fJ11b}, as will be explicitly verified below.

Since the proportionality constant $A$ appearing in Eqs.\ (\ref{Fmono}) and (\ref{FMJ}) is hard to control in physical terms, we will in many cases specify the asymptotic particle number density $n_\infty$ or the asymptotic energy density $\varepsilon_\infty$, instead. The value of $n_\infty$ can be computed from Eqs.~(\ref{Eq:Jmu}) and (\ref{Eq:nCovariant}). This can be achieved by using the asymptotic expressions\footnote{It is convenient to first perform the variable substitution $Q = r\tilde{Q}$ in the integrals and then use the explicit formulas~(\ref{Eq:QInt1}) and (\ref{Eq:QInt2}) for the resulting integrals to obtain these expressions. Note that $Q_c$ only appears in the $\mathcal{O}$-terms.}
\begin{subequations}
\begin{eqnarray}
&& \int\limits_0^{2\pi} d\chi\int\limits_0^{Q_c} \frac{Q dQ}{\sqrt{R}} = \mathcal{O}\left( \frac{1}{r^2} \right),\qquad
\int\limits_0^{2\pi} d\chi \sin\chi\int\limits_0^{Q_c} \frac{Q^2 dQ}{\sqrt{R}} = \mathcal{O}\left( \frac{1}{r^3} \right),\\
&& \int\limits_0^{2\pi} d\chi\int\limits_{Q_c}^{Q_\mathrm{max}} \frac{Q dQ}{\sqrt{R}} = 2\pi\sqrt{E^2 - m^2}\left[ 1 + \frac{M}{r}\frac{2E^2 - m^2}{E^2 - m^2} + \mathcal{O}\left( \frac{1}{r^2} \right) \right],
\\
&& \int\limits_0^{2\pi} d\chi\sin\chi\int\limits_{Q_c}^{Q_\mathrm{max}} \frac{Q^2 dQ}{\sqrt{R}} = -2\pi aE\sin\vartheta\sqrt{E^2-m^2}\left[ 1 + \frac{M}{r}\frac{4E^2 - 3m^2}{E^2 - m^2} + \mathcal{O}\left( \frac{1}{r^2} \right) \right],
\end{eqnarray}
\end{subequations}
for $r\to \infty$, which yields
\begin{subequations}
\begin{eqnarray}
J_t &=& -4\pi\int\limits_m^\infty dE E \sqrt{E^2 - m^2} F_\infty(E) + \mathcal{O}\left( \frac{1}{r} \right),\\
J_\varphi &=& -\frac{8\pi a M\sin^2\vartheta}{r}\int\limits_m^\infty dE E\sqrt{E^2 - m^2} F_\infty(E) + \mathcal{O}\left( \frac{1}{r^2} \right),
\end{eqnarray}
\end{subequations}
and $J^r = \mathcal{O}(r^{-2})$. Together with Eq.~(\ref{Eq:JTransform}), one also obtains $J^t = -J_t + \mathcal{O}(r^{-1})$, $J^\varphi = \mathcal{O}(r^{-4})$, and $J_r = \mathcal{O}(r^{-1})$, which implies that
\begin{equation}
n = -J_t + \mathcal{O}\left( \frac{1}{r} \right),\qquad
\ell_z = -\frac{2aM\sin^2\vartheta}{r} + \mathcal{O}\left( \frac{1}{r^2} \right),\qquad
\Omega = \mathcal{O}\left( \frac{1}{r^4} \right).
\end{equation}
In particular, one finds
\begin{equation}
n_\infty = 4 \pi m^3 A \varepsilon_0 \sqrt{\varepsilon_0^2 - 1},
\label{Eq:nInftyMonoE}
\end{equation}
for the monoenergetic model with $F_\infty$ given by Eq.\ (\ref{Fmono}) and
\begin{equation}
n_\infty = 4 \pi m^3 A \frac{K_2(z)}{z}, \end{equation}
for $F_\infty$ defined by Eq.\ (\ref{FMJ}), where $K_2$ denotes the modified Bessel function of the second kind (see, for instance, Ref.~\cite{DLMF}).

The asymptotic energy density is defined as
\begin{equation}
\varepsilon_\infty := \lim\limits_{r\to\infty} T_{\mu\nu}k^\mu k^\nu = \lim\limits_{r\to\infty} T_{tt}
 = 4\pi\int\limits_m^\infty dE E^2 \sqrt{E^2 - m^2} F_\infty(E),
\end{equation}
which leads to
\begin{equation}
\varepsilon_\infty = 4 \pi m^4 A \varepsilon_0^2 \sqrt{\varepsilon_0^2 - 1},
\end{equation}
for the monoenergetic model and
\begin{equation}
\varepsilon_\infty = 4 \pi m^4 A \left[ \frac{K_1(z)}{z} + 3 \frac{K_2(z)}{z^2} \right],
\end{equation}
in the Maxwell-J\"{u}ttner case.

\section{Accretion rates}
\label{sec:accretionrates}

In this section we discuss the accretion rates for our model. For this, we recall that the particle flux through a three-dimensional timelike surface $S$ with unit normal vector field $s^\mu\partial_\mu$ is given by
\begin{equation}
\mathcal{F}[S] = \int\limits_S J_\mu s^\mu \eta_S,
\label{Eq:Flux}
\end{equation}
with $\eta_S$ the induced volume form on $S$. Likewise, the energy and angular momentum fluxes through $S$ are defined by replacing $J_\mu$ with $J^{(k)}_\mu := -T_{\mu\nu} k^\nu = -T_{\mu t}$ and $J^{(\eta)}_\mu := T_{\mu\nu}\eta^\nu = T_{\mu\varphi}$, respectively, where $k$ and $\eta$ refer to the Killing vector fields defined in Eq.~(\ref{Eq:KVF}). In order to compute the accretion rate, we consider $S$ to be a constant $r$-surface enclosed between two spatial hypersurfaces $t=t_1$ and $t=t_2$, with $t_2 > t_1$. The unit normal in the outward radial direction is
\begin{equation}
s^\mu\partial_\mu = N g^{\mu\nu}(\nabla_\nu r)\partial_\mu 
 = \frac{N}{\rho^2}\left[ 
 \Delta\frac{\partial}{\partial r} + 2M r\frac{\partial}{\partial t} + a\frac{\partial}{\partial \varphi}
 \right] = \frac{N\Delta}{\rho^2} \frac{\partial}{\partial r_\mathrm{BL}},
\end{equation}
with $N = \rho/\sqrt{\Delta}$ a normalization constant. Furthermore, the induced volume element on $S$ is found to be
\begin{equation}
\eta_S = \rho\sqrt{\Delta}\sin\vartheta dt d\vartheta d\varphi,
\end{equation}
such that
\begin{equation}
J_\mu s^\mu\eta_S = \left( \Delta J_r + 2Mr J_t + a J_\varphi \right)  dt (\sin\vartheta d\vartheta) d\varphi
 = \rho^2 J^r dt (\sin\vartheta d\vartheta) d\varphi.
\end{equation}
Using Eqs.~(\ref{Eq:Jrabs}) and~(\ref{Eq:Jrscat}) one finds that the contribution of the scattered particles cancel, as expected, and hence
\begin{equation}
\rho^2 J^r = -\int_m^\infty dE F_\infty(E)\int\limits_0^{2\pi} d\chi \int\limits_0^{Q_c} dQ Q.
\label{Eq:JrBL}
\end{equation}
Inserting into Eq.~(\ref{Eq:Flux}) and dividing by $t_2-t_1$, one finds that the particle accretion rate through the surface $S$ is
\begin{equation}
\label{Eq:dotN}
\mathcal{\dot N} = -\pi\int\limits_m^\infty dE F_\infty(E) \int\limits_0^\pi d\vartheta \sin\vartheta \int\limits_0^{2\pi} d\chi (Q_c)^2.
\end{equation}

The energy and angular momentum accretion rates can be obtained in the same fashion by inserting the factors $-p_t = E$ and $p_\varphi = Q \sin \vartheta \sin \chi + a \sin^2 \vartheta E$ under the integral~(\ref{Eq:JrBL}), which yields
\begin{equation}
\mathcal{\dot E} = -\pi\int\limits_m^\infty dE E F_\infty(E) \int\limits_0^\pi d\vartheta \sin\vartheta \int\limits_0^{2\pi} d\chi (Q_c)^2,
\label{Eq:dotE}
\end{equation}
and
\begin{equation}
\mathcal{\dot J} = - 2 \pi \int\limits_m^\infty dE F_\infty(E) \int\limits_0^\pi d \vartheta \sin^2 \vartheta \int\limits_0^{2 \pi} d \chi \left[ \frac{1}{3} \sin \chi Q_c^3 + \frac{a}{2} \sin \vartheta E Q_c^2 \right].
\label{Eq:dotJ}
\end{equation}
Note that $\mathcal{\dot N}$, $\mathcal{\dot E}$, and $\mathcal{\dot J}$ are independent of $r$. In fact, they are independent of any continuous deformation of the surface $S$, as a consequence of the continuity equations for $J_\mu$, $J^{(k)}_\mu$, and $J^{(\eta)}_\mu$ and the fact that the gas distribution is stationary.

In the Schwarzschild limit $a=0$, $Q_c = L_c(E)$ is independent of $\chi$ and $\vartheta$ and the above expressions for the accretion rates  further simplify to
\begin{equation}
\left. \mathcal{\dot N} \right|_{a=0} = -4\pi^2\int\limits_m^\infty dE F_\infty(E) L_c(E)^2,\qquad
\left. \mathcal{\dot E} \right|_{a=0} = -4\pi^2\int\limits_m^\infty dE E F_\infty(E) L_c(E)^2, \qquad \left. \mathcal{\dot J} \right|_{a = 0} = 0,
\end{equation}
which is consistent with Eqs.~(76) and (77) in Ref.~\cite{pRoS16}.

Combining Eqs.~(\ref{Eq:dotN}) and (\ref{Eq:dotE}) with the asymptotic expressions in Section~\ref{SubSec:Asymptotic} one obtains
\begin{subequations}
\label{Eq:AccretionRateME}
\begin{eqnarray}
\frac{\dot{\mathcal N}}{n_\infty} & = & \frac{\dot{\mathcal{E}}}{\varepsilon_\infty} = -\frac{M^2}{4\varepsilon_0\sqrt{\varepsilon_0^2-1}}\int\limits_0^\pi d\vartheta \sin\vartheta \int\limits_0^{2 \pi} d\chi q_c(\alpha,\vartheta,\sin\chi,\varepsilon_0)^2, \\
\frac{\dot{\mathcal{J}}}{\varepsilon_\infty} & = & - \frac{M^3}{2 \varepsilon_0^2 \sqrt{\varepsilon_0^2 - 1}} \int\limits_0^\pi d \vartheta \sin^2 \vartheta \int_0^{2 \pi} d \chi \left[ \frac{1}{3} \sin \chi q_c(\alpha, \vartheta, \sin \chi, \varepsilon_0)^3 + \frac{\alpha}{2} \sin \vartheta \varepsilon_0 q_c(\alpha, \vartheta, \sin \chi, \varepsilon_0)^2 \right],
\end{eqnarray}
\end{subequations}
for the monoenergetic model and
\begin{subequations}
\label{Eq:AccretionRateMJ}
\begin{eqnarray}
\frac{\dot{\mathcal N}}{n_\infty} & = & - \frac{M^2 z}{4 K_2(z)} \int\limits_1^\infty d \varepsilon \exp(-z \varepsilon) \int\limits_0^\pi d \vartheta \sin \vartheta \int\limits_0^{2 \pi} d \chi q_c(\alpha,\vartheta,\sin\chi,\varepsilon)^2, 
\\
\frac{\dot{\mathcal{E}}}{\varepsilon_\infty} & = & - \frac{M^2 z}{4 \left[ K_1(z) + 3 K_2(z)/z \right]} \int\limits_1^\infty d \varepsilon \varepsilon \exp(-z \varepsilon) \int\limits_0^\pi d \vartheta \sin \vartheta \int\limits_0^{2 \pi} d \chi q_c(\alpha,\vartheta,\sin\chi,\varepsilon)^2, 
\\
\frac{\dot{\mathcal{J}}}{\varepsilon_\infty} & = &- \frac{M^3 z}{2 \left[ K_1(z) + 3 K_2(z)/z \right]}
\nonumber \\
&&\times \int\limits_1^\infty d \varepsilon \exp(-z \varepsilon) \int\limits_0^\pi d \vartheta \sin^2 \vartheta \int\limits_0^{2 \pi} d \chi \left[ \frac{1}{3} \sin \chi q_c(\alpha,\vartheta,\sin\chi,\varepsilon)^3 + \frac{\alpha}{2} \sin \vartheta \varepsilon q_c(\alpha,\vartheta,\sin\chi,\varepsilon)^2 \right],
\end{eqnarray}
\end{subequations}
for the Maxwell-J\"uttner model, where $Q_c = M m q_c$. It is also useful to consider the low and high-temperature limits of the Maxwell-J\"uttner model. Using the asymptotic behavior $K_\nu(z) \sim \sqrt{\pi/(2z)} e^{-z}$, performing the variable substitution $\varepsilon\mapsto \xi$ with $\varepsilon = 1 + \xi/z$, and taking the limit $z\to\infty$ one finds in the low temperature limit
\begin{subequations}
\label{MJaccretionrateslargez}
\begin{eqnarray}
\lim\limits_{z\to\infty} \frac{1}{\sqrt{2\pi z}} \frac{\dot{\mathcal N}}{n_\infty} & = & \lim\limits_{z\to\infty} \frac{1}{\sqrt{2\pi z}} \frac{\dot{\mathcal E}}{\varepsilon_\infty} = - \frac{M^2}{4\pi}  \int\limits_0^\pi d \vartheta \sin \vartheta \int\limits_0^{2 \pi} d\chi q_c(\alpha,\vartheta,\sin\chi,1)^2, \\
\lim\limits_{z\to\infty} \frac{1}{\sqrt{2\pi z}} \frac{\dot{\mathcal J}}{\varepsilon_\infty} & = &- \frac{M^3}{2 \pi} \int\limits_0^\pi d \vartheta \sin^2 \vartheta \int\limits_0^{2 \pi} d \chi \left[ \frac{1}{3} \sin \chi q_c(\alpha,\vartheta,\sin\chi,1)^3 + \frac{\alpha}{2} \sin \vartheta q_c(\alpha,\vartheta,\sin\chi,1)^2 \right].
\end{eqnarray}
\end{subequations}
In the high-temperature limit one uses the substitution $\varepsilon\mapsto \eta:= z\varepsilon$ and $K_\nu(z) \sim \frac{1}{2} \Gamma(\nu) (z/2)^{-\nu}$ in Eq. (\ref{Eq:AccretionRateMJ}) and takes the limit $z \to 0$ to obtain
\begin{subequations}
\label{MJaccretionratessmallz}  
\begin{eqnarray}
\lim\limits_{z\to 0}\frac{\dot{\mathcal N}}{n_\infty} & = & \lim\limits_{z\to 0} \frac{\dot{\mathcal{E}}}{\varepsilon_\infty}
 = -\frac{M^2}{4}\int\limits_0^\pi d\vartheta \sin \vartheta \int\limits_0^{2\pi} d\chi \zeta_c(\alpha,\vartheta,\sin\chi)^2, \\
\lim\limits_{z\to 0} \frac{\dot{\mathcal{J}}}{\varepsilon_\infty}
 & = &- \frac{M^3}{2} \int\limits_0^\pi d \vartheta \sin^2 \vartheta \int\limits_0^{2 \pi} d \chi \left[ \frac{1}{3} \sin \chi \zeta_c(\alpha, \vartheta, \sin \chi)^3 + \frac{\alpha}{2} \sin \vartheta \zeta_c(\alpha,\vartheta,\sin \chi)^2 \right],
\end{eqnarray}
\end{subequations}
where
\begin{equation}
\zeta_c(\alpha,\vartheta,\sin\chi) := \lim\limits_{\varepsilon\to\infty} \frac{1}{\varepsilon} q_c(\alpha,\vartheta,\sin\chi,\varepsilon),
\end{equation}
can be shown to be finite, see Appendix~\ref{App:QcAnalytic}. This relation reduces to $\zeta_c=3\sqrt{3}$ for the standard Schwarzschild black hole (see Eq.~(33) of \cite{mMoS25} and the related discussion).

\section{Slow-rotation approximation}
\label{sec:slowrotapprox}

Since $Q_c = Q_c(a,\vartheta,\chi,E)$ is only given in parametric form (see Eq.~(\ref{eq:QcEcParametricGeneralForm})), the integrals with respect to $\chi$ present in Eqs.~(\ref{Eq:Jmu}) cannot be computed analytically and one has to resort to numerical methods. Although we shall compute these integrals numerically in the upcoming section, here we circumvent this problem by considering a slowly rotating Kerr black hole and performing an expansion in the rotation parameter, up to cubic order. The slow-rotation approximation is an important astrophysically motivated case \cite{jFlM2019}, and it provides an effective strategy to overcome difficulties arising in the Kerr background spacetime due to the frame-dragging effect \cite{pPeBlG2013}. In principle, the analytic formulas presented in this section can be extended to higher order of the black hole spin parameter, to derive relations valid for more rapidly rotating black holes. However, as we will see, the cubic approximation already leads to an accurate description of the accretion process.

Up to the third order in $\alpha$, the critical value $Q_c$ can be expressed explicitly in terms of $E$, $\chi$, $\vartheta$, and $\alpha$ in the following way. First, we expand the critical radius
\begin{equation}
x_c = x_c^{(0)}+\alpha  x_c^{(1)} + \alpha^2 x_c^{(2)} + \alpha^3 x_c^{(3)} + \mathcal{O}(\alpha^4),
\label{Eq:xcExpansion}
\end{equation}
where the zeroth order term $x_c^{(0)}$ is known from the Schwarzschild case, see Eq.~(\ref{Eq:LcSchwarzschild}):
\begin{equation}
x_c^{(0)} = \frac{8}{\varepsilon\sqrt{9\varepsilon^2 - 8} + 4 - 3\varepsilon^2},\qquad
\varepsilon = \frac{E}{m},
\label{Eq:xc0}
\end{equation}
which is a function of energy only. To find the correction terms $x_c^{(1)}$, $x_c^{(2)}$, and $x_c^{(3)}$, we substitute the ansatz~(\ref{Eq:xcExpansion}) into the expression for $E_c$ given in Eq.~(\ref{eq:QcEcParametricGeneralForm}) and then solve the equation $E_c = E$ order by order in $\alpha$. After some manipulations, this yields
\begin{subequations}
\begin{eqnarray}
x_c^{(1)} &=& \frac{2 \sqrt{x_c^{(0)}}}{x_c^{(0)}-6}\sin\vartheta \sin\chi,
\\
x_c^{(2)} &=& \frac{4\cos^2\vartheta+2 x_c^{(0)} \sin^2\vartheta }{x_c^{(0)}\left(x_c^{(0)}-6\right)} - 3\frac{\left(x_c^{(0)}\right)^2 - 9 x_c^{(0)}+22}{\left(x_c^{(0)}-6\right)^3}\sin^2\vartheta \sin ^2\chi,
\\
x_c^{(3)} &=& \left(\frac{\sin \vartheta \sin \chi }{x_c^{(0)}-6}\right)^3 \left[ \frac{-7 \left(x_c^{(0)}\right)^2+36 x_c^{(0)}-12}{\sin ^2\vartheta \sin ^2\chi \left(x_c^{(0)}\right)^{3/2}} + \frac{-3 \left(x_c^{(0)}\right)^3+23 \left(x_c^{(0)}\right)^2-48 x_c^{(0)}+12}{\sin ^2\chi  \left(x_c^{(0)}\right)^{3/2}} \right. \notag \\
&& \left. + \frac{3 \left(x_c^{(0)}\right)^4-43 \left(x_c^{(0)}\right)^3+230 \left(x_c^{(0)}\right)^2-516 x_c^{(0)}+360}{\sqrt{x_c^{(0)}} \left(x_c^{(0)}-6\right)^2} \right],
\end{eqnarray}
\end{subequations}
and in principle this procedure can be followed to compute higher correction terms in $\alpha$. This provides $x_c$ as a function of $\alpha$, $E$, $\vartheta$, and $\chi$ as an expansion in $\alpha$ to the desired order.

Next, we substitute this function into the expression for $Q_c$ given in Eq.~(\ref{eq:QcEcParametricGeneralForm}) and expand the result in powers of $\alpha$. In terms of the dimensionless quantities $q_c = Q_c/(M m)$ this yields, up to the third order,
\begin{equation}
q_c = q_c^{(0)} + \alpha q_c^{(1)} +\alpha^2 q_c^{(2)}
 + \alpha^3 q_c^{(3)} + \mathcal{O}(\alpha^4),
\label{qcApp}
\end{equation}
where
\begin{subequations}
\label{qcApp2}
\begin{eqnarray}
q_c^{(0)}&=& \frac{x_c^{(0)}}{\sqrt{x_c^{(0)}-3}},\\
q_c^{(1)}&=& -\frac{q_c^{(0)}}{\sqrt{x_c^{(0)}}}\sin\vartheta \sin\chi,
 \\
q_c^{(2)}&=&\frac{q_c^{(0)}}{2 x_c^{(0)}} \left[ -\frac{\cos ^2\vartheta+x_c^{(0)} \sin ^2\vartheta}{x_c^{(0)}} +\frac{\left(x_c^{(0)}-5\right)  \sin ^2\vartheta  \sin ^2\chi}{x_c^{(0)}-6} \right],
 \\
q_c^{(3)}&=& \frac{\sin ^3\vartheta  \sin ^3\chi}{\left(x_c^{(0)}-6\right) \sqrt{x_c^{(0)}-3}} \left[ \frac{x_c^{(0)}+2 \cot ^2\vartheta}{\sin ^2\chi  \left(x_c^{(0)}\right)^{3/2}}-\frac{\left(x_c^{(0)}\right)^2-10 x_c^{(0)}+26}{\sqrt{x_c^{(0)}} \left(x_c^{(0)}-6\right)^2} \right].
\end{eqnarray}
\end{subequations}
Note that the zeroth order term again agrees with the value known from the Schwarzschild case, see Eq.~(\ref{Eq:LcSchwarzschild}). Together with Eq.~(\ref{Eq:xc0}), Eqs.\ (\ref{qcApp}) and (\ref{qcApp2}) provide the desired expansion for $Q_c$ as a function of $E$, $\chi$, $\vartheta$, and $\alpha$.

From these results one easily obtains
\begin{equation}
\int_0^\pi d\vartheta\sin\vartheta\int_0^{2\pi} d\chi q_c^2 = 4\pi\frac{\left(x_c^{(0)}\right)^2}{x_c^{(0)}-3}\left[ 1 - \frac{2\alpha^2}{\left(x_c^{(0)}\right)^2\left(6 - x_c^{(0)}\right)} + \mathcal{O}(\alpha^4) \right],
\label{Eq:IntSlowRot}
\end{equation}
from which the accretion rates can be computed by means of Eqs.~(\ref{Eq:AccretionRateME}) and (\ref{Eq:AccretionRateMJ}) for the monoenergetic and the Maxwell-J\"{u}ttner case, respectively. In particular, for the Maxwell-J\"{u}ttner model, using formulas (\ref{MJaccretionrateslargez}) and (\ref{MJaccretionratessmallz}), one obtains the particle and the energy accretion rates in the slow-rotation approximation for both the low ($z\to\infty$) and high ($z\to 0$) temperature limits:
\begin{subequations}
\label{Eq:Extreme-temperatureTE}
\begin{eqnarray}
\lim\limits_{z\rightarrow \infty }\frac{1}{\sqrt{2\pi z}}\frac{\mathcal{\dot{N}}}{n_{\infty }} &=& \lim\limits_{z\rightarrow \infty }\frac{1}{\sqrt{2\pi z}}\frac{\mathcal{\dot{E}}}{\varepsilon_{\infty }} = -(16-\alpha ^2)M^2 + \mathcal{O}(\alpha^4),\quad \lim\limits_{z\rightarrow \infty }\frac{1}{\sqrt{2\pi z}}\frac{\mathcal{\dot{J}}}{\varepsilon_{\infty }} = \alpha \left( \frac{32}{3}+\frac{2}{5} \alpha^2 \right) M^3+ \mathcal{O}(\alpha^4),\qquad
\label{low-temperatureTE}\\
\lim\limits_{z\rightarrow 0 }\frac{\mathcal{\dot{N}}}{n_{\infty }} &=& \lim\limits_{z\rightarrow 0 }\frac{\mathcal{\dot{E}}}{\varepsilon_{\infty }} = -\pi(27-2 \alpha ^2)M^2 + \mathcal{O}(\alpha^4),\quad
\lim\limits_{z\rightarrow 0 }\frac{\mathcal{\dot{J}}}{\varepsilon_{\infty }} = 36 \pi \alpha M^3+ \mathcal{O}(\alpha^4).
\label{high-temperatureTE}
\end{eqnarray}
\end{subequations}
These expressions show that $\mathcal{\dot N}/n_\infty$, $\mathcal{\dot E}/\varepsilon_\infty$, and $\mathcal{\dot J}/n_\infty$ scale as $\sqrt{2\pi z} = \sqrt{2\pi m/(k_\mathrm{B} T)}$ (the thermal wavelength divided by the reduced Compton wavelength of the gas particles) for $z \to \infty$, whereas these ratios are constant for $z \to 0$. In addition, the nonvanishing values of the rotation parameter reduce both the particle and energy accretion rates with respect to their values for $\alpha = 0$, and the first correction term in $\alpha$ is of second order. On the other hand, in the case of angular momentum accretion rate, the leading term is of first order and the black hole slows down. Note that for $\alpha = 0$, $\mathcal{\dot N}/n_\infty $ and $ \mathcal{\dot E}/\varepsilon_\infty$ in the relations (\ref{Eq:Extreme-temperatureTE}) reduce to the ones obtained for the Schwarzschild case in \cite{pRoS16}, whereas $\mathcal{\dot J}/n_\infty $ vanishes as it should be.

On the other hand, one can numerically perform the following integrations over the energy and compute the accretion rates for generic values of $z$ up to the third order in the rotation parameter $\alpha$. Using Eq.~(\ref{Eq:IntSlowRot}), one obtains
\begin{subequations}
\label{Eq:AccretionRateMJapp}
\begin{eqnarray}
\frac{\dot{\mathcal{N}}}{n_\infty}&=&-\frac{\pi z M^2}{K_2(z)}\int_1^\infty\frac{\left(x_c^{(0)}\right)^2}{x_c^{(0)}-3}\left[ 1 - \frac{2\alpha^2}{\left(x_c^{(0)}\right)^2\left(6 - x_c^{(0)}\right)} \right]e^{-z \varepsilon} d\varepsilon + \mathcal{O}(\alpha^4),\\
\frac{\dot{\mathcal{E}}}{\varepsilon_\infty}&=&-\frac{\pi z^2  M^2}{zK_1(z)+3K_2(z)}\int_1^\infty\frac{\left(x_c^{(0)}\right)^2}{x_c^{(0)}-3}\left[ 1 - \frac{2\alpha^2}{\left(x_c^{(0)}\right)^2\left(6 - x_c^{(0)}\right)} \right] \varepsilon e^{-z \varepsilon} d\varepsilon + \mathcal{O}(\alpha^4),\\
\frac{\dot{\mathcal{J}}}{\varepsilon_\infty}&=& \frac{4 \pi \alpha z^2 M^3}{3[zK_1(z)+3K_2(z)]}\int_1^\infty  \left(\frac{x_c^{(0)}}{x_c^{(0)}-3} \right)^{\frac{3}{2}} \left\{1+ \frac{3 \alpha ^2 \left(x_c^{(0)}-3\right) \left[\left(x_c^{(0)}\right)^2-24\right]}{5 \left(x_c^{(0)}-6\right)^3 \left(x_c^{(0)}\right)^2} \right\} e^{-z \varepsilon} d\varepsilon + \mathcal{O}(\alpha^4).
\end{eqnarray}
\end{subequations}

Therefore, within the slow-rotation approximation, the original triple integrals appearing in Eqs.\ (\ref{Eq:AccretionRateMJ}) can be replaced with simpler formulas~(\ref{Eq:AccretionRateMJapp}). Remarkably, for all numerical examples discussed in Sec.\ \ref{sec:numerics}, the difference between the results of the slow-rotation approximation and exact numerical values computed according Eqs.\ (\ref{Eq:AccretionRateMJ}) stays below $4\%$ for the mass accretion rate and below $5\%$ for the angular momentum accretion rate, even for rapidly rotating black holes with rotation parameters up to $\alpha\leq0.99$.

In a similar manner, one can approximate the integrals appearing in Eqs.\ (\ref{Eq:Jmu}), to compute the particle current and number densities around a slowly rotating Kerr black hole. To this end, we first expand Eq.~(\ref{Eq:Qmax}) up to the second order in $\alpha$ and obtain $Q_{\max} = M m q_{\max}$ with
\begin{eqnarray}
q_{\max} & = &\frac{x \sqrt{x \varepsilon ^2-x+2}}{\sqrt{x-2}}-\frac{\alpha \varepsilon x \sin \vartheta \sin \chi }{x-2} \nonumber \\
&& -\frac{\alpha ^2 \varepsilon ^2 x }{2 (x-2)^{3/2} \sqrt{x \varepsilon ^2-x+2}}\left(1-\frac{\left(2 x \varepsilon ^2-x+2\right) \left[ (x-2)\cos^2 \vartheta+x \sin^2 \vartheta \sin ^2\chi \right]}{x^2 \varepsilon ^2}\right) + \mathcal{O}(\alpha^3).
\label{qmaxApp}
\end{eqnarray}
Contrary to the exact expressions which always satisfy $q_{\max} \geq q_c$ with $q_{\max} = q_c$ only at $x = x_c$, when truncating the expansion at the second order, these conditions are no longer guaranteed to hold. In particular, it can occur that $q_{\max}$ given by Eq.~(\ref{qmaxApp}) can be smaller than $q_c$ defined in Eq.~(\ref{qcApp}). For this reason, when computing the expressions corresponding to scattered trajectories, we impose an additional condition, enforcing that $q_{\max}>q_c$ also in the slow-rotation approximation. Specifically, we numerically determine the largest root of the equation $q_{\max}(x) = q_c(x)$. Although this root depends on the angle $\chi$, in practice we found that it can be replaced with its maximum value corresponding to $\chi = 0$, without introducing significant errors.

In the next section, we apply these conditions and compute the integrals~(\ref{Eq:Jmu}) in the slow-rotation approximation. We also compute $n/n_{\infty}$ and the ratio $\Omega/\Omega_\mathrm{ZAMO}$. Then these approximate results are compared with the corresponding exact numerical values.

On the other hand, for the Maxwell-J\"{u}ttner model, it is also possible to compute the particle current density in the low temperature limit $z \to \infty$ and obtain analytic expressions for the particle density, up to second order in the rotation parameter. To do so, we first calculate the components of the particle current density (\ref{Eq:Jmu}) in the limit $z \to \infty$ as follows:
\begin{subequations}
\label{Eq:JmuLowT}
\begin{eqnarray}
\lim\limits_{z\rightarrow \infty }\frac{1}{\sqrt{2\pi z}} \frac{J_t^\mathrm{(tot)}}{n_{\infty}} &=& -\frac{1}{\pi }\sqrt{\frac{2}{x}} \left(1+ \frac{\alpha ^2}{8 x}\right) + \mathcal{O}(\alpha^3),
\label{Eq:JtTot}\\
\lim\limits_{z\rightarrow \infty }\frac{1}{\sqrt{2\pi z}} \frac{J^r_\mathrm{(tot)}}{n_{\infty}} &=& -\frac{M}{4 \pi x^2} \left[16- \left(1+ \frac{16 \cos^2\vartheta}{x^2}\right) \alpha^2 \right] + \mathcal{O}(\alpha^3),
\label{Eq:JrTot}\\
\lim\limits_{z\rightarrow \infty } \frac{1}{\sqrt{2\pi z}} \frac{J_\varphi^\mathrm{(tot)}}{n_{\infty}} &=& 0+ \mathcal{O}(\alpha^3), 
\label{Eq:JphiTot}
\end{eqnarray}
\end{subequations}
where up to second order in the rotation parameter, $J_\varphi^\mathrm{(tot)}$ vanishes in the low-temperature regime.

Employing Eqs.\ (\ref{Eq:JTransform}) and (\ref{Eq:nCovariant}) and performing some straightforward calculations, one can write the expression for the particle number density in the low-temperature limit as
\begin{equation}
\lim\limits_{z\rightarrow \infty }\frac{1}{\sqrt{2\pi z}}\frac{n(x,\vartheta)}{n_{\infty }}=\frac{1}{4 \pi  x^3}\sqrt{\frac{x (x+2)+4}{2 x}} \left[8 x^2+\frac{\alpha ^2 (x+2)}{x (x+2)+4} \left(8 \sin ^2\vartheta+x^2-\frac{16}{x+2}\right) \right]+\mathcal{O}(\alpha^3).
\label{eq:nRatioLargeZ}
\end{equation}
It is worth mentioning that Eq.~(\ref{eq:nRatioLargeZ}) reduces to $\sqrt{3}/\pi$ at $x_+=2$ in the standard Schwarzschild limit $\alpha \to 0$ as follows from~\cite{pRoS16}. We would like to stress that other observables and physical quantities can be computed within the slow-rotation approximation in an analogous way; the limited examples provided in this work are intended to demonstrate this technique.

\section{Numerical results}
\label{sec:numerics}

In this section, we provide results for the numerical calculation of the observables, compression ratios, angular velocity, and accretion rates and compare them with the corresponding analytic results from the slow-rotation approximation. All results shown in this section were obtained using \textit{Wolfram Mathematica} \cite{Wolfram}. Integrals with respect to $Q$ appearing in Eqs.\ (\ref{Eq:Jmu}) are computed using explicit, analytic formulas collected in Appendix \ref{App:Qintegrals}. Integration limits $Q_c$ and $Q_\mathrm{max}$ are computed as described in Appendices~\ref{App:QcAnalytic} and \ref{App:Qmax}. In particular, since $Q_c$ is given in a parametric form, the relation $Q_c = Q_c(\alpha,\vartheta,\sin \chi,E)$ is obtained numerically. Integration with respect to $\chi$ is also performed numerically. This is essentially straightforward, except for possible cases in which the integrands with respect to $\chi$ are not smooth, which occurs in the range $r_\mathrm{ph} \le r \le r_\mathrm{mb}$. This behavior is further explained in Appendix \ref{App:smoothness}. Here we only note that numerical difficulties associated with integration with respect to $\chi$ can be removed by directly integrating the sums of expressions corresponding to absorbed and scattered trajectories, which remain regular.

\subsection{Monoergetic models}

\begin{figure}
\begin{center}
\includegraphics[width=0.49\linewidth]{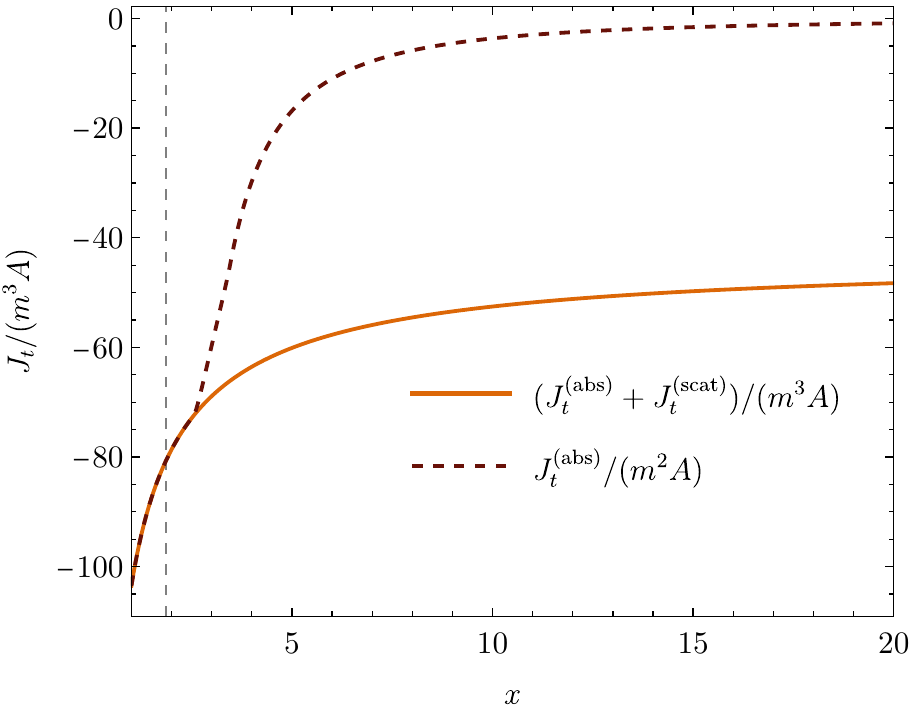}
\includegraphics[width=0.49\linewidth]{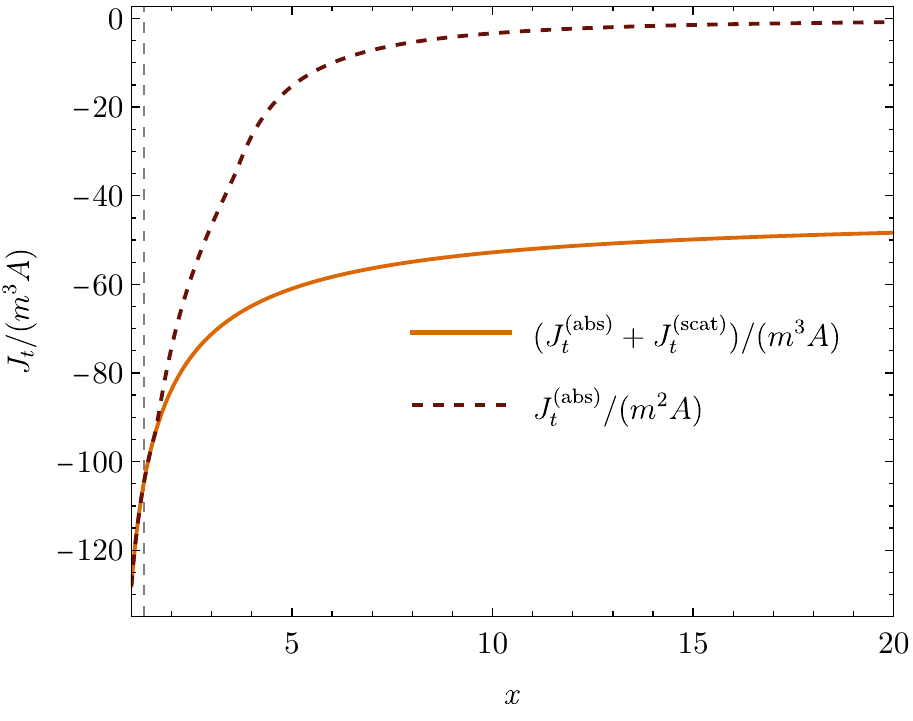}
\end{center}
\caption{\label{fig:Jtmono} Time components of the particle current density $J_t$ for a monoenergetic model with $\varepsilon_0 = 2$. Left: $\alpha = 1/2$. Right: $\alpha = 0.95$. Both graphs are plotted for $\vartheta = \pi/4$. Dashed vertical lines mark locations of the outer horizon.}
\end{figure}

\begin{figure}
\begin{center}
\includegraphics[width=0.49\linewidth]{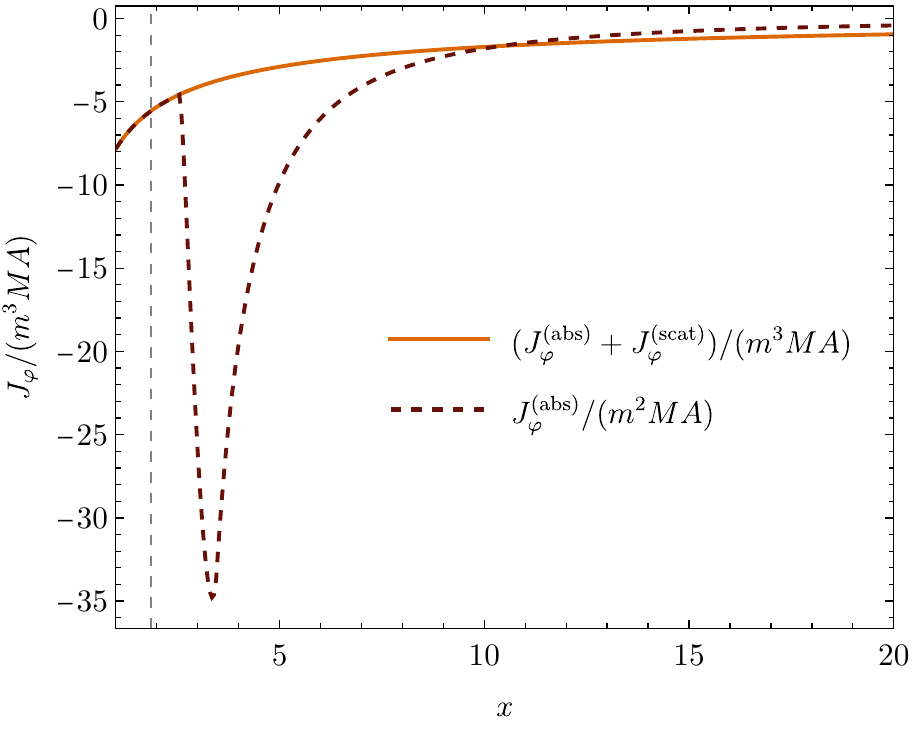}
\includegraphics[width=0.49\linewidth]{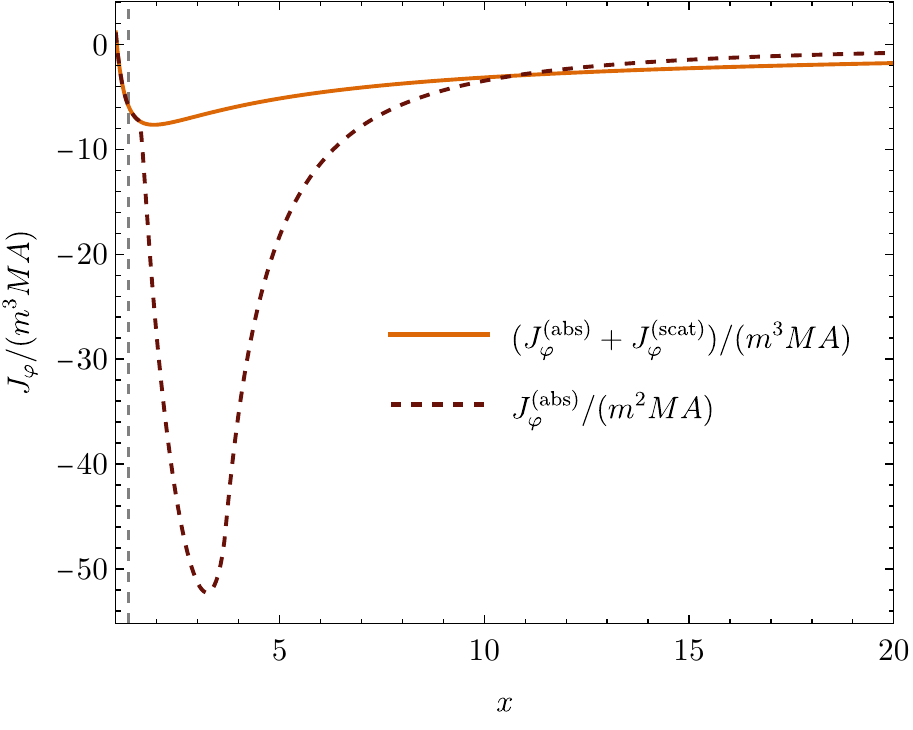}
\end{center}
\caption{\label{fig:Jphimono} Angular components of the particle current density $J_\varphi$ for a monoenergetic model with $\varepsilon_0 = 2$. Left: $\alpha = 1/2$. Right: $\alpha = 0.95$. Both graphs are plotted for $\vartheta = \pi/4$. Dashed vertical lines mark locations of the outer horizon.}
\end{figure}

\begin{figure}
\begin{center}
\includegraphics[width=0.49\linewidth]{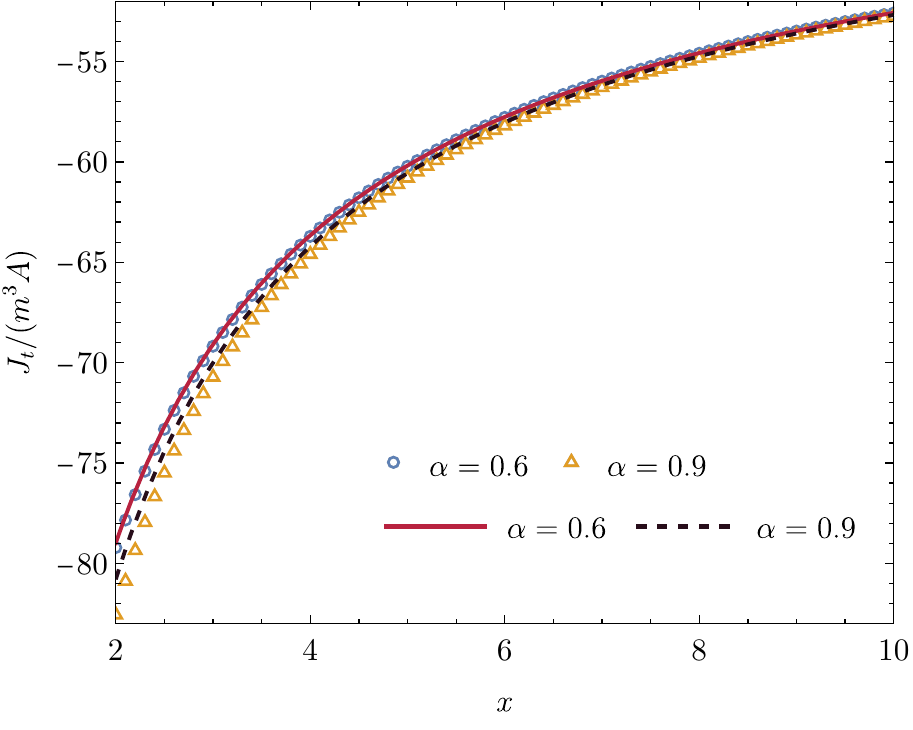}
\includegraphics[width=0.49\linewidth]{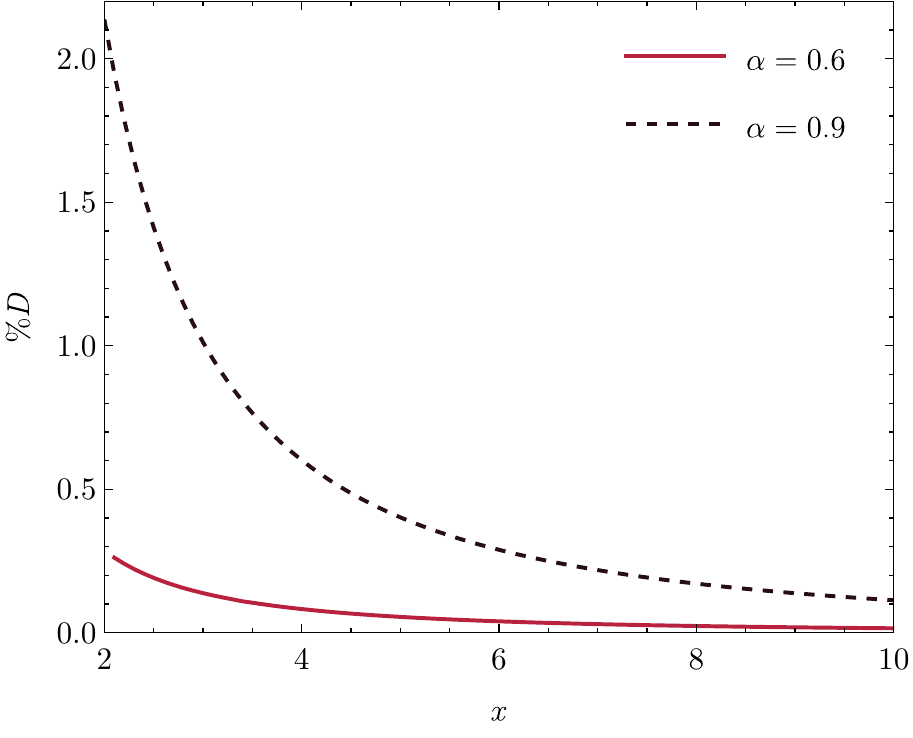}
\end{center}
\caption{\label{fig:JtmonoAppNum} Time component of the total particle current density $J_t=J_t^\mathrm{(abs)}+J_t^\mathrm{(scat)}$ for the monoenergetic model with $\varepsilon_0 = 2$ at the fixed polar angle $\vartheta=\pi/4$. In the left panel, exact numerical values are plotted with discrete points. The lines illustrate results of the slow-rotation approximation. The right panel shows the percentage difference $\%D$ between the results of the slow-rotation approximation and exact numerical values. Surprisingly, $\%D$ stays below $2.2\%$, even for rapidly rotating black holes with $\alpha = 0.9$ for $x \geq 2$.}
\end{figure}

\begin{figure}
\begin{center}
\includegraphics[width=0.49\linewidth]{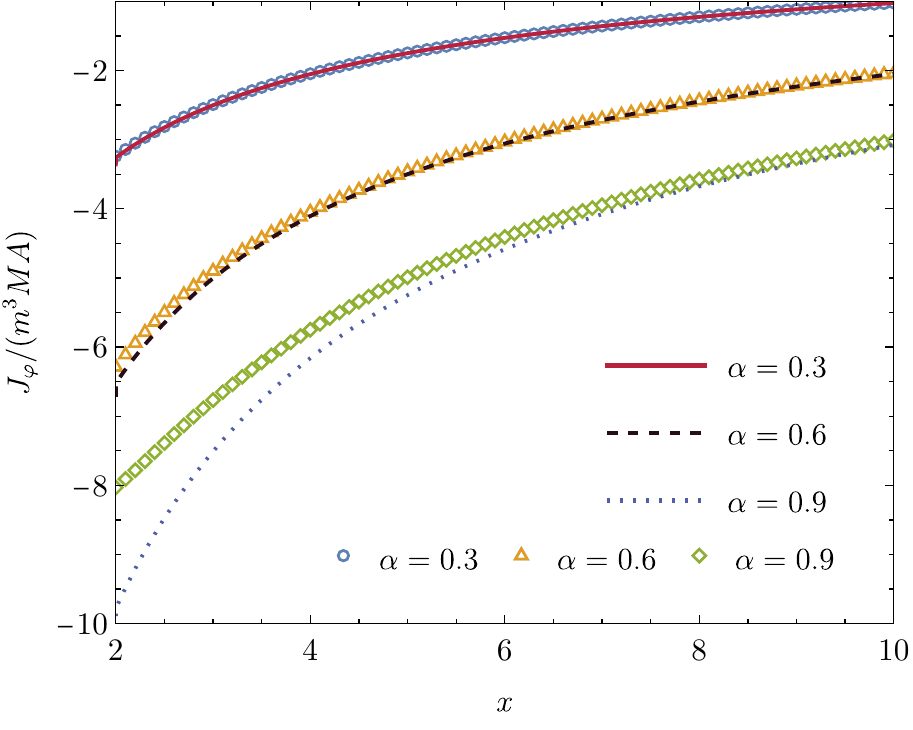}
\includegraphics[width=0.49\linewidth]{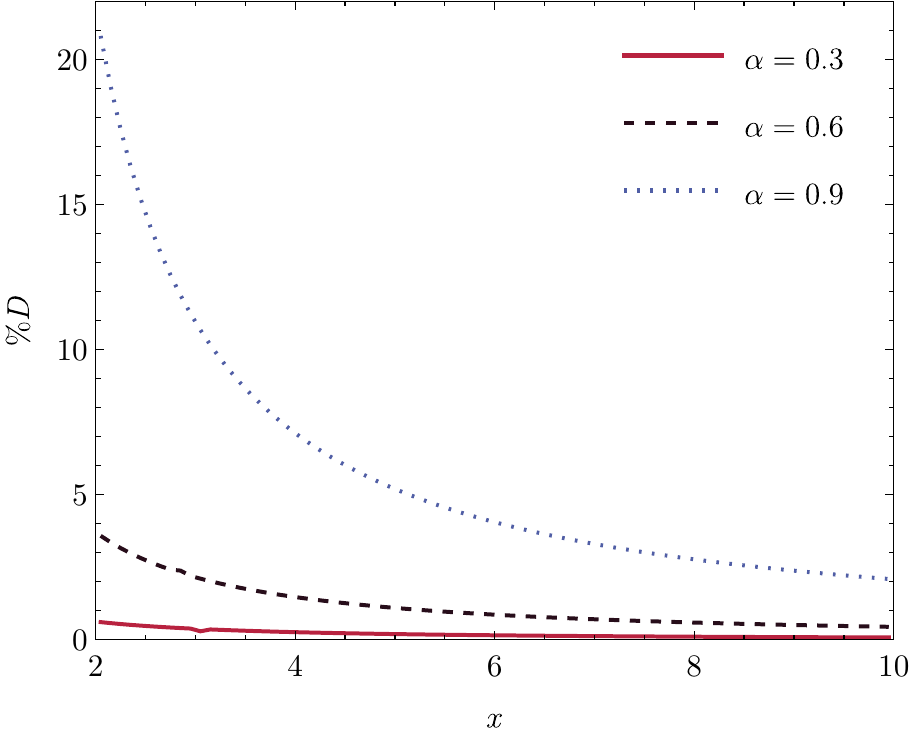}
\end{center}
\caption{\label{fig:JphimonoAppNum} 
Same as in Fig.\ \ref{fig:JtmonoAppNum}, but for the azimuthal component of the total particle current density $J_\varphi=J_\varphi^\mathrm{(abs)}+J_\varphi^\mathrm{(scat)}$. We plot data for $\alpha = 0.3$, 0.6, and 0.9.}
\end{figure}

We start with monoenergetic models defined by Eq.\ (\ref{Fmono}) and illustrate our method by numerically computing the momentum integrals appearing in Eqs.\ (\ref{Eq:Jmu}).

Figures~\ref{fig:Jtmono}--\ref{fig:JphimonoAppNum}
illustrate the components of the particle current density $J_\mu$ for $\varepsilon_0 = 2$. In Figs.~\ref{fig:Jtmono} and \ref{fig:Jphimono} we show the contribution of absorbed particles (dashed lines) and the total contribution of both absorbed and scattered particles (solid lines) to $J_t$ and $J_\varphi$ as a function of the dimensionless radius $x$ for $\vartheta = \pi/4$ and two different values of $\alpha$. Both $J_t$ and $J_\varphi$ remain regular on the event horizon, as expected. The effects of rotation on the behavior of these observables are manifest when comparing the plots on the left panels ($\alpha=1/2)$ with those on the right panels ($\alpha=0.95$) in Figs.~\ref{fig:Jtmono} and \ref{fig:Jphimono}. As discussed in Appendix \ref{App:smoothness}, while $J_t^\mathrm{(abs)}$ and $J_\varphi^\mathrm{(abs)}$ are non-smooth functions of the radius $r$, the total components $J_t = J_t^\mathrm{(abs)} + J_t^\mathrm{(scat)}$ and $J_\varphi = J_\varphi^\mathrm{(abs)} + J_\varphi^\mathrm{(scat)}$ remain smooth. We omit the plots of radial profiles of  $J^r = J^r_\mathrm{(abs)}$, as they behave simply as $J^r \propto 1/\rho^2$.

Figures~\ref{fig:JtmonoAppNum} and \ref{fig:JphimonoAppNum} provide a comparison of radial profiles of $J_t$ and $J_\varphi$ obtained within the slow-rotation approximation with the corresponding exact numerical values. We restrict ourselves to the range $2 \le x$ and assume $\varepsilon_0 = 2$. The slow-rotation approximation turns out to work remarkably well, even when truncated to second order in $\alpha$. The deviation from the exact solution (shown on the right panels in Figs.\ \ref{fig:JtmonoAppNum} and \ref{fig:JphimonoAppNum}) remain smaller than $2.2\%$ for $J_t$ and black hole spin parameters $\alpha \le 0.9$ and smaller than $4\%$ in the case of $J_\varphi$, for $\alpha \leq 0.6$. In the vicinity of the horizon, the error of the slow-rotation approximation of $J_\varphi$ becomes larger for $\alpha > 0.6$. On the other hand, the second-order terms in $\alpha$ vanish identically in the expression for $J_\varphi$. Consequently, in the expansion of $J_\varphi$ conducted up to the second order in $\alpha$, the effects of rotation are described by first-order terms only.

\begin{figure}
\begin{center}
\includegraphics[width=0.49\linewidth]{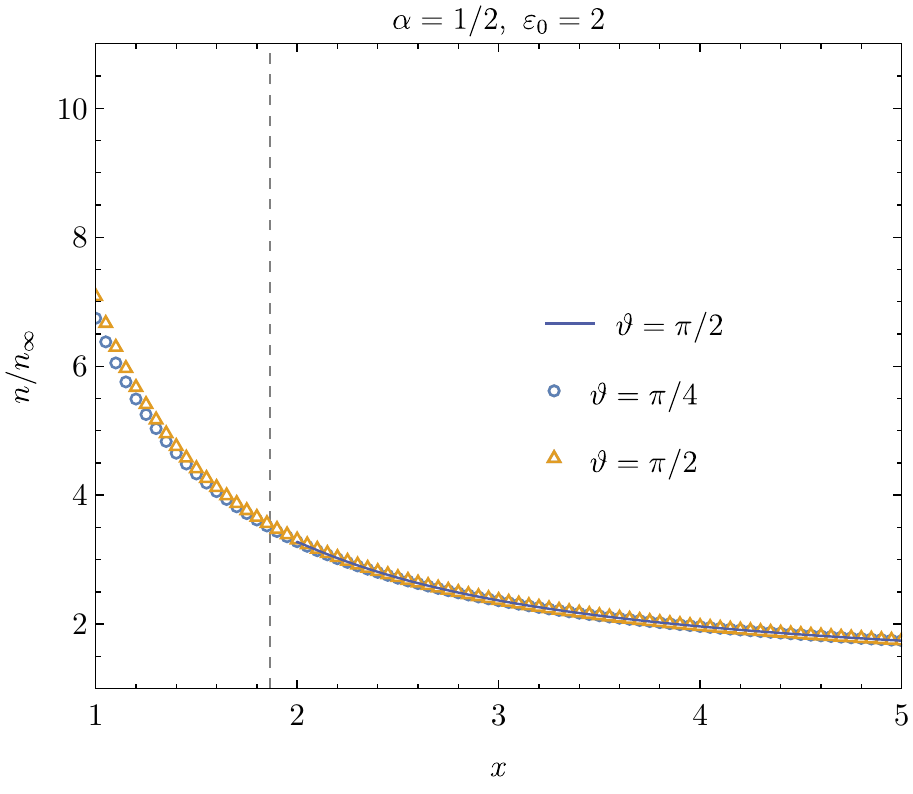}
\includegraphics[width=0.49\linewidth]{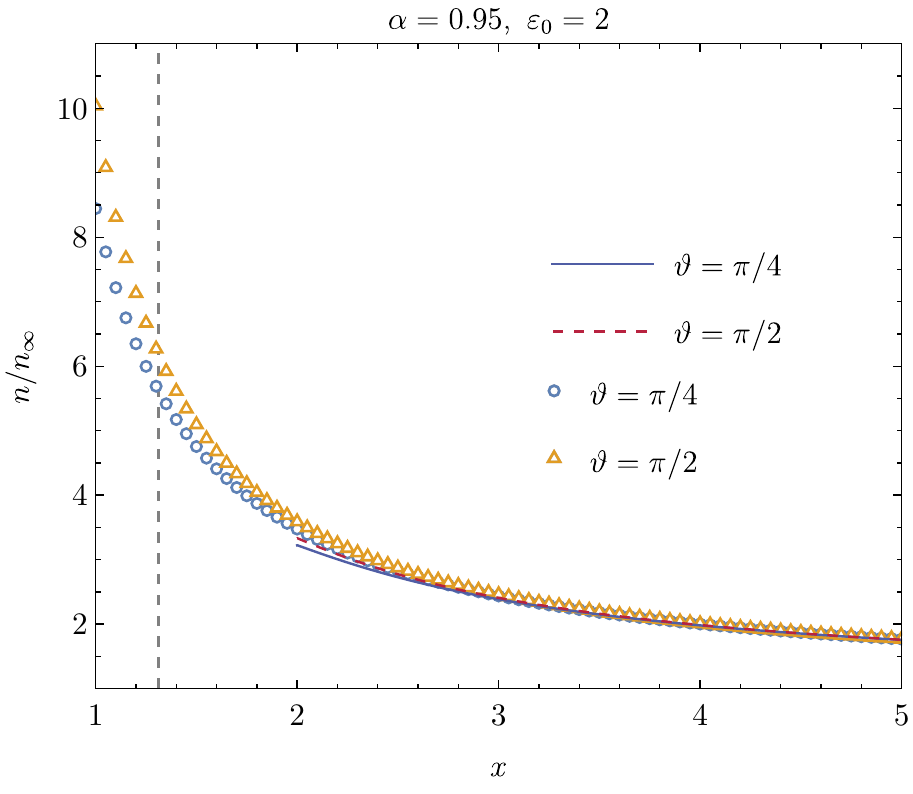}
\end{center}
\caption{\label{fig:newCR} Compression ratio $n/n_\infty$ versus radius $x$ for the monoenergetic model with $\varepsilon_0 = 2$. Vertical dashed lines mark locations of the event horizon $x_+$. Geometric shapes correspond to exact numerical values for $x \geq 1$; results of the slow-rotation approximation are plotted with lines for $x \geq 2$.
For $x \geq 2$, the difference between approximated and exact values is smaller than $1.1\%$ for $\alpha=0.5$ and smaller than $7.2\%$ for $\alpha=0.95$.}
\end{figure}

\begin{figure}
\begin{center}
\includegraphics[width=0.49\linewidth]{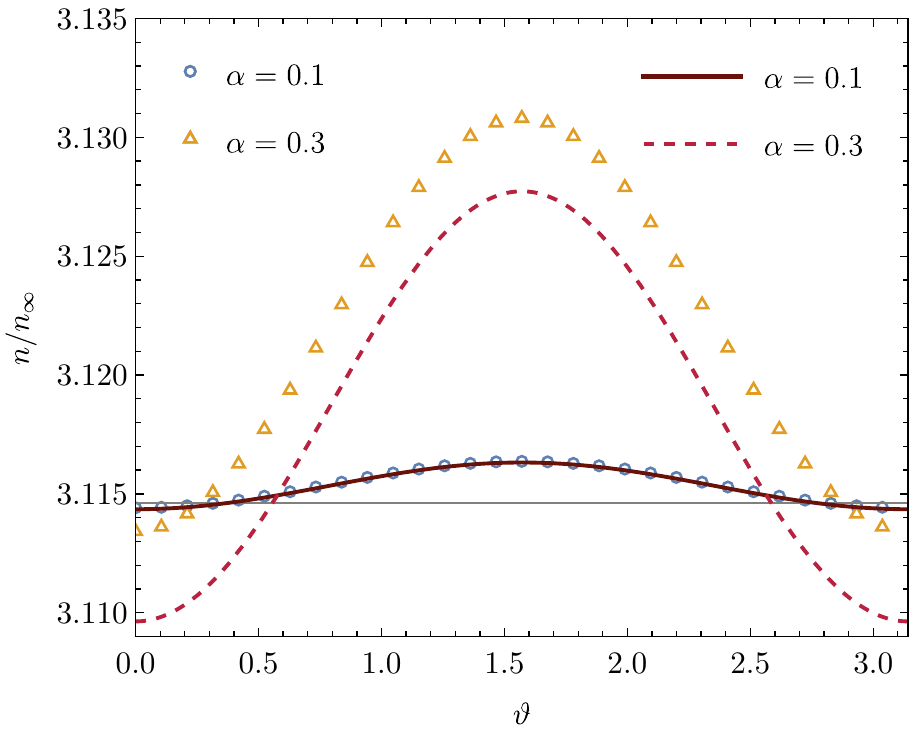}
\includegraphics[width=0.49\linewidth]{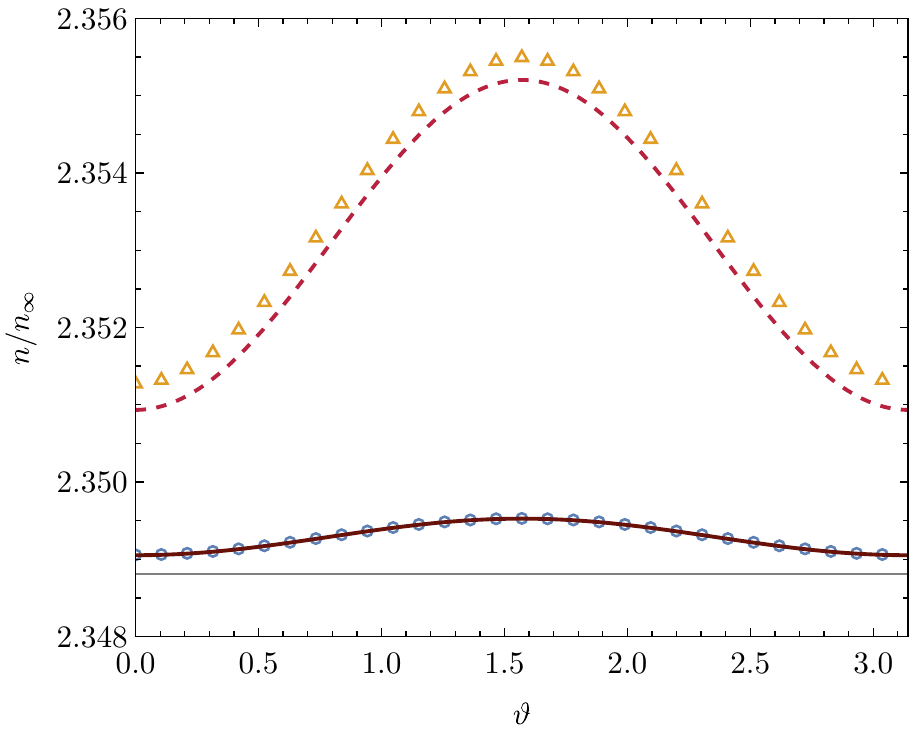}
\end{center}
\caption{\label{fig:CompRatAppTheta} Compression ratio $n/n_\infty$ versus $\vartheta$ for the monoenergetic model with $\varepsilon_0=2$. Left panel: $x=2.1$; right panel: $x=3$. Exact numerical values are plotted with discrete points. The lines illustrate results of the slow-rotation approximation. Close to the black hole (at $x=2.1$), the compression ratio $n/n_{\infty}$ is a decreasing function of $\alpha$ at the axis ($\vartheta=0,\pi$), whereas it increases with $\alpha$ on the equatorial plane $\vartheta=\pi/2$. Far from the black hole, the compression ratio is an increasing function of $\alpha$ in the whole range of $\vartheta$. Continuous gray horizontal lines depict the corresponding value in the  Schwarzschild case ($\alpha = 0$).}
\end{figure}

\begin{figure}
\begin{center}
\includegraphics[width=0.49\linewidth]{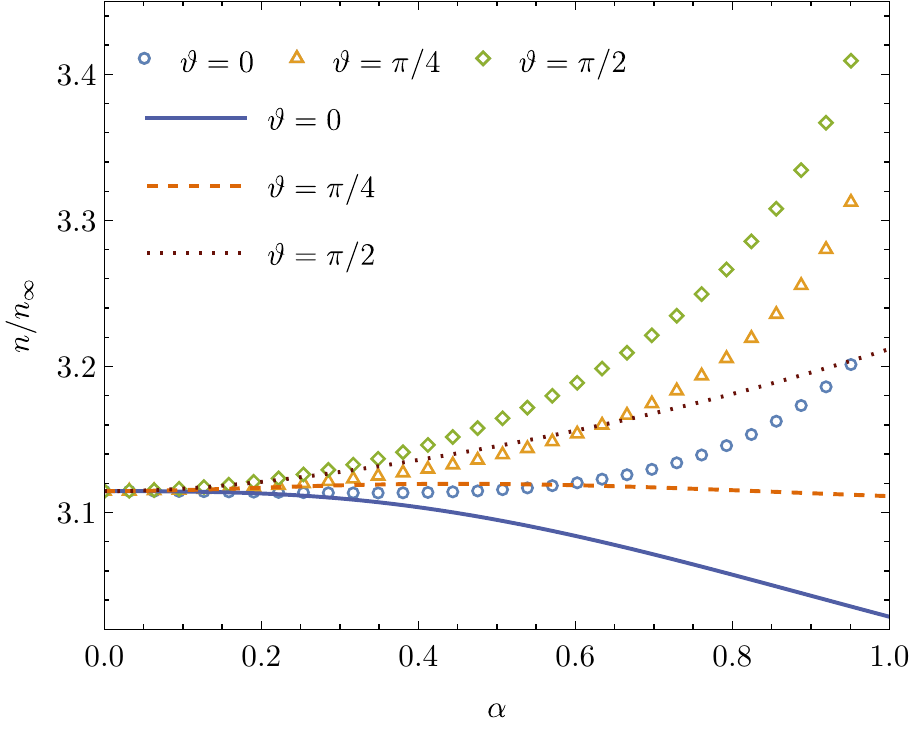}
\includegraphics[width=0.49\linewidth]{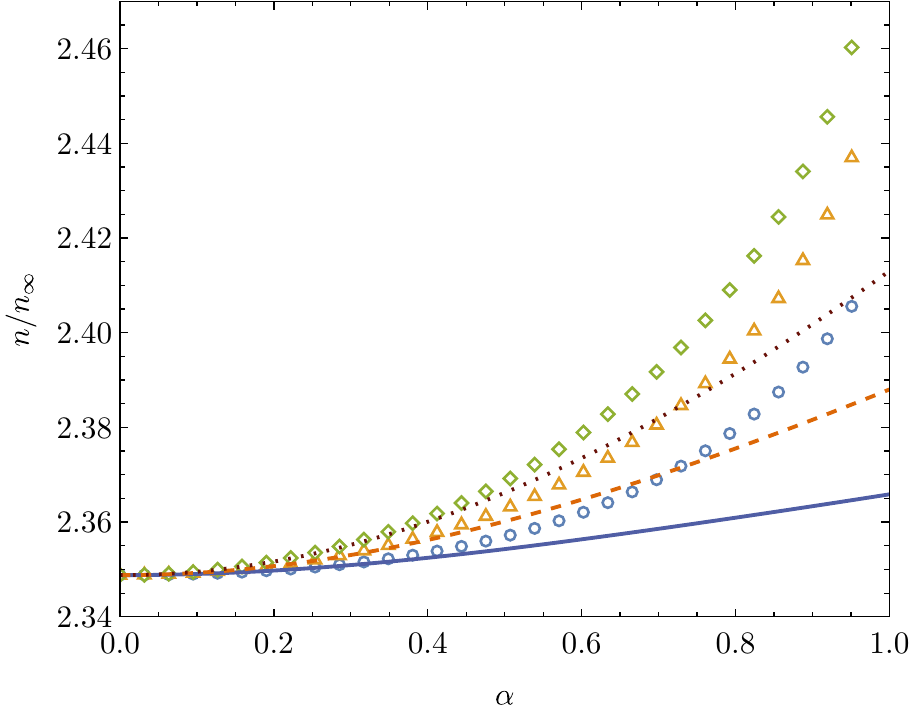}
\end{center}
\caption{\label{fig:CompRatAppAlpha}  Compression ratio $n/n_\infty$ versus $\alpha$ for the monoenergetic model with $\varepsilon_0=2$. Left panel: $x=2.1$; right panel: $x=3$. Exact numerical values are plotted with discrete points. The lines illustrate results of the slow-rotation approximation. We plot values of $n/n_\infty$ corresponding to three polar angles $\vartheta=0,\pi/4,\pi/2$.}
\end{figure}

\begin{figure}
\begin{center}
\includegraphics[height=0.46\textwidth]{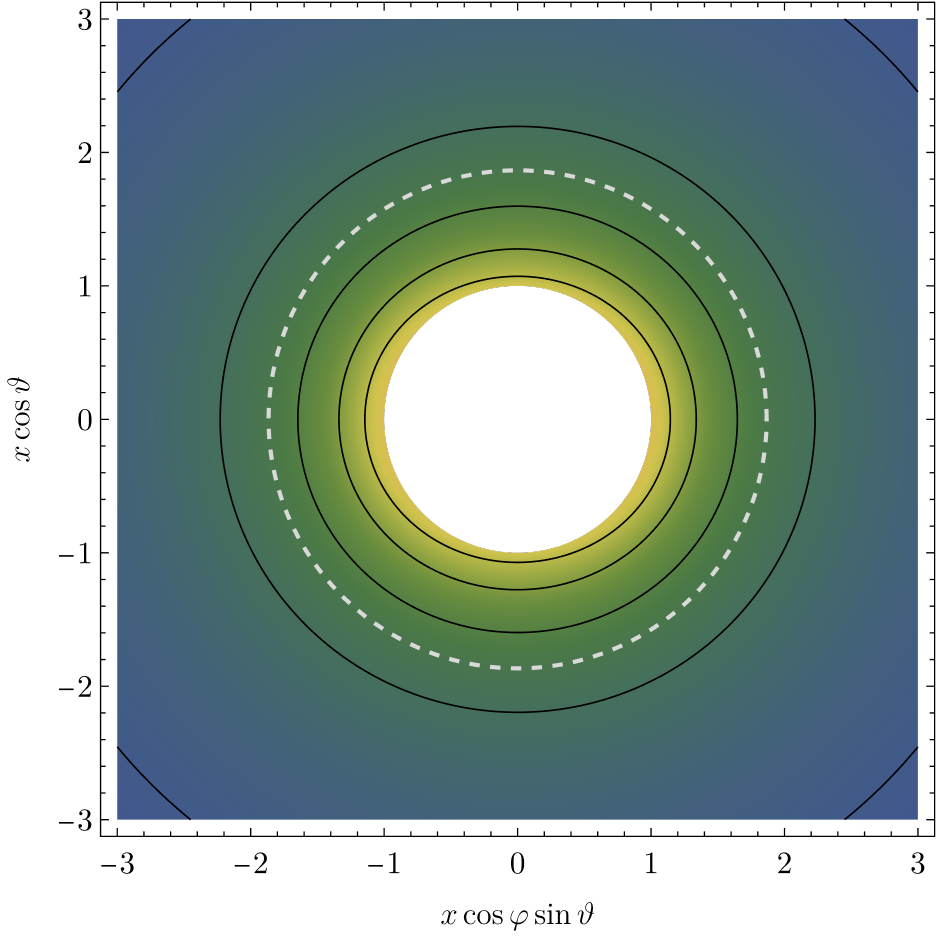}
\includegraphics[height=0.46\textwidth]{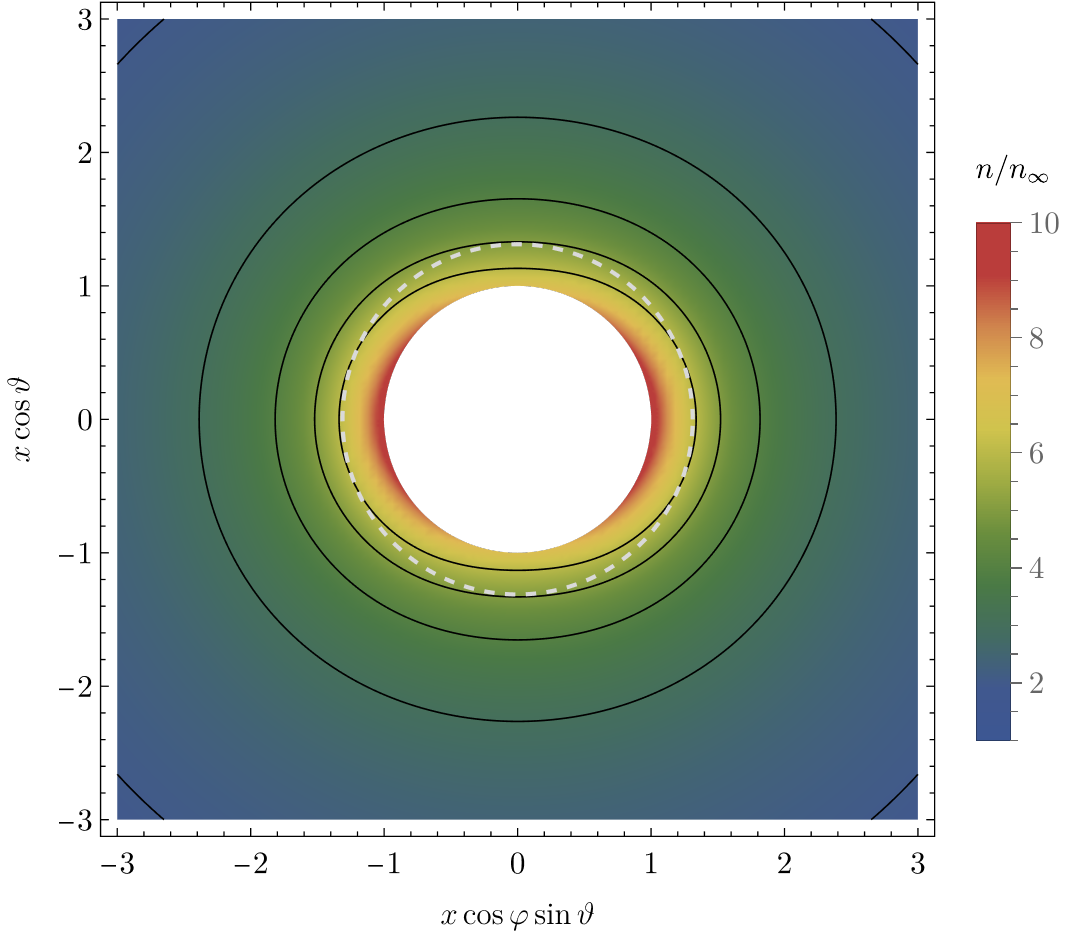}
\end{center}
\caption{\label{fig:nmeridplane} Compression ratio $n/n_\infty$ at the meridional plane ($\varphi = 0, \pi$) close to the black hole horizon. Left: $\alpha = 0.5$, right: $\alpha = 0.95$. In both cases $\varepsilon_0 = 2$. Contours of constant $n$ are marked with solid dark lines. Dashed lines mark the locations of the event horizon. Both plots share the same color scale.}
\end{figure}

Figures~\ref{fig:newCR}--\ref{fig:nmeridplane} illustrate the behavior of the compression ratio $n/n_\infty$. In Fig.\ \ref{fig:newCR} we plot radial profiles of $n$ at the equatorial plane ($\vartheta = \pi/2$) and at $\vartheta = \pi/4$, assuming $\varepsilon_0 = 2$ and $\alpha = 1/2$ or $\alpha = 0.95$.

The dependence of the compression ratio on $\vartheta$ and $\alpha$ is shown in Figs.~\ref{fig:CompRatAppTheta} and \ref{fig:CompRatAppAlpha}, respectively. For $\varepsilon_0 = 2$ and fixed values of the radius $x$ and the spin parameter $\alpha$, the particle density attains a maximum at the equatorial plane and reaches its lowest values at the axis. The left panels in Figs.~\ref{fig:CompRatAppTheta} and \ref{fig:CompRatAppAlpha} show that $n/n_{\infty}$ is a decreasing function of $\alpha$ at the poles, whereas it increases with $\alpha$ in the equatorial plane for sufficiently small radius and rotation parameters, say $x=2.1$ and $\alpha \leq 0.3$. For black hole spin parameters $\alpha > 0.6$, the compression ratio becomes an increasing function of $\alpha$ for all values of $\vartheta \in [0,\pi]$. The right panels of these figures show that $n/n_\infty$ is an increasing function of $\alpha$ for all values of $\alpha \in [0,1)$ and $\vartheta \in [0,\pi]$, and for sufficiently large radii, say $x \geq 3$. 

In all plots in Figs.\ \ref{fig:newCR}--\ref{fig:CompRatAppAlpha} discrete marks refer to exact numerical values. Results of the slow-rotation approximation are shown with lines. For $\varepsilon_0 = 2$ and $x \geq 2.1$, the difference between the values of $n/n_\infty$ obtained within the slow-rotation approximation and the exact numerical values is smaller than $0.7\%$ for $\alpha \leq 0.5$ and smaller than $6.1\%$ for $\alpha \leq 0.95$. For $x \geq 3$, the approximation is even better: the difference between the approximated and exact values drops below $0.2\%$ for $\alpha \leq 0.5$ and below $2.2\%$ for $\alpha \leq 0.95$.

Finally, in Fig.~\ref{fig:nmeridplane} we plot the exact numerical value of the particle number density in the meridional plane for $\varepsilon_0 = 2$ and the same two values of $\alpha = 0.5$ and $\alpha = 0.95$.

\begin{figure}
\begin{center}
\includegraphics[width=0.49\linewidth]{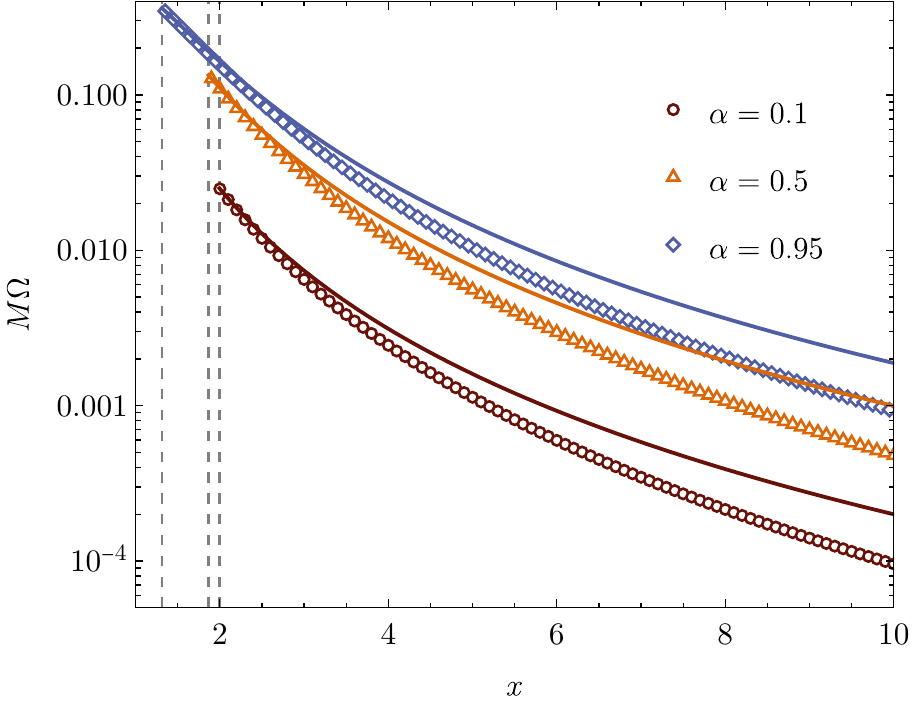}
\includegraphics[width=0.49\linewidth]{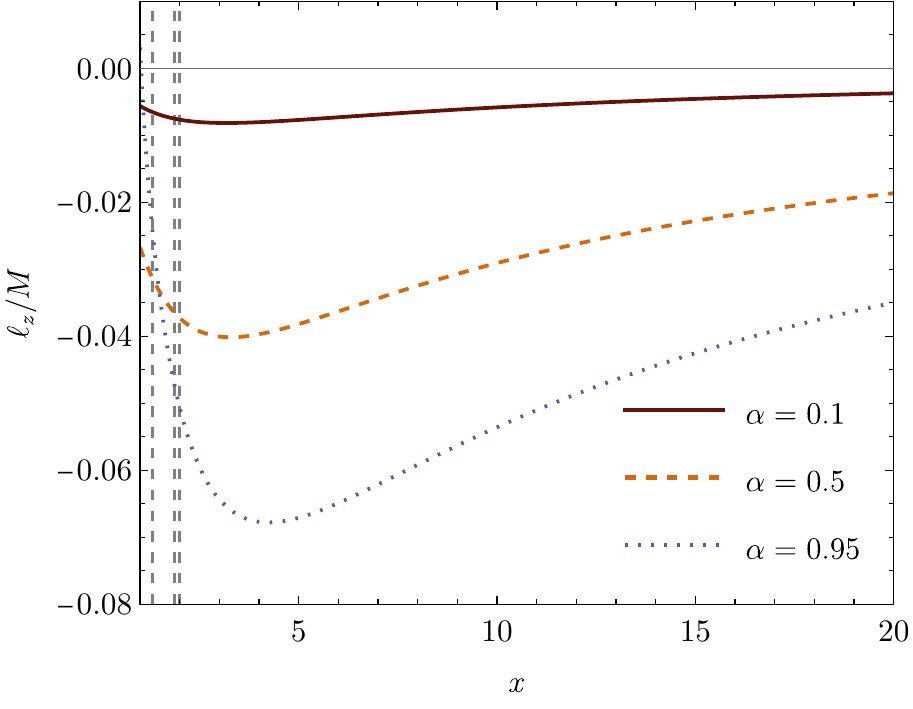}
\end{center}
\caption{\label{fig:OmegaLz} The angular velocity $\Omega$ (left) and the azimuthal angular momentum associated with an observer co-moving with the mean particle flow $\ell_z$ (right). In both plots $\varepsilon_0 = 2$ and $\vartheta = \pi/4$. In the left plot, solid lines depict the values of $\Omega_\mathrm{ZAMO}$. Dots represent numerical values obtained for accretion models with $\alpha = 0.1$, $0.5$, and $0.95$. Vertical dashed lines mark locations of black hole horizons ($x_+$).}
\end{figure}

\begin{figure}
\begin{center}
\includegraphics[width=0.49\linewidth]{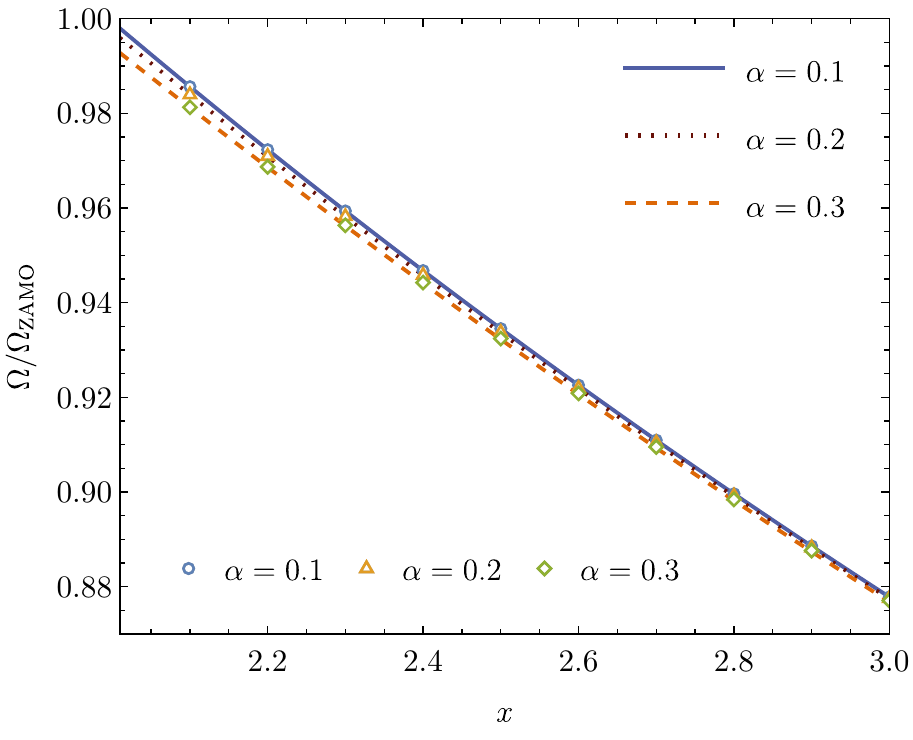}
\includegraphics[width=0.49\linewidth]{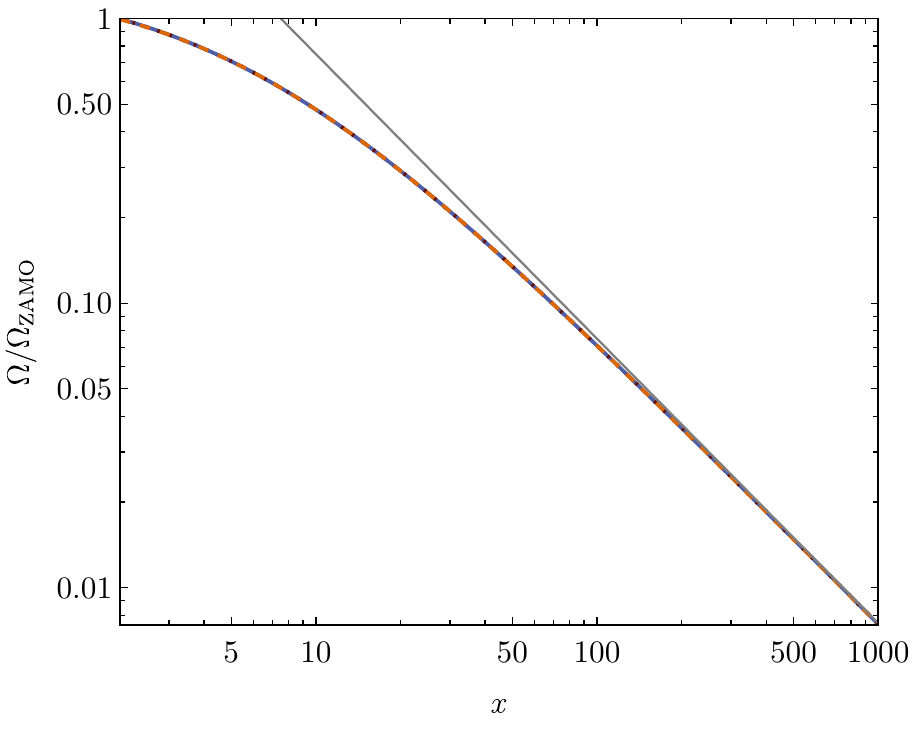}
\includegraphics[width=0.49\linewidth]{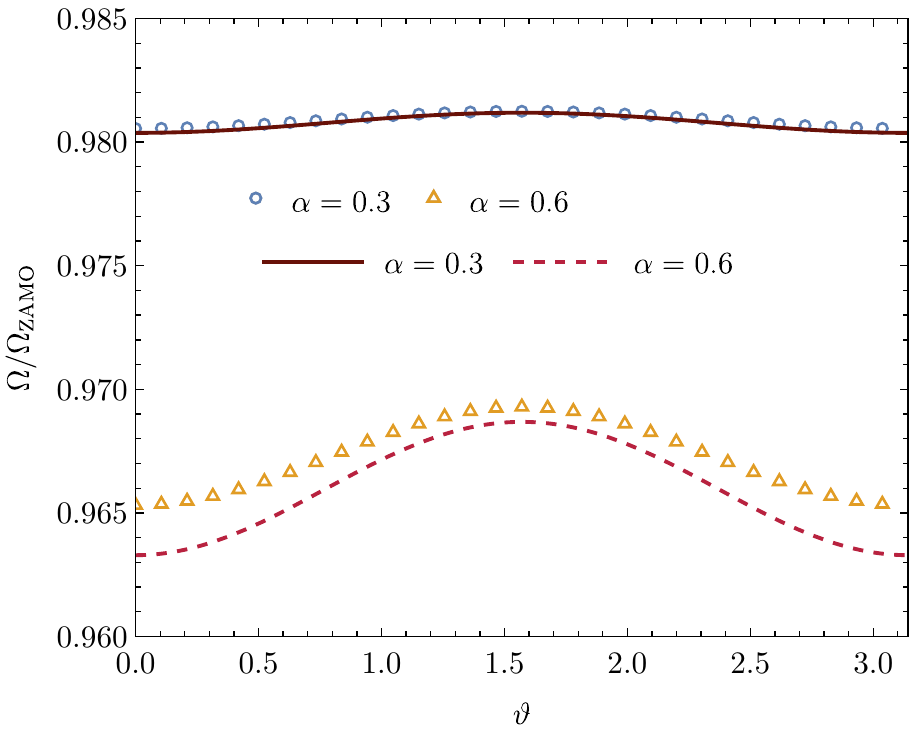}
\includegraphics[width=0.49\linewidth]{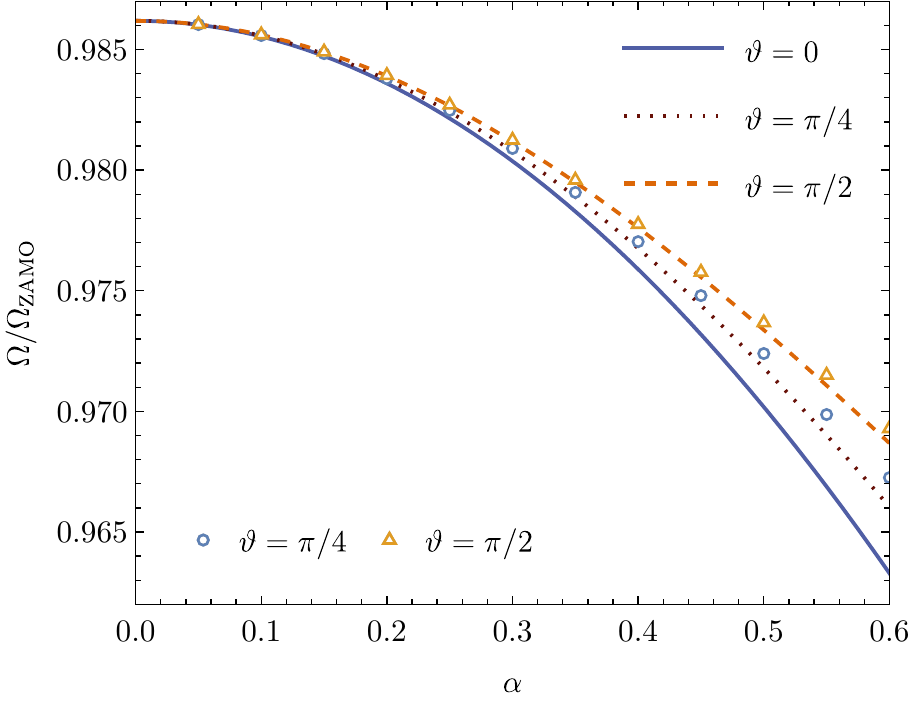}
\end{center}
\caption{\label{fig:OmegaApp} The profile of $\Omega/\Omega_\mathrm{ZAMO}$ versus $x$, $\vartheta$, and $\alpha$ for monoenergetic particles with $\varepsilon_0=2$. The upper panels show the values obtained at the equatorial plane ($\vartheta = \pi/2$). In lower panels we assume $x=2.1$. The continuous narrow line in the upper right panel shows $7.5/x$, which indicates the asymptotic behavior of the ratio $\Omega/\Omega_\mathrm{ZAMO}$ at infinity. Exact numerical results are plotted with discrete marks. Results of the slow-rotation approximation are shown with lines. The difference between the slow-rotation aproximation and the exact numerical values is highest at the axis $\vartheta = 0, \pi$. For $\vartheta \in [0,\pi]$ and $x \geq 2.1$, this difference is smaller than $0.21\%$ for $\alpha \leq 0.6$ and smaller than $2.7\%$ for $\alpha \leq 0.95$.}
\end{figure}

\begin{figure}
\begin{center}
\includegraphics[width=0.49\textwidth]{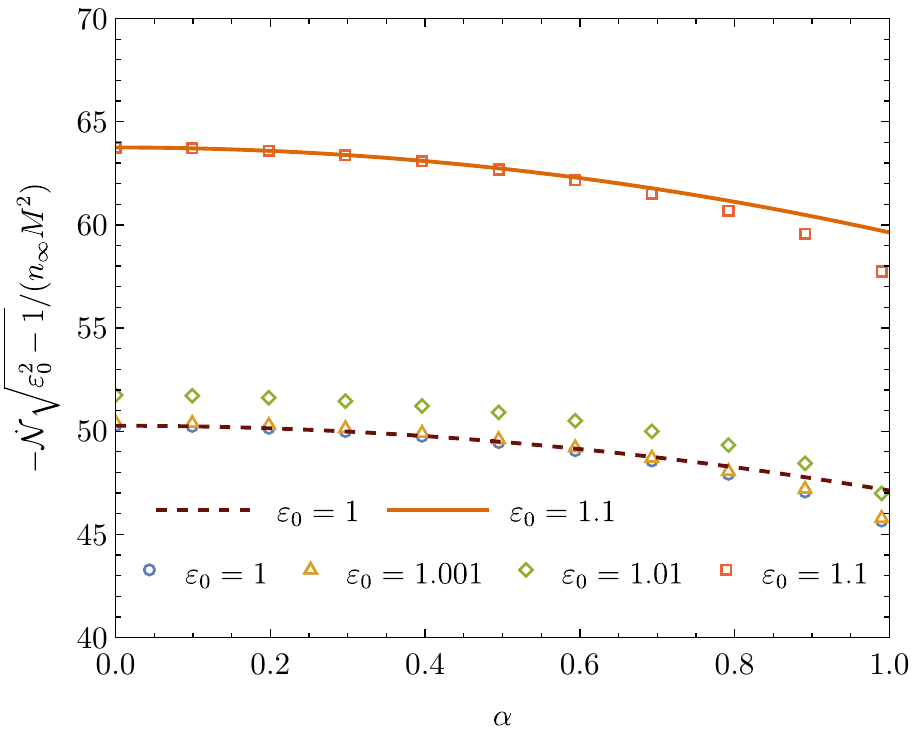}
\includegraphics[width=0.49\textwidth]{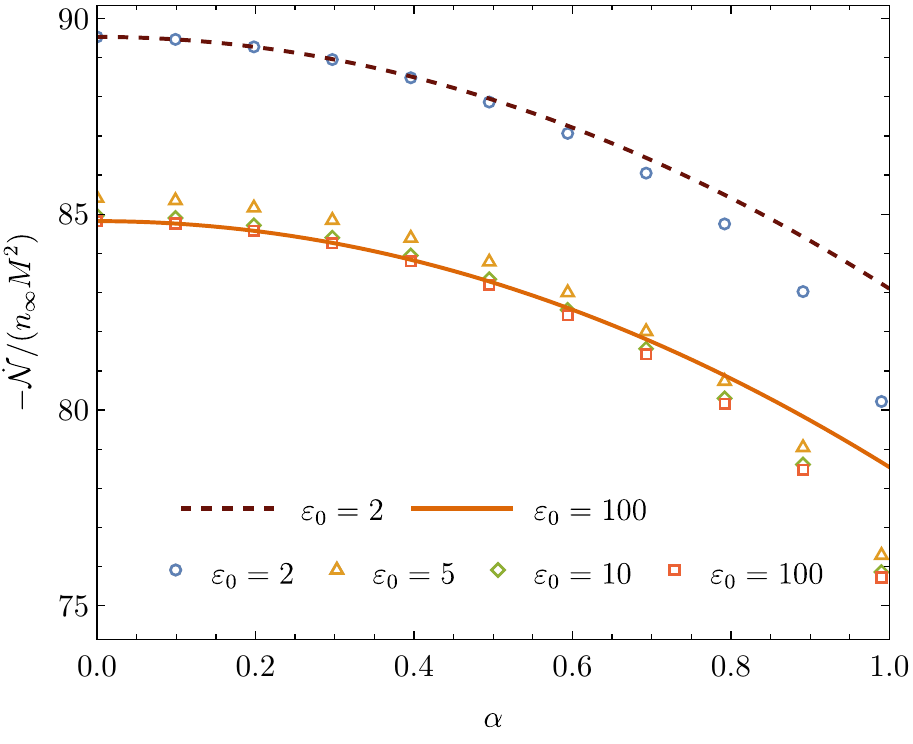}
\end{center}
\caption{\label{figNdot} Particle number accretion rates for monoenergetic models. We plot $- \dot{\mathcal N} \sqrt{\varepsilon_0^2 - 1}/(n_\infty M^2)$ for the energies $\varepsilon_0$ close to $\varepsilon_0 = 1$ (left panel) and  $-\dot{\mathcal N}/(n_\infty M^2)$ for large energies (right panel). Discrete points denote exact values obtained numerically. Results of the slow-rotation approximation are shown with lines.}
\end{figure}

\begin{figure}
\begin{center}
\includegraphics[width=0.49\textwidth]{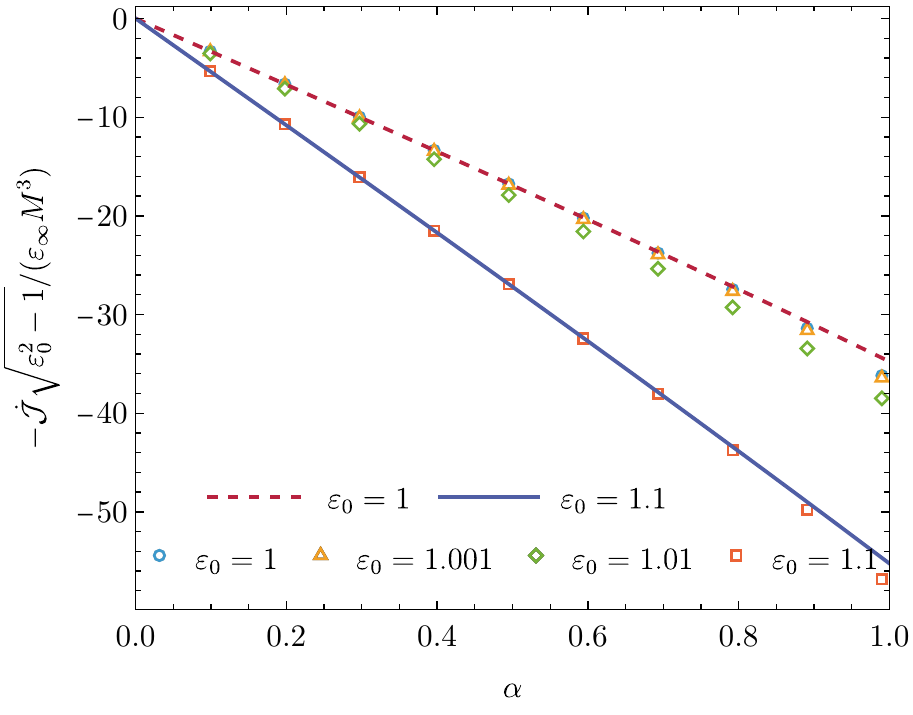}
\includegraphics[width=0.49\textwidth]{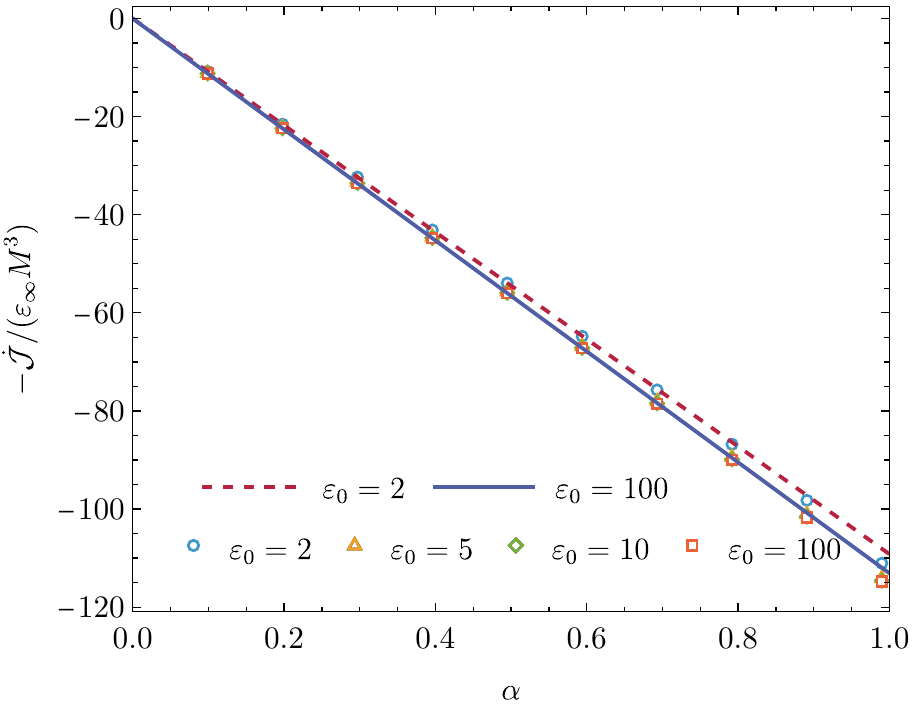}
\end{center}
\caption{\label{figJdot} Angular momentum accretion rates for monoenergetic models. We plot $-\dot{\mathcal J} \sqrt{\varepsilon_0^2 - 1}/(\varepsilon_\infty M^3)$ for the energies $\varepsilon_0$ close to $\varepsilon_0 = 1$ (left panel) and  $-\dot{\mathcal J}/(\varepsilon_\infty M^3)$ for large energies (right panel). For a rapidly rotating Kerr black hole with $\alpha \leq 0.99$, the error is less than $5\%$ in the left panel and less than $2.7\%$ in the right panel.}
\end{figure}

The angular velocity $\Omega$ and the specific angular momentum $\ell_z$ are illustrated in Fig.\ \ref{fig:OmegaLz}. The left panel shows a deviation of $\Omega$ from $\Omega_\mathrm{ZAMO}$ for fixed $\varepsilon_0 = 2$ and $\vartheta = \pi/4$, and three different values of the spin parameter $\alpha$. A comparison of $\Omega$ and $\Omega_\mathrm{ZAMO}$ is also provided in Fig.\ \ref{fig:OmegaApp}, both within the slow-rotation approximation and for exact numerical values. For $\varepsilon_0 = 2$, the ratio $\Omega/\Omega_\mathrm{ZAMO}$ decreases with the rotational parameter $\alpha$. For fixed values of $x$ and $\alpha$, $\Omega/\Omega_\mathrm{ZAMO}$ has a maximum on the equatorial plane and is the smallest on the axis. The exact numerical values are shown in Fig.\ \ref{fig:OmegaApp} with discrete marks. For $x \geq 2.1$, the difference between the approximate and exact values of $\Omega/\Omega_\mathrm{ZAMO}$ is smaller than $0.21\%$ for $\alpha \leq 0.6$ and remains below $2.7\%$ for $\alpha \leq 0.95$ in the entire range of $\vartheta$.

In Figs.~\ref{figNdot}--\ref{figJdot}, we show the dependence of $\dot{\mathcal{N}}$ and $\dot{\mathcal{J}}$, computed according to Eq.\ (\ref{Eq:AccretionRateME}), on the black hole spin parameter $\alpha$. The left panels show the behavior characteristic for the energies $\varepsilon_0$ close to one. In this case, we plot $- \dot{\mathcal N} \sqrt{\varepsilon_0^2 - 1}/(n_\infty M^2)$ and $- \dot{\mathcal J} \sqrt{\varepsilon_0^2 - 1}/(\varepsilon_\infty M^3)$, which tends to a finite limit as $\varepsilon_0 \to 1$. The right panels illustrate the behavior of $- \dot{\mathcal N}/(n_\infty M^2)$ and $- \dot{\mathcal J}/(\varepsilon_\infty M^3)$ for large energies $\varepsilon_0$. In all cases, when $\alpha=0$, we recover the values known from spherical models corresponding to the Schwarzschild background spacetime.

\subsection{Maxwell-J\"{u}ttner models}

\begin{figure}
\begin{center}
\includegraphics[width=0.49\textwidth]{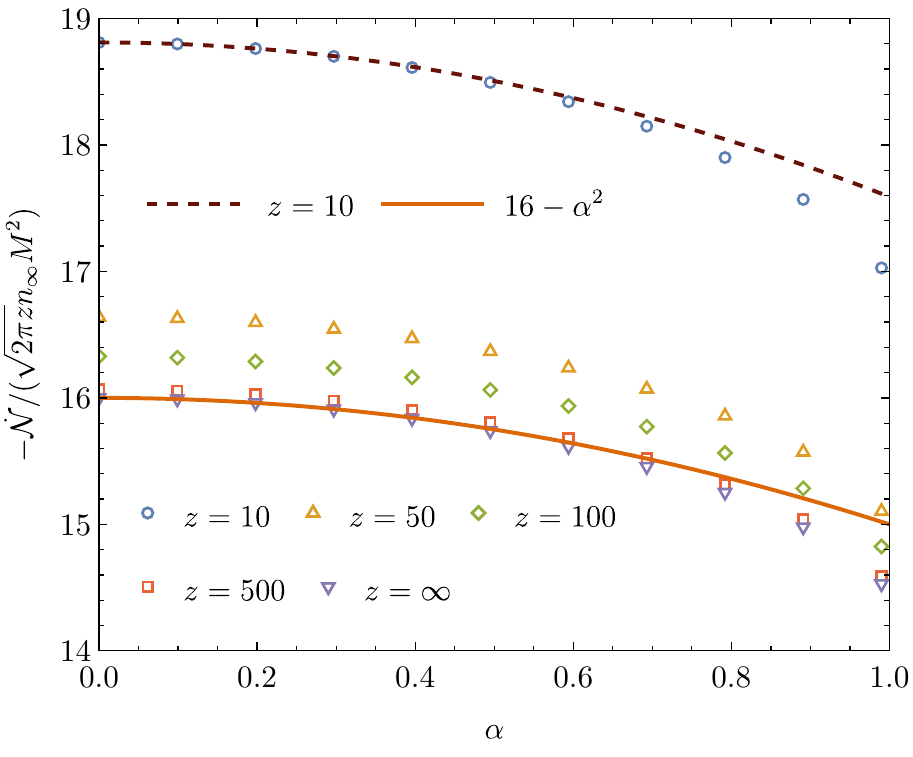}
\includegraphics[width=0.49\textwidth]{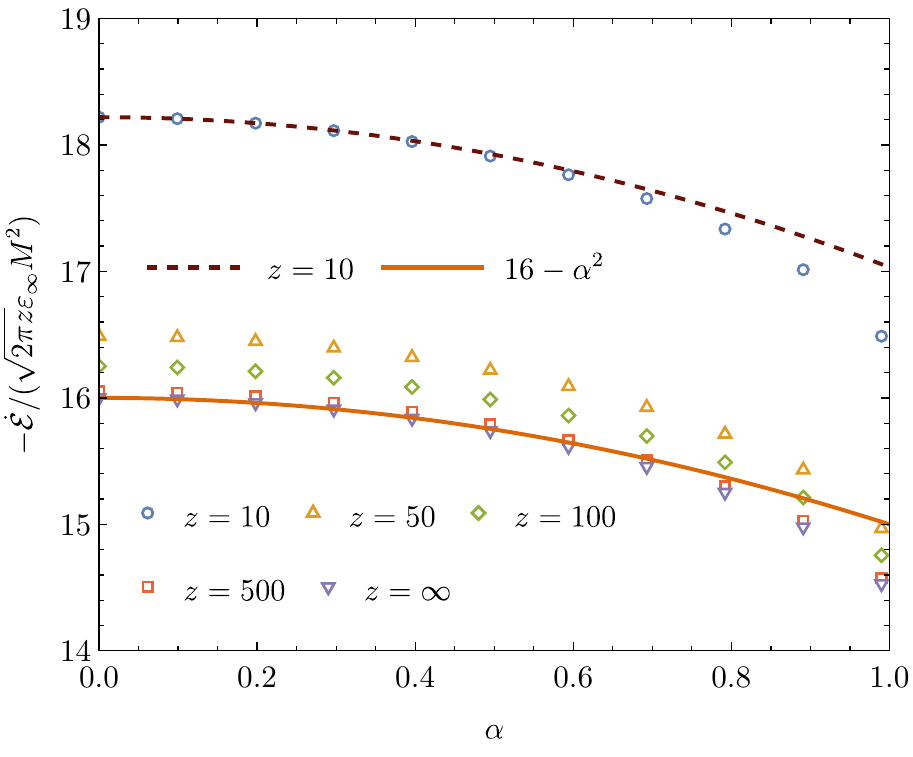}
\end{center}
\caption{\label{fig:accretionratesMJlargez} Particle number and energy accretion rates for the Maxwell-J\"{u}ttner model with large values of $z$. We plot $-\dot{\mathcal N}/(\sqrt{2 \pi z} n_\infty M^2)$ and $-\dot{\mathcal E}/(\sqrt{2 \pi z} \varepsilon_\infty M^2)$. The numerical values denoted as $z = \infty$ are computed according to Eq.\ (\ref{MJaccretionrateslargez}) and the corresponding continuous orange curves come from the relation~(\ref{low-temperatureTE}) obtained in the slow-rotation approximation. }
\end{figure}

\begin{figure}
\begin{center}
\includegraphics[width=0.49\textwidth]{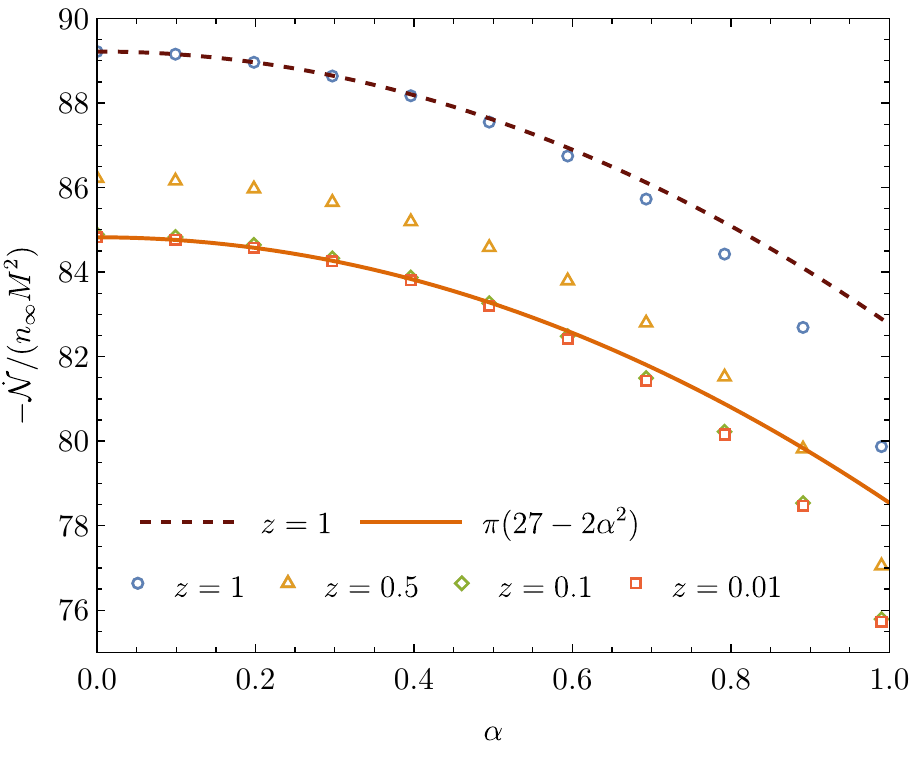}
\includegraphics[width=0.49\textwidth]{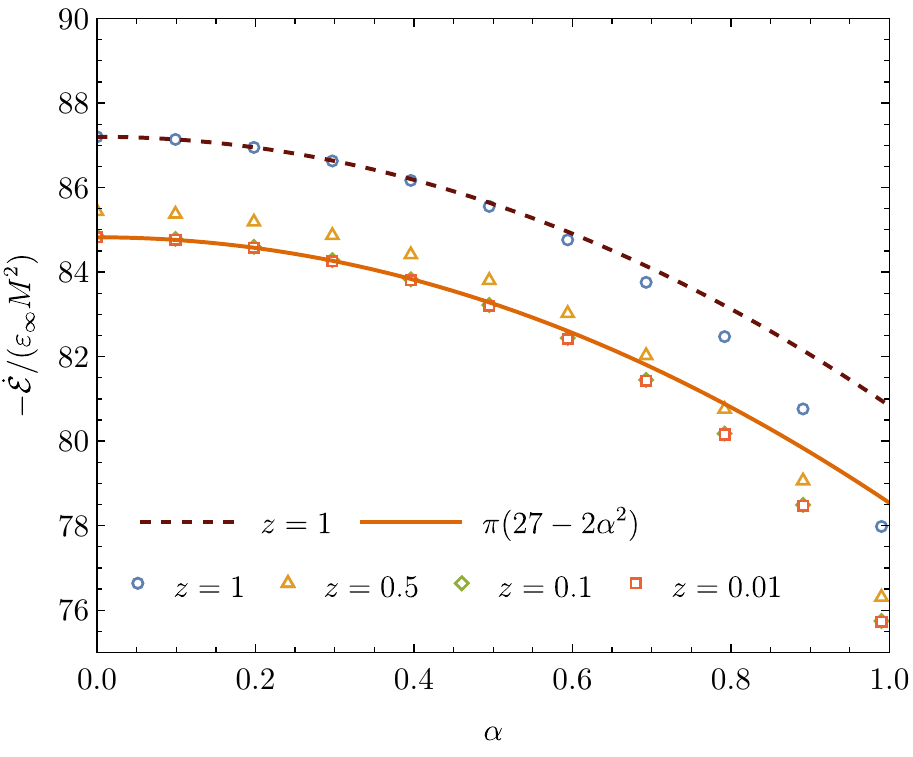}
\end{center}
\caption{\label{fig:accretionratesMJsmallz} Particle number and energy accretion rates for the Maxwell-J\"{u}ttner model with small values of $z$. The continuous orange curves are the accretion rates in the high-temperature limit derived in Eq.~(\ref{high-temperatureTE}) within the slow-rotation approximation.}
\end{figure}

\begin{figure}
    \begin{center}
    \includegraphics[width=0.49\textwidth]{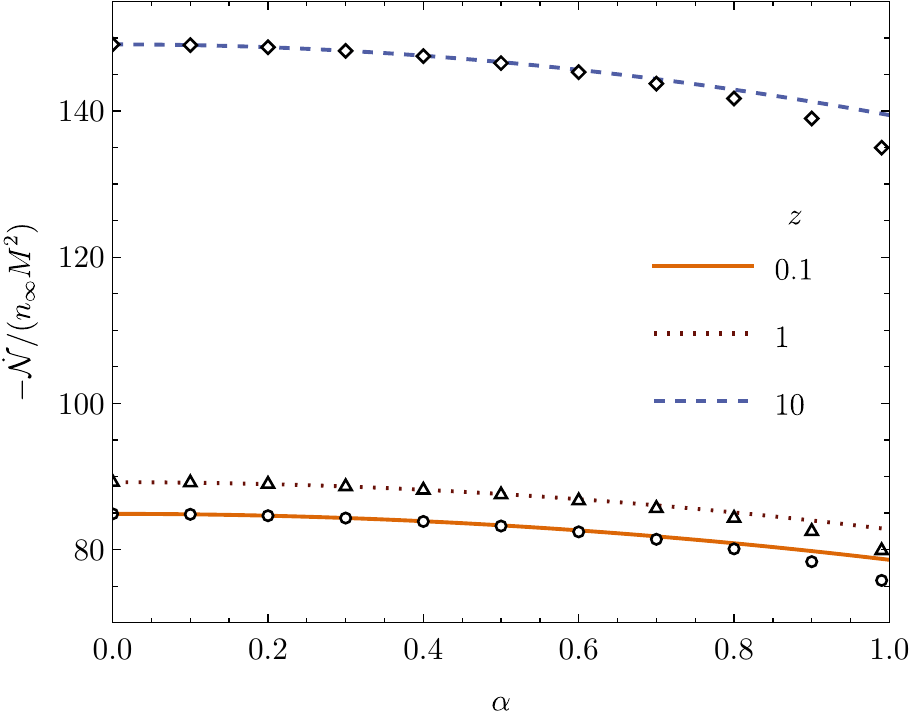}
    \includegraphics[width=0.49\textwidth]{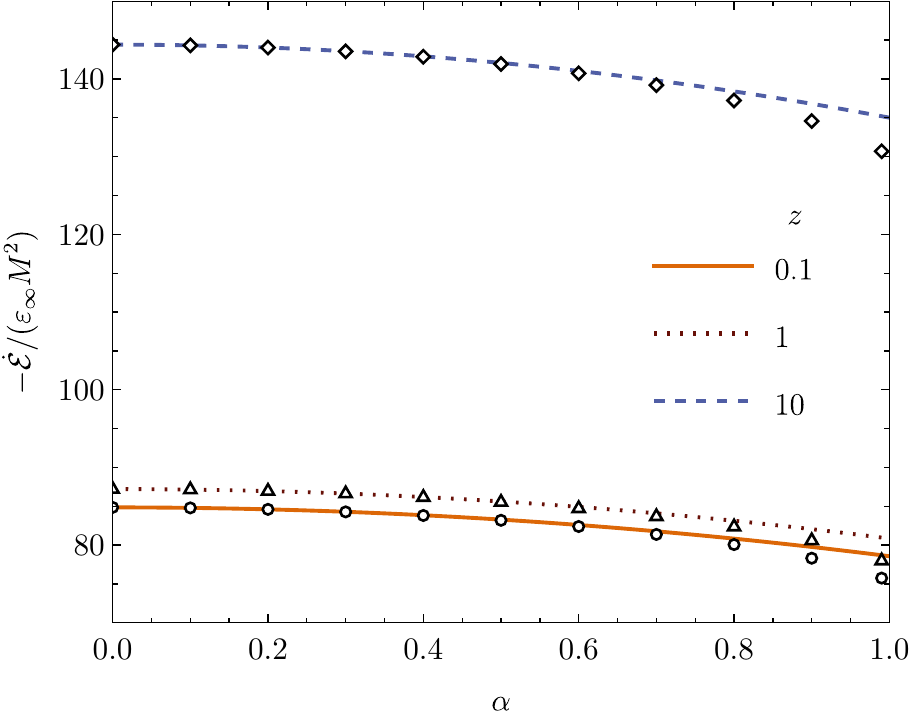}

    \end{center}
    \caption{\label{AccrRatFigs}  Particle number and energy accretion rates for the Maxwell-J\"{u}ttner distribution and various values of $z$ as a function of $\alpha$. The continuous curves represent the analytic values in the slowly rotating limit, and the discrete marks denote the corresponding numerical results. Surprisingly, the slow-rotation approximation remains in remarkable agreement with the numerical values: the deviation percentage is below $1\%$ for all rotating black holes with $\alpha \leq 0.8$ and less than $4\%$ even in the rapidly rotating case up to values of $\alpha \leq 0.99$.}
\end{figure}

\begin{figure}
\begin{center}
\includegraphics[width=0.49\textwidth]{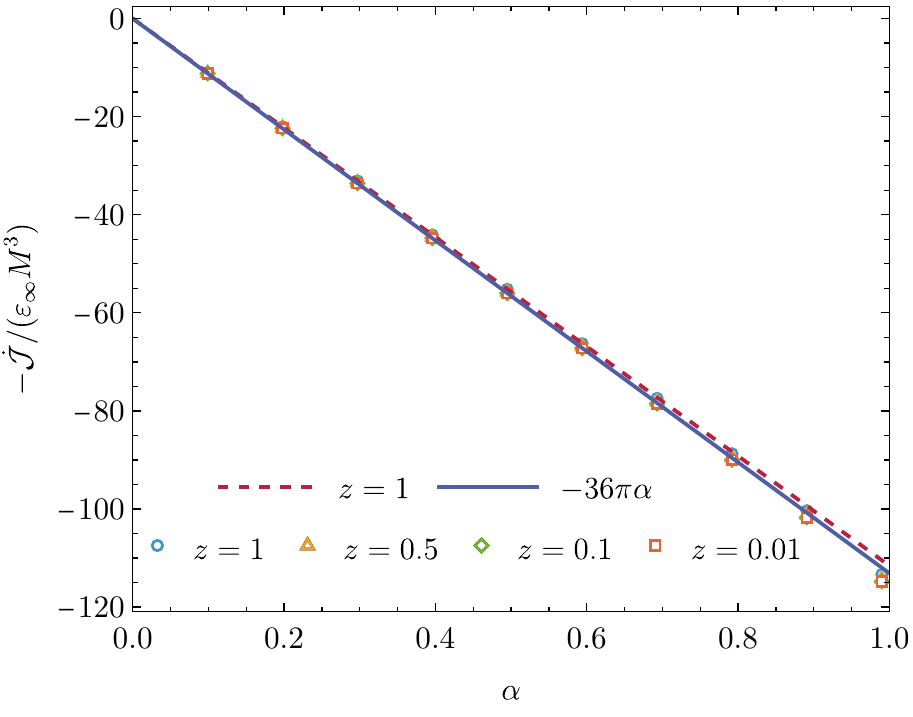}
\includegraphics[width=0.49\textwidth]{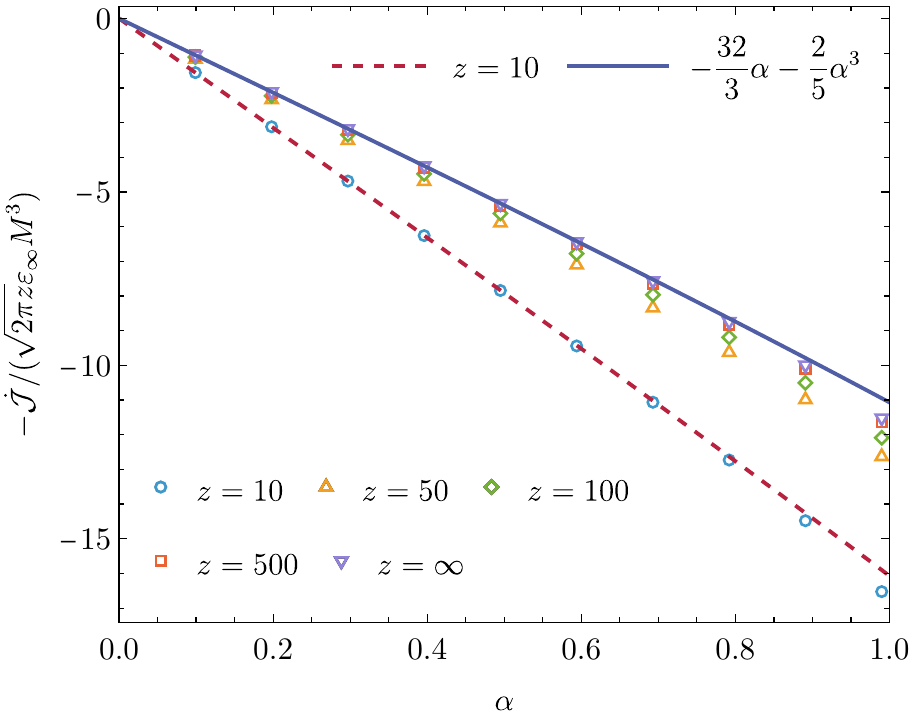}
\end{center}
\caption{\label{JdotFig}  The angular momentum accretion rates for the Maxwell-J\"{u}ttner distribution and various values of $z$ as a function of $\alpha$. For all rotating Kerr black holes with $\alpha \leq 0.99$, the error is less than $2.6\%$ in the left panel and less than $5\%$ in the right panel.}
\end{figure}

\begin{figure}
\begin{center}
\includegraphics[width=0.5\linewidth]{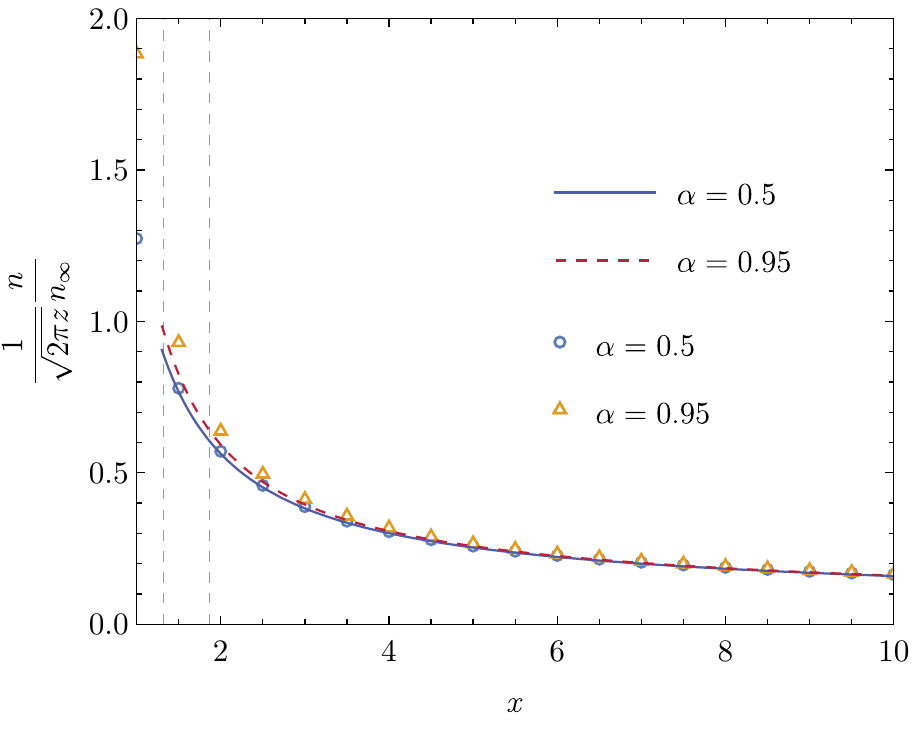}
\end{center}
\caption{\label{fig:nRateNumAppLargeZ} Compression ratio $n/n_\infty$ versus radius $x$ in the low-temperature limit at the equatorial plane  ($\vartheta = \pi/2$). Discrete marks (plotted for $x > 1$) correspond to exact numerical values obtained for $z = 200$. The analytic relation (\ref{eq:nRatioLargeZ}) of the slow-rotation approximation is plotted with lines for $x \geq 1.3$. Vertical dashed lines mark locations of the event horizon $x_+$.
At $x = 1.5$, the difference between approximated and exact values is less than $1.5\%$ (inside the horizon) for $\alpha=0.5$ and smaller than $12\%$ (outside the horizon) for $\alpha=0.95$.}
\end{figure}

As the next example, we consider Maxwell-J\"{u}ttner models defined by the distribution (\ref{FMJ}). In Figures~\ref{fig:accretionratesMJlargez}--\ref{JdotFig}, we show the particle, energy, and angular momentum accretion rates versus the spin parameter of the black hole for several values of the parameter $z$. Figure~\ref{fig:accretionratesMJlargez} and the right panel of Fig.~\ref{JdotFig} depict accretion rates $\dot{\mathcal N}$, $\dot{\mathcal E}$, and $\dot{\mathcal J}$ for large values of $z$ (small asymptotic temperatures). To illustrate the limit of $z \to \infty$, we plot the values of $-\dot{\mathcal N}/(\sqrt{2 \pi z} n_\infty M^2)$, $-\dot{\mathcal E}/(\sqrt{2 \pi z} \varepsilon_\infty M^2)$, and $-\dot{\mathcal J}/(\sqrt{2 \pi z} \varepsilon_\infty M^3)$. The values marked as $z = \infty$ correspond to Eq.\ (\ref{MJaccretionrateslargez}). Figure~\ref{fig:accretionratesMJsmallz} and the left panel of Fig.~\ref{JdotFig} show the values of $-\dot{\mathcal N}/(n_\infty M^2)$, $-\dot{\mathcal E}/(\varepsilon_\infty M^2)$, and $-\dot{\mathcal J}/(\varepsilon_\infty M^3)$ for small values of $z$, and the limiting behavior of Eq.\ (\ref{MJaccretionratessmallz}).

In all cases, $-\dot{\mathcal N}/(n_\infty M^2)$ and $-\dot{\mathcal E}/(\varepsilon_\infty M^2)$ decrease with $\alpha$, whereas $-\dot{\mathcal J}/(\varepsilon_\infty M^3)$ increases. This remains in agreement with the behavior observed for razor-thin accretion disks confined to the equatorial plane and investigated in~\cite{aCpMaO22}, or with the properties of accretion flows in the Reissner-Nordstr\"{o}m spacetime, understood as a toy-model for the Kerr geometry~\cite{aCpM20}. The same behavior is also shared by hydrodynamical models~\cite{aAeToSdL21}. In contrast to planar models of \cite{aCpMaO22}, both quantities $-\dot{\mathcal N}/(n_\infty M^2)$ and $-\dot{\mathcal E}/(\varepsilon_\infty M^2)$ decrease with increasing asymptotic temperature (they increase with $z$). This is again to be expected, as this behavior agrees with the properties of spherically symmetric flows in the Schwarzschild spacetime~\cite{pRoS16,pRoS17}. However, $-\dot{\mathcal J}/(\varepsilon_\infty M^3)$ is a decreasing function of $z$.

The curves presented in Figs.~\ref{figNdot}--\ref{JdotFig} correspond to the approximated particle, energy, and angular momentum accretion rates given by Eqs.\ (\ref{Eq:Extreme-temperatureTE}) and (\ref{Eq:AccretionRateMJapp}) for monoenergetic and Maxwell-J\"{u}ttner models. The difference between the results obtained within the slow-rotation approximation and the exact numerical values is smaller than $1\%$ ($1.1\%$ in case of $\mathcal{\dot{J}}$) for rotating Kerr black holes with $\alpha \leq 0.8$ and remains smaller than $4\%$ ($5\%$ in case of $\mathcal{\dot{J}}$) even for the rapidly rotating case with $\alpha \leq 0.99$.\footnote{For better visibility, some curves corresponding to the slow-rotation approximation are omitted in Figs.~\ref{figNdot}--\ref{fig:accretionratesMJsmallz} and~\ref{JdotFig}. We have checked that the statement regarding the difference between the slow-rotation approximation and the exact numerical values applies to the missing curves as well.} Thus, in practical applications it should be possible to replace the original triple integrals appearing in Eqs.\ (\ref{Eq:AccretionRateMJ}) with simpler formulas (\ref{Eq:AccretionRateMJapp}) obtained in the slow-rotation approximation, without introducing a significant error.

Finally, in Fig.~\ref{fig:nRateNumAppLargeZ}, we plotted the compression ratio $n/n_\infty$ versus radius $x$ in the low-temperature limit given by the analytic formula~(\ref{eq:nRatioLargeZ})  along with the corresponding exact numerical values. This figure shows that for points in the equatorial plane which have $x\geq 1.5$, the difference in $n/n_\infty$ between the results obtained within the slow-rotation approximation and the exact numerical values is smaller than $1.5\%$ for $\alpha=0.5$ and smaller than $12\%$ for $\alpha=0.95$.

\section{Conclusions}
\label{sec:conclusions}

In this work, we obtained new solutions of the Vlasov equation describing the Bondi-type accretion of a collisionless gas onto a rotating Kerr black hole. Deriving such solutions requires good control of the regions in the phase space available for the motion of the gas particles. Results presented in this work are based on a characterization of these regions in terms of specially adapted ``momentum'' coordinates $E$, $Q$, and $\chi$.\footnote{When this work was nearly completed, Ref.~\cite{yL25} appeared which treats the Bondi-type accretion problem in a Kerr-Newman black hole and uses a parametrization which, in the Kerr limit, reduces to the same variables $Q$ and $\chi$ used in this work. However, the author in~\cite{yL25} reaches different conclusions regarding the accretion rates.}

As particular examples, we considered two models, differing in the assumed asymptotic form of the one-particle distribution function: a monoenergetic distribution and a Maxwell-J\"{u}ttner distribution function characterized by constant temperature and constant particle density at infinity. In both cases, the particle and energy accretion rates decrease with the growing black hole spin parameter. We also find that the angular momentum accretion rate is roughly proportional to the spin parameter, slowing the black hole rotation down. This stays in agreement with the model of thin accretion disks confined to the equatorial plane, investigated in \cite{aCpMaO22,gKpM25}. In the quasistationary approximation (i.e., with the evolution of the system approximated by a sequence of stationary solutions corresponding to different black hole parameters), this effect, together with an increase of the black hole mass, should lead to a decrease in the black hole spin parameter $\alpha$.

Particle, energy, and the angular momentum accretion rates can also be computed assuming a slow-rotation approximation. Many of the results obtained within the slow-rotation approximation agree well with the exact solution, even for large values of the black hole spin parameters. This applies not only to the accretion rates, but also for components of the particle current density, except for a region close to the black hole horizon.

Many natural problems associated with the model presented in this paper are left for future studies. They include addressing questions about mass inflation at the Cauchy horizon or taking into account the effects of self-gravity of the gas. Taking into account the self-gravity (i.e., solving the full Einstein--Vlasov problem) requires addressing several issues. On one hand, different types of particle orbits can no longer be analyzed independently. In particular an inclusion of particles on bound orbits would affect particles associated with unbound ones. On the other hand, in a Bondi-type model a self-gravitating gas cannot extend to infinity. In spherical symmetry, hydrodynamical Bondi-type accretion of a self-gravitating fluid was investigated in~\cite{eMalec1999,jKarkowskia2006,pMeM2022,viDynE2011}, assuming boundary conditions imposed at a finite radius. Imposing boundary conditions at a finite boundary introduces a modification even in kinetic models with a fixed spacetime background (see, e.g.,~\cite{aGetal21,gKpM25}).

Since in this work we consider an unmagnetized gas of neutral particles our results cannot be applied to astrophysical plasmas. However, they do apply to the accretion of dark matter or distributions of stars. Some examples of dark matter models based on the kinetic approach can be found in~\cite{cAjN14,DarioGRG2017}. We discuss two short applications to the accretion of dark matter on black holes in the accompanying paper~\cite{letter}, both for hot (in the early universe) and cold dark matter. In this regard, another natural generalization of our model involves considering a uniform velocity of the gas at infinity (accretion onto a moving black hole), especially in the case in which the asymptotic velocity of the gas is misaligned with the rotation axis. That, in turn, implies natural questions about drag effects (see, e.g., \cite{lPsSrSsT1989,cDjRmMvC24}).

\begin{acknowledgments}
P.\ M.\ acknowledges a support of the Polish National Science Centre Grant No.\ 2017/26/A/ST2/00530. M.\ M.\ was supported by SECIHTI through Estancias
Posdoctorales por M\'exico Convocatoria 2023(1) under the postdoctoral Grant No.~1242413. O.\ S.\ was partially supported by CIC Grant No.~18315 to Universidad Michoacana and by CONAHCyT Network Project No.~376127 ``Sombras, lentes y ondas gravitatorias generadas por objetos compactos astrof\'isicos". Finally, M.\ M.\ and O.\ S.\ acknowledge financial support from SECIHTI-SNII. The data that support the findings of this article are
openly available \cite{articleDataFile2026}. 
\end{acknowledgments}

\appendix
\section{Existence and properties of $Q_c$}
\label{App:Qc}

In this appendix we discuss the definition of the critical function $Q_c$ introduced in Section~\ref{SubSec:NewParametrization}, which separates scattered trajectories from absorbed one in the $(\chi,Q,E)$-space. For simplicity, we shall work with the dimensionless variables $(s,q,\varepsilon)$ instead, where $s = \sin\chi$, $q = Q/(Mm)$, and $\varepsilon =E/m$. Likewise, we set $q_c := Q_c/(Mm)$.

To proceed, we recall from Section~\ref{SubSec:AbsorbedScattered} that in the $(\beta,\lambda,\varepsilon)$-space, scattered trajectories are separated from the absorbed ones by a critical surface $S_c$ which can be represented as
\begin{equation}
S_c = \left\{ \left( \beta,\lambda_\mathrm{sph}(\alpha,\beta,x),\varepsilon_\mathrm{sph}(\alpha,\beta,x) \right) \colon -1\leq \beta\leq 1, \, x_\mathrm{ph}(\alpha,\beta) < x < x_\mathrm{mb}(\alpha,\beta)
\right\},
\label{Eq:ScDef}
\end{equation}
where $\lambda_\mathrm{sph}$ and $\varepsilon_\mathrm{sph}$ are given in Eqs.~(\ref{lsph}), (\ref{esph}) and we have added the dependence on $\alpha$ and $\beta$ for convenience. Two linearly independent vectors tangent to $S_c$ are
\begin{equation}
v_1 := \begin{pmatrix}
    1 \\ \frac{\partial\lambda_\mathrm{sph}}{\partial\beta} \\
    \frac{\partial\varepsilon_\mathrm{sph}}{\partial\beta}
\end{pmatrix},
\qquad
v_2 := \begin{pmatrix}
    0 \\ -\frac{\partial \lambda_\mathrm{sph}}{\partial x} \\
    -\frac{\partial\varepsilon_\mathrm{sph}}{\partial x}
\end{pmatrix},
\label{Eq:TangentVectors}
\end{equation}
where the minus signs in $v_2$ have been introduced for convenience, taking into account that both $\frac{\partial \lambda_\mathrm{sph}}{\partial x}$ and $\frac{\partial\varepsilon_\mathrm{sph}}{\partial x}$ are negative in the interval of interest $x_\mathrm{ph} < x < x_\mathrm{mb}$. A vector normal to the surface $S_c$ is
\begin{equation}
n:=v_1\wedge v_2 = \begin{pmatrix}
-D \\ \frac{\partial\varepsilon_\mathrm{sph}}{\partial x} \\
-\frac{\partial \lambda_\mathrm{sph}}{\partial x}
\end{pmatrix},
\end{equation}
where
\begin{equation}
D:=\frac{\partial\varepsilon_\mathrm{sph}}{\partial x}\frac{\partial \lambda_\mathrm{sph}}{\partial\beta} - \frac{\partial\varepsilon_\mathrm{sph}}{\partial\beta}\frac{\partial \lambda_\mathrm{sph}}{\partial x}.
\end{equation}
Since $n\neq 0$ everywhere on $S_c$, it follows that $S_c$ is a smooth hypersurface in the $(\beta,\lambda,\varepsilon)$-space.

For later use we note that $D$ can be represented as
\begin{equation}
D = -\alpha\frac{\lambda_\mathrm{sph}}{x^2}\frac{\partial\lambda_\mathrm{sph}}{\partial x},
\label{Eq:DExplicit}
\end{equation}
which can be deduced with the help of Eqs.~(B10)--(B13) and (B23) in~\cite{pRoS24} and the identity
\begin{equation}
\frac{\partial\varepsilon_\mathrm{sph}}{\partial\beta} = \frac{\alpha\lambda_\mathrm{sph}}{x^2} + \frac{\sqrt{ x- \alpha^2(1-\beta^2)}}{x^2}\frac{\partial \lambda_\mathrm{sph}}{\partial\beta}.
\label{Eq:BetaDerivId}
\end{equation}
In particular, $D$ has a fixed sign on $S_c$ as long as $\alpha\neq 0$.

The relation between the variables $(L,L_z)$ and $(Q,\chi)$ introduced in Section~\ref{SubSec:NewParametrization} can be described by the following map: For each fixed $\alpha\in [0,1)$ and $\vartheta\in (0,\pi)$ one considers
\begin{equation}
T_{\alpha,\vartheta} \colon U_0\to \mathbb{R}^3,\quad
(s,q,\varepsilon)\mapsto (\beta,\lambda,\varepsilon),
\end{equation}
where $U_0:=(-1,1)\times (0,\infty)\times (0,\infty)$, and
\begin{align}
\beta &=\frac{q s\sin\vartheta - \alpha\varepsilon\cos^2\vartheta}{\sqrt{q^2 + \alpha^2\cos^2\vartheta}},\\
\lambda &= \sqrt{q^2 + \alpha^2\cos^2\vartheta},\\
\varepsilon &= \varepsilon.
\end{align}
Then, the relation between $(Q,\chi)$ and $(L,\beta)$ in terms of dimensionless variables is given by 
\begin{equation}
\left( \beta,\lambda,\varepsilon \right) = T_{\alpha,\vartheta}\left( \sin\chi,q,\varepsilon \right).
\end{equation}

It is easy to verify that $T_{\alpha,\vartheta}$ is a smooth and injective map, such that it is invertible when restricted to its image $V_0 := T_{\alpha,\vartheta}(U_0)$. In order to determine $V_0$ we first note that for $\vartheta=\pi/2$ this map reduces to the identity map. Likewise, when $\alpha=0$ the map simplifies to $T_{\alpha,\vartheta}(s,q,\varepsilon) = (s\sin\vartheta,q,\varepsilon)$. In both cases, we have $V_0 = U_0$.

\begin{figure}
    \begin{center}
    \includegraphics[width=0.49\textwidth]{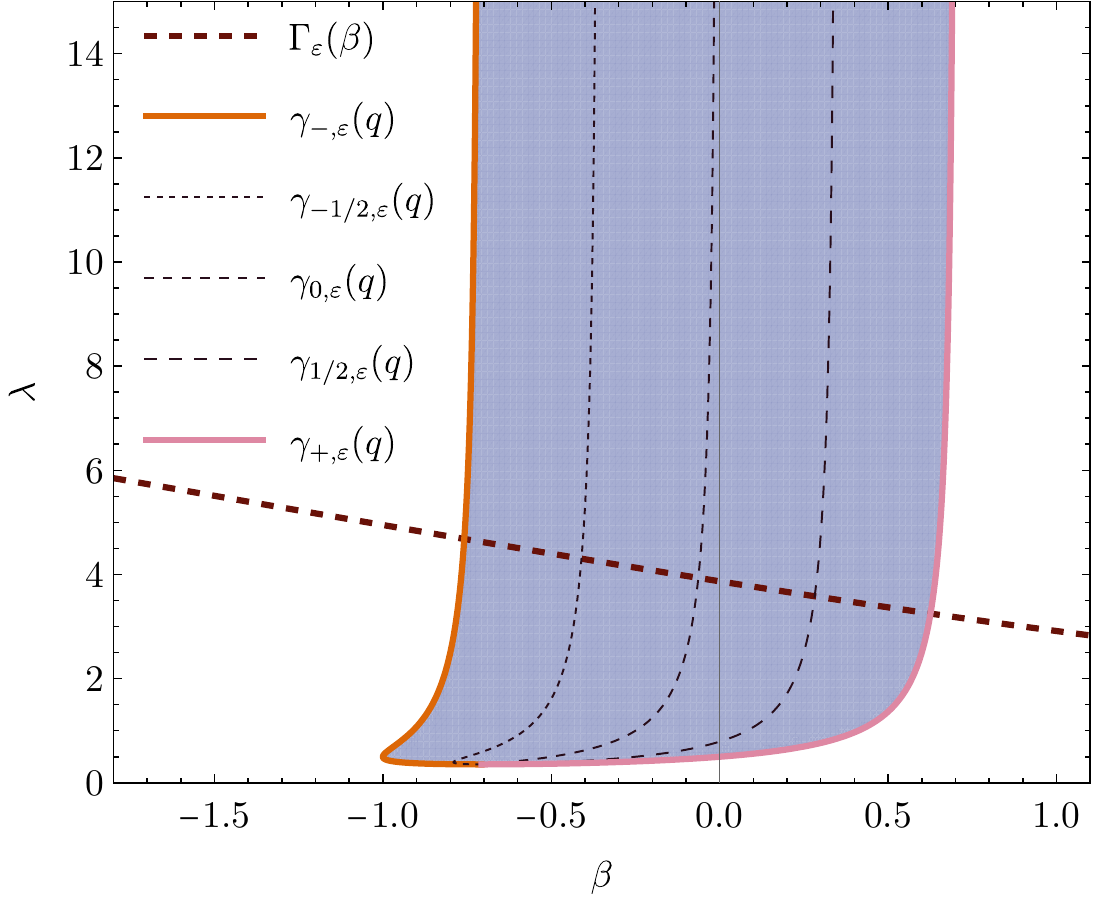}
    \includegraphics[width=0.49\textwidth]{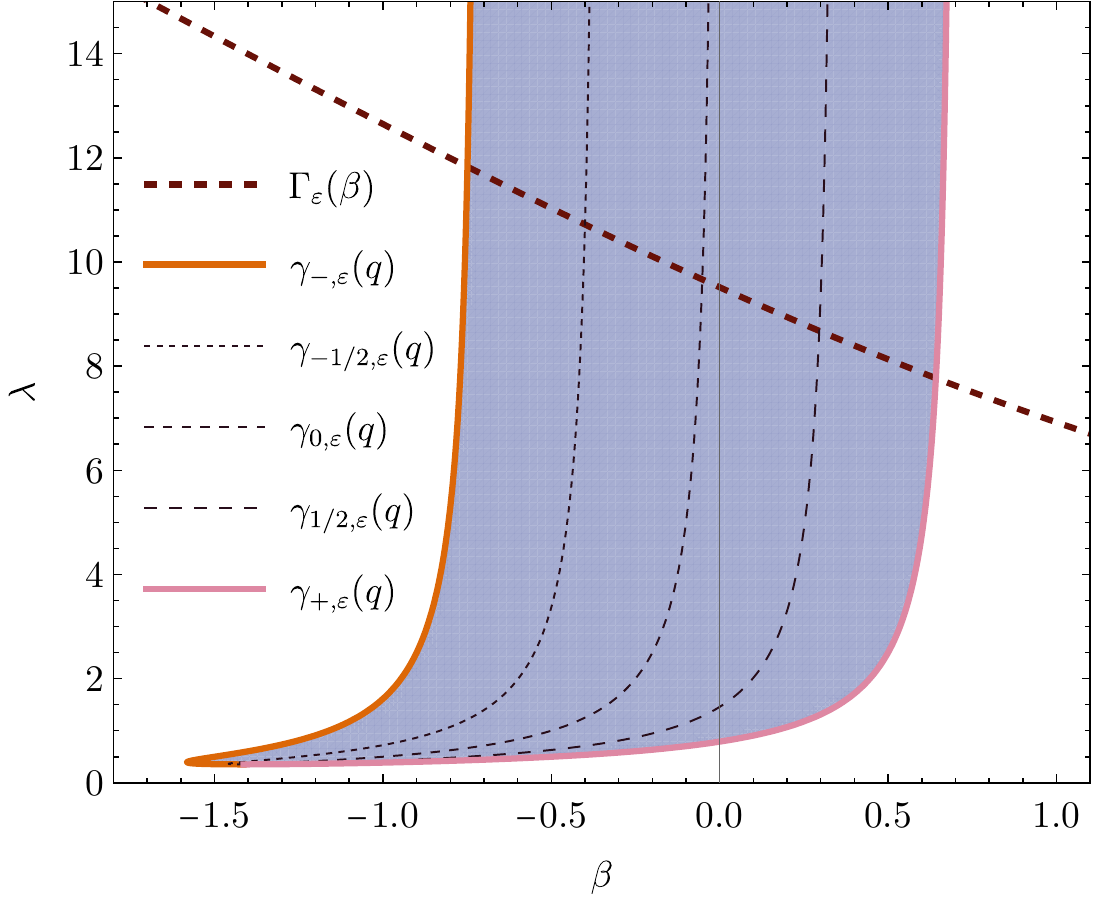}
    \end{center}
    \caption{\label{figV0} Illustrations of the set $V_0$ (blue shaded regions) and the curves $\Gamma_\varepsilon(\beta)$, $\gamma_{s,\varepsilon}(q)$, and $\gamma_{\pm,\varepsilon}(q)$. Left panel: $V_0 \cap \{ \varepsilon = 1\}$, right panel: $V_0 \cap \{ \varepsilon = 2 \}$. In both cases $\alpha = 1/2$, $\vartheta = \pi/4$.}
\end{figure}

It remains to determine $V_0$ when $\alpha > 0$ and $\vartheta\neq \pi/2$. For this, note first that the boundary portion corresponding to $q=0$ is mapped onto the line
\begin{equation}
(-\varepsilon|\cos\vartheta|,\alpha|\cos\vartheta|,\varepsilon),\qquad\varepsilon \geq 0.
\end{equation}
Next, we look at the images of the boundary segments $(\pm 1,q,\varepsilon)$ for constant $\varepsilon$, which are mapped onto the curves
\begin{equation}
\gamma_{\pm,\varepsilon}(q) := \left( \beta_{\pm,\varepsilon}(q),\sqrt{q^2 + \alpha^2\cos^2\vartheta},\varepsilon \right),\qquad
q \geq 0,
\end{equation}
with
\begin{equation}
\beta_{\pm,\varepsilon}(q) := \frac{\pm q \sin\vartheta - \alpha\varepsilon\cos^2\vartheta}{\sqrt{q^2 + \alpha^2\cos^2\vartheta}}.
\end{equation}
By differentiating the functions $\beta_{\pm,\varepsilon}$ one finds that $\beta_{+,\varepsilon}$ is monotonously increasing from $-\varepsilon|\cos\vartheta|$ to $\sin\vartheta$. Meanwhile, $\beta_{-,\varepsilon}$ decreases on the interval $0 < q <  \alpha\sin\vartheta/\varepsilon$, and increases monotonically to $-\sin\vartheta$ as $q$ increases from $\alpha\sin\vartheta/\varepsilon$ to $\infty$. Its minimum value is $-\sqrt{1+(\varepsilon^2-1)\cos^2\vartheta}$, which is smaller than $-1$ when $\varepsilon > 1$. Furthermore, one can show that
\begin{equation}
\left. \frac{d\beta_{\pm,\varepsilon}}{dq} \right|_{q=0} = \pm \frac{|\tan\vartheta|}{\alpha}.
\end{equation}
Two different cross sections of the boundary of $V_0$ are shown in the plots of Fig.~\ref{figV0} for the parameter values $\alpha=1/2$ and $\vartheta=\pi/4$.

For the accretion problem of interest in this article, $\varepsilon > 1$, and hence we can restrict our attention to the corresponding subsets $U_1\subset U_0$ and $V_1\subset V_0$, where $\varepsilon$ is restricted to be larger than one. In the following, we prove that the critical surface $S_c$ divides $V_1$ into two connected sets, describing the absorbed and scattered particles, respectively. For this, it is sufficient to show that the boundary curves
\begin{equation}
\Gamma_\pm(x) := \left( \pm 1, \lambda_\mathrm{sph}(\alpha,\pm 1, x), \varepsilon_\mathrm{sph}(\alpha,\pm 1,x ) \right),\qquad
x_\mathrm{ph}(\alpha,\pm 1) < x < x_\mathrm{mb}(\alpha,\pm 1),
\end{equation}
of $S_c$ lie outside $V_1$. Clearly, this is the case for $\Gamma_+$ since $\beta_{+,\varepsilon} < 1$. Hence, it remains to show that $\Gamma_-$ lies outside $V_1$. This is the case if
\begin{equation}
\beta_-(x):= \left. \beta_{-,\varepsilon}(q) \right|_{\varepsilon=\varepsilon_\mathrm{sph}(\alpha,-1,x), q = \sqrt{\lambda_\mathrm{sph}(\alpha,-1,x)^2 - \alpha^2\cos^2\vartheta}},
\end{equation}
is larger than $-1$ for all $x_\mathrm{ph}(\alpha,-1) < x < x_\mathrm{mb}(\alpha,-1)$. Setting $\lambda(x) := \lambda_\mathrm{sph}(\alpha,-1,x)$ and $\varepsilon(x):=\varepsilon_\mathrm{sph}(\alpha,-1,x)$ to simplify the notation, the condition $\beta_-(x) > -1$ is equivalent to
\begin{equation}
\lambda(x) > \alpha\cos^2\vartheta\varepsilon(x)
\quad\hbox{and}\quad
\lambda(x)^2 - 2\alpha\lambda(x)\varepsilon(x) + \alpha^2\left[ 1 + \cos^2\vartheta(\varepsilon(x)^2-1) \right] > 0.
\end{equation}
This needs to be satisfied independent of the value of $\vartheta$. In particular, we 
see that when $\varepsilon(x)\geq 1$ these conditions are implied by the stronger conditions
\begin{equation}
\lambda(x) > \alpha\varepsilon(x)
\quad\hbox{and}\quad
\lambda(x)^2 - 2\alpha\lambda(x)\varepsilon(x) + \alpha^2 > 0.
\end{equation}
Using Eqs.~(B11)--(B13) in~\cite{pRoS24} it is easy to check that these conditions are always satisfied.

After these comments, it is a simple task to determine $q_c$ and to prove that it gives rise to a well-defined smooth function of $s$ and $\varepsilon$. For this, we first note that for fixed $\varepsilon > 1$, the critical curve
\begin{equation}
\Gamma_\varepsilon(\beta) := (\beta,\lambda_c(\alpha,\beta,\varepsilon),\varepsilon),\qquad
\lambda_c(\alpha,\beta,\varepsilon) = \lambda_\mathrm{sph}(\alpha,\beta,x(\alpha,\beta,\varepsilon)),
\end{equation} 
where $x = x(\alpha,\beta,\varepsilon)\in (x_\mathrm{ph}(\alpha,\beta),x_\mathrm{mb}(\alpha,\beta))$ is implicitly determined from the condition
\begin{equation}
\varepsilon_\mathrm{sph}(\alpha,\beta,x) = \varepsilon,
\end{equation}
satisfies
\begin{equation}
\frac{\partial\lambda_c}{\partial \beta} = \left( \frac{\partial\varepsilon_\mathrm{sph}}{\partial x} \right)^{-1} D < 0,
\label{Eq:dlambdacdbeta}
\end{equation}
which is true since $D > 0$ and $\frac{\partial\varepsilon_\mathrm{sph}}{\partial x} < 0$. Using Eq.~(\ref{Eq:DExplicit}) and the identity
\begin{equation}
    \left( \frac{\partial \varepsilon_\mathrm{sph}}{\partial x} \right)^{-1} \frac{\partial \lambda_\mathrm{sph}}{\partial x} = \frac{x^2}{\sqrt{x + \alpha^2 (\beta^2 - 1)}},
\end{equation}
one can express Eq.~(\ref{Eq:dlambdacdbeta}) in explicit form and obtain
\begin{equation}
    \frac{\partial\lambda_c}{\partial \beta} = - \alpha \frac{\lambda_\mathrm{sph}}{x^2} \left( \frac{\partial \varepsilon_\mathrm{sph}}{\partial x} \right)^{-1} \frac{\partial \lambda_\mathrm{sph}}{\partial x} = - \frac{\alpha \lambda_\mathrm{sph}}{\sqrt{x + \alpha^2 (\beta^2 - 1)}} < 0.
\end{equation}

On the other hand, for fixed $s\in (-1,1)$, the curve
\begin{equation}
\gamma_{s,\varepsilon}(q) := \left( \beta_{s,\varepsilon}(q),\sqrt{q^2 + \alpha^2\cos^2\vartheta},\varepsilon \right),\qquad
\beta_{s,\varepsilon}(q) := \frac{q s\sin\vartheta - \alpha\varepsilon\cos^2\vartheta}{\sqrt{q^2 + \alpha^2\cos^2\vartheta}},
\end{equation}
has $\lambda$ growing for increasing $q$. Therefore, for each $s\in (-1,1)$ and $\varepsilon > 1 $ there is a unique intersection between the two curves $\Gamma_\varepsilon$ and $\gamma_{s,\varepsilon}$, which determines $q$ as a function of $s$ and $\varepsilon$. By definition, for each $s$ and $\varepsilon$, this particular value of $q$ yields the value of $q_c$. For an illustration showing these curves and their intersection, see Fig.~\ref{figV0}.

An alternative way to characterize $q_c$ is to consider the inverse image $S_c':=T_{\alpha,\vartheta}^{-1}(S_c\cap V_1)$ of the critical surface $S_c$ defined in Eq.~(\ref{Eq:ScDef}) and to prove that it defines a smooth hypersurface in $U_0$ whose normal vector $N$ has an everywhere non-vanishing $q$-component. The fact that $S_c'$ is smooth follows from the smoothness of $S_c$ and $T_{\alpha,\vartheta}^{-1}$. To compute $N$, we first map the two tangent vectors $v_1$ and $v_2$ defined in Eq.~(\ref{Eq:TangentVectors}) using the inverse Jacobian $(DT_{\alpha,\vartheta})^{-1}$, i.e., 
$v_1' := (DT_{\alpha,\vartheta})^{-1} v_1$, $v_2' := (DT_{\alpha,\vartheta})^{-1} v_2$ and then define $N:=v_1'\wedge v_2'$. The inverse Jacobian reads
\begin{equation}
(DT_{\alpha,\vartheta})^{-1} = \begin{pmatrix}
\frac{\lambda}{q\sin\vartheta} &
-\frac{\alpha\cos^2\vartheta(\alpha\beta + \lambda\varepsilon)}{q^3\sin\vartheta} & \frac{\alpha\cos^2\vartheta}{q\sin\vartheta} \\
0 & \frac{\lambda}{q} & 0 \\
0 & 0 & 1
\end{pmatrix}.
\end{equation}
This yields
\begin{align}
v_1' &= \begin{pmatrix}
    \frac{1}{q_\mathrm{sph}\sin\vartheta}\left[ \lambda_\mathrm{sph} - \frac{\alpha\cos^2\vartheta(\alpha\beta+\lambda_\mathrm{sph}\varepsilon_\mathrm{sph})}{q_{sph}^2}\frac{\partial\lambda_\mathrm{sph}}{\partial\beta} + \alpha\cos^2\vartheta\frac{\partial\varepsilon_\mathrm{sph}}{\partial\beta}\right] \\ \frac{\lambda_\mathrm{sph}}{q_\mathrm{sph}}\frac{\partial\lambda_\mathrm{sph}}{\partial\beta} \\
    \frac{\partial\varepsilon_\mathrm{sph}}{\partial\beta}
\end{pmatrix},\\
v_2' &= \begin{pmatrix}
    \frac{\alpha\cos^2\vartheta}{q_\mathrm{sph}\sin\vartheta}\left[ \frac{\alpha\beta + \lambda_\mathrm{sph}\varepsilon_\mathrm{sph}}{q_\mathrm{sph}^2}\frac{\partial\lambda_\mathrm{sph}}{\partial x} - \frac{\partial\varepsilon_\mathrm{sph}}{\partial x} \right] \\ -\frac{\lambda_\mathrm{sph}}{q_\mathrm{sph}}\frac{\partial \lambda_\mathrm{sph}}{\partial x} \\
    -\frac{\partial\varepsilon_\mathrm{sph}}{\partial x}
\end{pmatrix}.
\end{align}
Some simplifications can be achieved by using identities~(\ref{Eq:BetaDerivId}) and
\begin{equation}
\alpha\beta+\lambda_\mathrm{sph}\varepsilon_\mathrm{sph} = \frac{\sqrt{x-\alpha^2(1-\beta^2)}}{x^2}(x^2 + \lambda_\mathrm{sph}^2).
\end{equation}
The result for $N$ can be written in the form
\begin{equation}
N = \begin{pmatrix}
N_s \\ N_q \\ N_\varepsilon
\end{pmatrix}
 =
\frac{1}{q_\mathrm{sph}^3\sin\vartheta}\frac{\lambda_\mathrm{sph}^2(\alpha,\beta,x)}{x^2}\frac{\partial\lambda_\mathrm{sph}(\alpha,\beta,x)}{\partial x}
\begin{pmatrix}
  \alpha q_\mathrm{sph}^2\sin\vartheta \\
 \frac{\lambda_\mathrm{sph}(x,\alpha,\beta)}{x^2}\sqrt{x - \alpha^2(1-\beta^2)}(x^2 + \alpha^2\cos^2\vartheta) \\
    -q_\mathrm{sph}(x^2 + \alpha^2\cos^2\vartheta)
\end{pmatrix},
\end{equation}
where $q_\mathrm{sph} = \sqrt{\lambda_\mathrm{sph}(\alpha,\beta,x)^2 - \alpha^2\cos^2\vartheta}$.
It is clear from this expression that $N_q$ is negative on $S_c'$.

\section{Analytic expression for $Q_c$ in parametric form}
\label{App:QcAnalytic}

The critical value $Q_c = Q_c(\alpha,\vartheta,\sin\chi,E)$ described in the previous appendix can be given analytically in a parametric form. This can be achieved by writing down the two conditions $R = 0$ and $dR/dr = 0$, which, when expressed in terms of $Q$ and $\chi$ yield
\begin{eqnarray}
0 &=& \left( \rho^2 E - Q a\sin\vartheta\sin\chi \right)^2 - \Delta\left( Q^2 + \rho^2 m^2 \right),
\label{Eq:RZero}\\
0 &=& 2\left(\rho^2 E - Q a\sin\vartheta\sin\chi \right) E r - (r-M)\left( Q^2 + \rho^2 m^2 \right) r - \Delta m^2 r.
\label{Eq:dRZero}
\end{eqnarray}
By eliminating $Q$ and $E$ from these two equations and choosing the correct branch one obtains
\begin{equation}
Q_c = \frac{M m}{4}\sqrt{\frac{N_1 N_2}{D_1}}, \quad E_c =  m\sqrt{\frac{N_3}{D_2}},
\label{eq:QcEcParametricGeneralForm}
\end{equation}
where
\begin{eqnarray*}
N_1 & = & 2x^2 + 2 \alpha ^2\cos^2\vartheta, \\
N_2 & = & 8(x - 3) x^4 + \alpha^2\left\{ -8 x^2\overline{\Delta} \sin^2\vartheta\cos(2\chi) + 4\left[ \alpha^2 \left( x^2 - x - 1 \right) - (3x - 4) x^2 \right]\cos(2\vartheta)  - \alpha^2(x-1)^2\cos(4 \vartheta) \right\}\\
&+& 4\alpha^2 (3 x + 4) x^2 - 3 \alpha^4 ( x + 1)^2 + X_1, \\
D_1 & = & \left[ x \left( x^2 - 3x + 2 \alpha^2 \right) - \alpha^2 (x-1) \cos^2\vartheta \right]^2 - 4\alpha^2 x^2\overline{\Delta} \sin
   ^2\vartheta \sin^2\chi, \\
X_1 & = & -8\alpha x^{3/2}\overline{\Delta}\sin\vartheta \sin\chi \sqrt{4 x^2 - (2+x)\alpha^2 + \alpha ^2 \cos(2 \vartheta)\left[ x \cos (2 \chi )+x-2 \right] - \alpha ^2 x \cos (2\chi )}, \\
N_3 & = & x^4\overline{\Delta}^2\left( x^2 - 3x + 2\alpha^2 \right) + \alpha^2\left\{ 2\alpha^2 x^2\overline{\Delta} \sin^4\vartheta\sin^4\chi\left( x^2 + \alpha^2\cos^2\vartheta \right) \right. \\
&& -\sin^2\vartheta\sin^2\chi\left[ \alpha^2 x^4 \cos^2\vartheta (6\alpha ^2+5
   x^2+15)+x^2 (2 \alpha^6+x^4 (4x^2 - 15x +15)+\alpha^2 x^3 (7 x-23) \right. \\
&& \left. +\cos(2\vartheta) (2
   \alpha^6 - 8\alpha^2 x^3-7 \alpha^4 x)+\alpha^4 x (4 x-7)) + \alpha^4 (x-1) x(2x^2 -x - \alpha^2)\cos^4\vartheta + \alpha^6 (x-1)^2 \cos^6\vartheta  \right] \\
&& \left. -x \overline{\Delta}^2\cos^2\vartheta\left[ \alpha^2 (x-1) \cos ^2\vartheta + 2x(x-\alpha^2) \right] \right\} - 2 X_2, \\
D_2 &=& \left( x^2 + \alpha^2 \cos^2\vartheta \right)^2 \left\{ \left[ x \left( x^2 - 3x + 2\alpha ^2 \right)-\alpha ^2
   (x-1) \cos^2\vartheta\right]^2 - 4\alpha ^2 x^2\overline{\Delta} \sin^2\vartheta\sin^2\chi \right\}, \\
X_2 &=& \alpha\overline{\Delta}\sin\vartheta\sin\chi \left( x^2 + \alpha^2\cos^2\vartheta \right) \sqrt{x} \left[x \left(\alpha ^2 \sin ^2\vartheta\sin^2\chi - \alpha^2 + x\right)+\alpha^2 (x-1)\cos^2\vartheta \right]^{3/2},
\end{eqnarray*}
where we have set $\overline{\Delta} = x^2 - 2x + \alpha^2$. In the Schwarzschild limit $\alpha = 0$ one gets
\begin{equation}
Q_c = \frac{M m x}{\sqrt{x - 3}}, \quad 
E_c = \frac{m(x-2)}{\sqrt{x(x-3)}},
\end{equation}
which correctly reproduces Eq.~(\ref{Eq:LcSchwarzschild}).

In many applications, we numerically invert the function $E_c = E_c(x)$ given by Eq.\ (\ref{eq:QcEcParametricGeneralForm}). To do this, it is convenient to first identify the appropriate zero of the denominator $D_2$, for which $E_c$ diverges. The first numerical guess for the solution of the equation $E = E_c(x)$ is then set to be slightly larger than this zero.

In the high-energy limit one writes $Q = M\zeta E$, and after dividing by $E^2$ and taking the limit $E\to \infty$ the system of equations (\ref{Eq:RZero}), (\ref{Eq:dRZero}) leads to the following sixth-order equation for $x$ (which coincides with the equation $D_2=0)$:
\begin{equation}
\left[ x(x^2-3x+2\alpha^2) - \alpha^2(x-1)\cos^2\vartheta \right]^2 - 4\alpha^2 x^2\overline{\Delta}\sin^2\vartheta\sin^2\chi = 0,
\label{Eq:SixthOrder}
\end{equation}
from which $\zeta$ can be computed according to
\begin{equation}
\zeta = (x^2 + \alpha^2\cos^2\vartheta)\sqrt{\frac{x}{x^2 - \alpha^2\cos^2\vartheta - \alpha^2 x\sin^2\vartheta\cos^2\chi}}.
\label{Eq:zeta}
\end{equation}
For equatorial orbits ($\vartheta=\pi/2$ and $\chi = \pi/2$ or $\chi = 3\pi/2$) Eqs.~(\ref{Eq:SixthOrder}) and (\ref{Eq:zeta}) simplify to $x(x-3)^2 - 4\alpha^2 = 0$ and $\zeta = x^{3/2}$. In the Schwarzschild limit $\alpha=0$, this yields $x=3$ and $\zeta = \sqrt{27}$ (see Eq. (33) of \cite{mMoS25} and the related discussion).

\section{Computation and properties of $Q_\mathrm{max}$}
\label{App:Qmax}

In this appendix we compute $Q_\mathrm{max}$, the maximum value of $Q$ for scattered orbits. It is determined by the maximum value of $Q$ for which $W_+(r)\leq E$. When $L$ and $L_z$ are rewritten in terms of $Q$ and $\chi$ this inequality is equivalent to
\begin{equation}
\label{ineq:HQ}
H(Q) := a\sin\vartheta\sin\chi Q + \sqrt{\Delta}\sqrt{Q^2 + m^2\rho^2} - \rho^2 E \leq 0,
\end{equation}
so we need to analyze the behavior of $H(Q)$ (for fixed values of $a$, $r$, $\vartheta$, and $E$). Before we do so, we claim that
\begin{equation}
\sqrt{\Delta} + a\sin\vartheta\sin\chi > 0,
\end{equation}
for all $r > r_c$. This is clearly the case if $\sin\chi\geq 0$. Hence, let us assume that $\sin\chi < 0$. It then follows that
\begin{equation}
\beta = \frac{\hat{L}_z}{L} = \frac{Q\sin\vartheta\sin\chi - a\cos^2\vartheta E}{\sqrt{Q^2 + a^2 m^2\cos^2\vartheta}} \le 0,
\end{equation}
such that the orbit is retrograde. On the other hand, when $\beta\le 0$ it follows from Eq.~(\ref{Eq:hDef}) that $h(2) \leq -2 + 2\alpha^2(1-\beta^2) < 0$; consequently $x_\mathrm{ph} > 2$ when $\beta \le  0$. Since $r_c > r_\mathrm{ph}$, it follows that
\begin{equation}
\sqrt{\Delta} + a\sin\vartheta\sin\chi
 = M\left[ \sqrt{x^2 - 2x + \alpha^2} + \alpha\sin\vartheta\sin\chi \right]
 > M(\alpha - \alpha) = 0,
\end{equation}
for $r > r_c$, as claimed.

After this observation we return to the behavior of $H(Q)$. First, we note that
\begin{equation}
H(0) = -\rho(\rho E - \sqrt{\Delta} m) < 0,
\end{equation}
since $E > m$ and $\rho^2 > \Delta$. Next, we note that the asymptotic behavior of $H$ is given by
\begin{equation}
\lim\limits_{Q\to\infty} \frac{H(Q)}{Q} = \sqrt{\Delta} + a\sin\vartheta\sin\chi > 0,
\qquad
\lim\limits_{Q\to -\infty} \frac{H(Q)}{Q} = -(\sqrt{\Delta} - a\sin\vartheta\sin\chi).
\end{equation}
Finally, a short computation reveals that
\begin{equation}
\frac{d^2 H}{dQ^2}(Q) = \sqrt{\Delta}\frac{m^2\rho^2}{(Q^2 + m^2\rho^2)^{3/2}} > 0,
\end{equation}
such that $H$ is strictly convex. It follows from these observations that $H(Q)$ has a unique root $Q_\mathrm{max}$ in the interval $(0,\infty)$. When $\sqrt{\Delta}-a\sin\vartheta\sin\chi < 0$ (as could occur for prograde orbits), the function $H(Q)$ has a second root in the interval $(-\infty,0)$; otherwise, it has no other roots.

To find explicit expressions for these roots, one can square the equation $H(Q) = 0$ and obtain a quadratic equation for $Q$ which is equivalent to finding the zeros of $R = -A Q^2 + B Q + C$, see Eq.~(\ref{Eq:RQPoly}). The discriminant is
\begin{equation}
D = B^2 + 4AC = 4\rho^2\Delta\left[ \rho^2 E^2 - \Delta m^2 + a^2 m^2\sin^2\vartheta\sin^2\chi \right],
\end{equation}
which is always positive for $r > r_+$, and hence $R$ has two real roots. Noting that the standard formula for quadratic equations can be rewritten in the form
\begin{equation}
Q_\pm = \frac{B\pm \sqrt{D}}{2A} = \frac{2C}{-B\pm \sqrt{D}},
\end{equation}
one finds the two equivalent representations
\begin{eqnarray}
Q_\pm &=& \frac{-\rho^2 E a \sin\vartheta \sin\chi \pm \rho\sqrt{\Delta}\sqrt{\rho^2 E^2 - \Delta m^2 + a^2 m^2 \sin^2\vartheta \sin^2\chi}}{\Delta - a^2\sin^2\vartheta\sin^2\chi}
\nonumber\\
 &=& \frac{\rho(\rho^2 E^2 - \Delta m^2)}{\rho E a\sin\vartheta\sin\chi \pm \sqrt{\Delta}\sqrt{\rho^2 E^2 - \Delta m^2 + m^2 a^2\sin^2\vartheta\sin^2\chi}},
\label{Eq:Qpm}
\end{eqnarray}
where the second one has the advantage of being manifestly regular where $A$ changes its sign. When $A = \Delta - a^2\sin^2\vartheta\sin^2\chi > 0$, these roots satisfy $Q_- < 0 < Q_+$ and hence it follows directly that $Q_\mathrm{max} = Q_+$. In contrast, when $A < 0$ (which implies $\sqrt{\Delta} < a\sin\vartheta\sin\chi$ since $\sqrt{\Delta} + a\sin\vartheta\sin\chi >  0$ for $r > r_c$), then both roots are positive and satisfy $0 < Q_+ < Q_-$. However, the simple estimate
\begin{eqnarray}
H(Q_-) & \geq & a\sin\vartheta\sin\chi Q_- + \sqrt{\Delta} Q_- - \rho^2 E
\nonumber \\
& = & -\frac{\rho\sqrt{\Delta}}{\sqrt{\Delta} - a\sin\vartheta\sin\chi}\left[ \sqrt{\rho^2 E^2 - \Delta m^2 + m^2 a^2\sin^2\vartheta\sin^2\chi} + \rho E \right] > 0,
\end{eqnarray}
reveals that $Q_-$ cannot be a root of $H$, and thus it follows again that $Q_\mathrm{max} = Q_+$. Summarizing, $Q_\mathrm{max}$ is always given by Eq.~(\ref{Eq:Qpm}) with the plus sign.

When $r$ is not restricted to $r > r_c$, then $Q_\mathrm{max}$ is not relevant for our discussion. However, it is still interesting to analyze the behavior of the roots $Q_\pm$ defined in Eq.~(\ref{Eq:Qpm}). As long as $r > r_+$ and $A=\Delta - a^2\sin^2\vartheta\sin^2\chi > 0$ one still has $Q_- <  0 < Q_+$. When $r > r_+$ and $A < 0$ one can have both $0 < Q_+ < Q_-$ or $Q_+ < Q_- < 0$, depending on whether  $a\sin\vartheta\sin\chi$ is positive or negative.\footnote{The two roots $Q_\pm$ both converge to $\rho^2 E/(a\sin\vartheta\sin\chi)$ in the limit $r\to r_+$.} In the limit $A = 0$ there is the single root
\begin{equation}
Q_0 := -\frac{C}{B} = \frac{\rho^2 E^2 - \Delta m^2}{2E a\sin\vartheta\sin\chi},
\end{equation}
when $r > r_+$ and $\sin\chi\neq 0$. Finally, when $r_- < r < r_+$, the discriminant $D$ is negative, which implies that $R$ has no real roots. This is already clear from Eq.~(\ref{Eq:RQchi}), which shows that $R > 0$ when $\Delta <  0$.

\section{On the smoothness of the observables}
\label{App:smoothness}

\begin{figure}
\begin{center}
\includegraphics[width=0.49\linewidth]{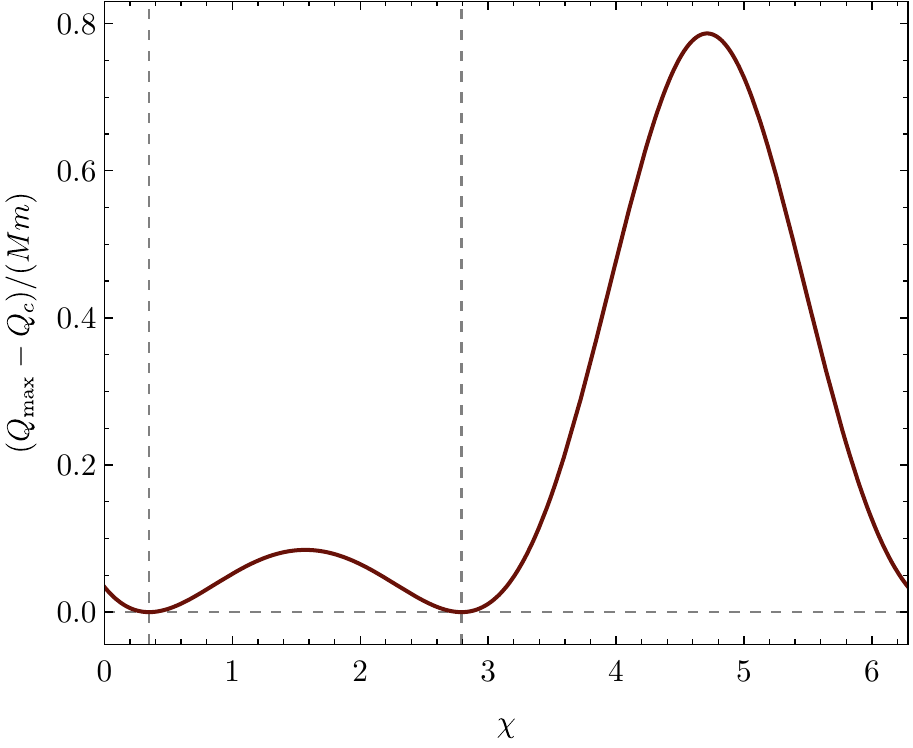}
\includegraphics[width=0.49\linewidth]{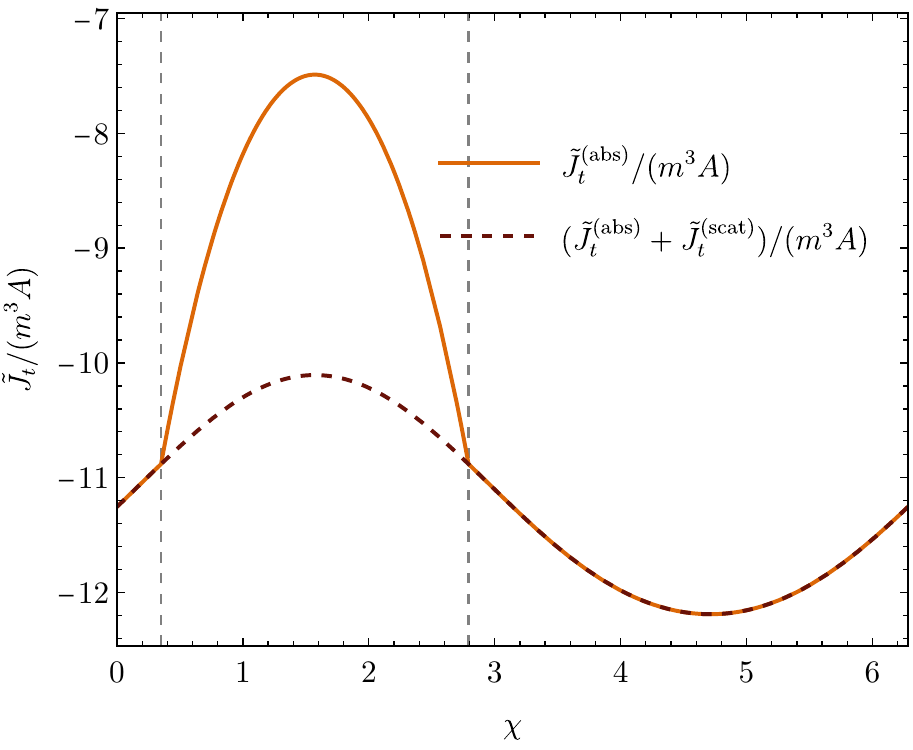}
\end{center}
\caption{\label{fig:nonsmoothness} Left: The difference $Q_\mathrm{max} - Q_c$ vs. $\chi$ for $\alpha = 1/2$, $\varepsilon = 2$, $\vartheta = \pi/4$, $x = 2.8$. Right: $\tilde J_t^\mathrm{(abs)}$ and $\tilde J_t = \tilde J_t^\mathrm{(abs)} + \tilde J_t^\mathrm{(scat)}$ vs. $\chi$ for the monoenergetic model with the same parameters. Dashed vertical lines mark the values of $\chi$ for which $E_0 = E_c$ or, equivalently, $r = r_c$.}
\end{figure}

As illustrated in the plots in Figs.~\ref{fig:Jtmono} and \ref{fig:Jphimono}, the radial profiles associated with the contributions to $J_t$ and $J_\varphi$ from the absorbed and scattered particles exhibit a discontinuity in their first-order derivative. This occurs in the region $r_\mathrm{ph} < r < r_\mathrm{mb}$ located between the photon sphere and the radius associated with marginally bound orbits. Interestingly, despite this fact, that total contributions $J_t = J_t^\mathrm{(abs)} + J_t^\mathrm{(scat)}$ and $J_\varphi = J_\varphi^\mathrm{(abs)} + J_\varphi^\mathrm{(scat)}$ do seem to have a smooth radial profile. In this appendix we show that this is indeed the case, based on a more detailed analysis of analytic expressions~(\ref{Eq:Jtabs}), (\ref{Eq:Jtscat}), (\ref{Eq:Jphiabs}), and (\ref{Eq:Jphiscat}).

To achieve this, we first note that by definition, $Q_c\leq Q_\mathrm{max}$ and $Q_c = Q_\mathrm{max}$ at $r=r_c$. Next, we fix the values of $a$, $M$, $m$, $\vartheta$, $E$, and $\chi$ and consider the function $R(r,Q)$ defined by Eq.~(\ref{Eq:RQchi}). Likewise, we denote by $Q_\mathrm{max}(r)$ the maximum value of $Q$ discussed in the previous appendix. By definition of $Q_c$ and $Q_\mathrm{max}(r)$ we have
\begin{equation}
R(r_c,Q_c) = 0,\qquad
\frac{\partial R}{\partial r}(r_c,Q_c) = 0,
\label{Eq:RQc}
\end{equation}
and
\begin{equation}
R(r,Q_\mathrm{max}(r)) = 0,
\label{Eq:RQmax}
\end{equation}
for all $r\geq r_c$. Differentiating Eq.~(\ref{Eq:RQmax}) with respect to $r$ leads to
\begin{equation}
\frac{\partial R}{\partial r}(r,Q_\mathrm{max}(r)) + \frac{\partial R}{\partial Q}(r,Q_\mathrm{max}(r)) \frac{dQ_\mathrm{max}}{dr}(r) = 0.
\label{Eq:FirstRQId}
\end{equation}
A short calculation reveals that
\begin{equation}
\frac{\partial R}{\partial Q} = -2\left( AQ + E\rho^2 a\sin\vartheta\sin\chi \right),
\end{equation}
where $A$ is defined by Eq.\ (\ref{Eq:RQPoly}) and thus
\begin{equation}
\label{Eq:dRdQaux}
    \frac{\partial R}{\partial Q}(r,Q_\mathrm{max}(r)) =  -2\rho\sqrt{\Delta}\sqrt{\rho^2 E^2 - \Delta m^2 + m^2 a^2\sin^2\vartheta\sin^2\chi},
\end{equation}
where we used Eq.\ (\ref{Eq:Qpm}). Evaluating Eq.\ (\ref{Eq:dRdQaux}) at $r = r_c$ we get $\partial R (r_c,Q_\mathrm{max}(r_c))/\partial Q = - \kappa^2$, where
\begin{equation}
    \kappa^2 :=  \left. 2\rho\sqrt{\Delta}\sqrt{\rho^2 E^2 - \Delta m^2 + m^2 a^2\sin^2\vartheta\sin^2\chi} \right|_{r = r_c}.
\end{equation}
We assume that $\kappa > 0$. Therefore, evaluating Eq.~(\ref{Eq:FirstRQId}) at $r = r_c$ yields
\begin{equation}
\left. \frac{dQ_\mathrm{max}}{dr} \right|_{r=r_c} = 0,
\label{Eq:QmaxPrime}
\end{equation}
where we take into account that $Q_c$ and $Q_\mathrm{max}$ coincide at $r = r_c$. Next, differentiating Eq.~(\ref{Eq:FirstRQId}) with respect to $r$, evaluating at $r = r_c$, and using Eq.~(\ref{Eq:QmaxPrime}) gives
\begin{equation}
\frac{\partial^2 R}{\partial r^2}(r_c,Q_c) + \frac{\partial R}{\partial Q}(r_c,Q_c) \left. \frac{d^2 Q_\mathrm{max}}{dr^2} \right|_{r_c} = 0.
\label{Eq:SecondRQId}
\end{equation}
In order to compute the second partial derivative of $R$ with respect to $r$, we use the identity
\begin{equation}
r\frac{\partial^2 R}{\partial r^2}(r,Q_\mathrm{max}(r)) - \frac{\partial R}{\partial r}(r,Q_\mathrm{max}(r)) = 8r^3(E^2- m^2) + 8 M m^2 r^2 - 2M\left[Q_\mathrm{max}(r)^2 + m^2\rho^2 \right].
\end{equation}
Evaluating both sides at $r=r_c$ and using Eq.~(\ref{Eq:RQc}) yields
\begin{equation}
\frac{\partial^2 R}{\partial r^2}(r_c,Q_c) = 8r_c(r_c-M)(E^2 - m^2) + \frac{2M}{r_c}\left[ 4E^2 r_c^2 - X_c^2 \right],
\label{Eq:ddR}
\end{equation}
where we have introduced $X_c := \left. \sqrt{Q_c^2 + m^2\rho^2} \right|_{r=r_c}$. Dropping the subindex $c$ for what follows in this paragraph we observe that Eqs.~(\ref{Eq:RQc}), or equivalently, Eqs.~(\ref{Eq:RZero}) and (\ref{Eq:dRZero}) and inequality (\ref{ineq:HQ}), imply that $X$ satisfies the quadratic equation
\begin{equation}
(r-M)X^2 - 2E r\sqrt{\Delta} X + m^2 r\Delta = 0,
\end{equation}
such that
\begin{equation}
X = X_\pm = \sqrt{Q^2 + m^2 \rho^2} = \frac{\sqrt{\Delta}}{r-M}\left[ E r \pm \sqrt{(Er)^2 - m^2 r(r-M)} \right].
\end{equation}
However, using the inequalities $\Delta < (r-M)^2$ and $0 < b - \sqrt{b^2-c^2} < c$ for all $0 < c < b$ we note that the solution belonging to the minus sign has $X_-\leq m\sqrt{r(r-M)} < m r$ which contradicts $X = \sqrt{Q^2 + m^2\rho^2}\geq m r$. Therefore, the correct solution is $X = X_+$, and it satisfies $X = X_+ < 2 E r$. 

It now follows from Eq.~(\ref{Eq:ddR}) that
\begin{equation}
\frac{\partial^2 R}{\partial r^2}(r_c,Q_c) > 8r_c(r_c-M)(E^2-m^2) > 0,
\end{equation}
which implies, using Eqs.~(\ref{Eq:dRdQaux}) and (\ref{Eq:SecondRQId}), that
\begin{equation}
\Lambda^2 := \left. \frac{1}{2}\frac{d^2 Q_\mathrm{max}}{dr^2} \right|_{r=r_c} > 0,
\end{equation}
where we assume $\Lambda > 0$ in what follows. As a consequence, the function $Q_\mathrm{max}(r)$ has a strict minimum at $r= r_c$ and there exists a smooth function $g(r)$ such that
\begin{equation}
\sqrt{Q_\mathrm{max}(r) - Q_c} = \Lambda|r-r_c| g(r),
\end{equation}
with $g(r)\to 1$ as $r\to r_c$.

After these remarks, we come back to the observables. First, we consider the integral
\begin{equation}
\int\limits_{Q_c}^{Q_\mathrm{max}} \frac{Q dQ}{\sqrt{R}},
\end{equation}
appearing in Eq.~(\ref{Eq:Jtscat}). By performing the variable substitution $Q\mapsto \eta := \sqrt{Q_\mathrm{max} - Q}$, recalling that $R = -A(Q-Q_+)(Q-Q_-)$ with $Q_+ = Q_\mathrm{max}$ and introducing $c_1:=1/Q_+$ and $c_2:=1/(Q_+-Q_-)$, one finds
\begin{equation}
\int\limits_{Q_c}^{Q_\mathrm{max}} \frac{Q dQ}{\sqrt{R}}
 = \frac{2 Q_+ \sqrt{c_2}}{\sqrt{A}}\int\limits_0^{\Lambda|r-r_c| g(r)}
 \frac{1  - c_1\eta^2}{\sqrt{1 - c_2 \eta^2}}  d\eta 
 = \mbox{sign}(r-r_c)\frac{2Q_+ \sqrt{c_2}}{\sqrt{A}}\int\limits_0^{\Lambda(r-r_c) g(r)} 
 \frac{1  - c_1 \eta^2}{\sqrt{1 - c_2 \eta^2}} d\eta,
\label{Eq:QIntId1}
\end{equation}
for $r$ lying close to $r_c$, where in the second step we have used the fact that the integrand is an even function of $\eta$. Note that both $c_1$ and $c_2$ remain finite. Moreover, at $r = r_c$, we have $A(Q_+ - Q_-) = A/c_2 = \kappa^2$. Since the right-hand side of Eq.\ (\ref{Eq:QIntId1}) has the form of a sign function multiplied with a smooth integral, it follows that its first derivative must jump at $r=r_c$. 

On the other hand, the combination of integrals that appear in $J_t = J_t^\mathrm{(abs)} + J_t^\mathrm{(scat)}$ is (see Eqs.~(\ref{Eq:Jtabs}) and (\ref{Eq:Jtscat}))
\begin{equation}
\int\limits_{0}^{Q_c} \frac{Q dQ}{\sqrt{R}} + 2 \times 1_{r_c < r}
\int\limits_{Q_c}^{Q_\mathrm{max}} \frac{Q dQ}{\sqrt{R}},
\end{equation}
which can be rewritten as
\begin{equation}
\int\limits_{0}^{Q_\mathrm{max}} \frac{Q dQ}{\sqrt{R}} + \left( 2 \times 1_{r_c < r} - 1 \right)
\int\limits_{Q_c}^{Q_\mathrm{max}} \frac{Q dQ}{\sqrt{R}}
 = \int\limits_{0}^{Q_\mathrm{max}} \frac{Q dQ}{\sqrt{R}} + \mbox{sign}(r - r_c)\int\limits_{Q_c}^{Q_\mathrm{max}} \frac{Q dQ}{\sqrt{R}}.
\label{Eq:QIntId2}
\end{equation}
The first integral on the right-hand side is regular, as a consequence of Eq.~(\ref{Eq:QInt1}) and the fact that $R$ and $S$ vanish at $Q = Q_\mathrm{max}$. The second term on the right-hand side of Eq.~(\ref{Eq:QIntId2}) is regular as a consequence of the identity~(\ref{Eq:QIntId1}) and the fact that $\mbox{sign}^2 = 1$. The argument for proving the smoothness of $J_\varphi$ is analogous.

The non-smooth radial behavior of $J_t^\mathrm{(abs)}$ and $J_t^\mathrm{(scat)}$ is closely related to the lack of smoothness of the expressions (integrands) appearing in the integrals with respect to $\chi$ in Eqs.\ (\ref{Eq:Jmu}). To be more precise, consider the monoenergetic model and write the time component $J_t$ as 
\begin{equation}
\label{eq:jtilde}
    J_t^\mathrm{(abs)} = \int\limits_0^{2 \pi} d \chi \tilde J_t^\mathrm{(abs)}, \qquad J_t^\mathrm{(scat)} = \int\limits_0^{2 \pi} d \chi \tilde J_t^\mathrm{(scat)},
\end{equation}
where
\begin{equation}
\tilde J_t^\mathrm{(abs)} = - A m E_0 \int\limits_0^{Q_c} \frac{Q dQ}{\sqrt{R}}, \qquad \tilde J_t^\mathrm{(scat)} = -2A m E_0 1_{r_c < r}
\int\limits_{Q_c}^{Q_\mathrm{max}} \frac{Q dQ}{\sqrt{R}},
\end{equation}
and where $E_0 \ge m$. Clearly, $J_t = J_t^\mathrm{(abs)} + J_t^\mathrm{(scat)} = \int_0^{2 \pi} \left( \tilde J_t^\mathrm{(abs)} + \tilde J_t^\mathrm{(scat)}  \right) d \chi.$ Neither $\tilde J_t^\mathrm{(abs)}$, nor $\tilde J_t^\mathrm{(scat)}$ has to be a smooth function of $\chi$, and an example of such behavior is illustrated in Fig.\ \ref{fig:nonsmoothness}. In this case, an efficient and accurate numerical evaluation of integrals (\ref{eq:jtilde}) requires dividing the interval $\chi \in [0,2\pi]$ into regions in which the integrand is regular and performing the integration in each region separately. If $J_t^\mathrm{(abs)}$ and $J_t^\mathrm{(scat)}$ are not required separately, one can directly integrate the sum $\tilde J_t^\mathrm{(abs)} + \tilde J_t^\mathrm{(scat)}$, which remains sufficiently regular. The azimuthal components $J_\varphi$ can be treated analogously.

\section{Explicit forms of integrals with respect to $Q$.}
\label{App:Qintegrals}

In this appendix, we provide explicitly real expressions for the integrals of the type
\begin{equation}
    \int \frac{Q dQ}{\sqrt{R}}, \qquad \int \frac{Q^2 dQ}{\sqrt{R}}.
\end{equation}
They were used to obtain numerical results reported in Section~\ref{sec:numerics}. We assume that
\begin{equation}
    R = -A Q^2 + BQ + C = -A (Q - Q_-)(Q - Q_+),
\end{equation}
where the coefficients $A$, $B$, and $C$ are given by Eq.~(\ref{Eq:RQPoly}), and the roots $Q_\pm$ can be expressed by Eq.~(\ref{Eq:Qpm}). For simplicity, if $Q_\pm$ are real, we will assume that $Q_- \le Q_+$ in this appendix. This convention allows us to write the relevant integrals in a consistent and simple way, but the labeling of the roots might be reversed with respect to Eq.~(\ref{Eq:Qpm}).

Suppose that $A > 0$ and $Q_+ > Q_-$. The relevant range of $Q$ is $Q_- \le Q \le Q_+$. We have
\begin{subequations}
\begin{eqnarray}
\int_{Q_-}^Q \frac{q dq}{\sqrt{R}} & = & \frac{1}{\sqrt{A}} \int_{Q_-}^Q \frac{q dq}{\sqrt{(q - Q_-)(Q_+ - q)}} \nonumber \\
& = & \frac{1}{\sqrt{A}} \left[ - \sqrt{(Q - Q_-)(Q_+ - Q)} + (Q_- + Q_+) \mathrm{arcsin} \sqrt{\frac{Q - Q_-}{Q_+ - Q_-}} \right], \\
\int_{Q_-}^Q \frac{q^2 dq}{\sqrt{R}} & = & \frac{1}{\sqrt{A}} \int_{Q_-}^Q \frac{q^2 dq}{\sqrt{(q - Q_-)(Q_+ - q)}} \nonumber \\
& = & \frac{1}{4 \sqrt{A}} \left\{ - \sqrt{(Q - Q_-)(Q_+ - Q)} \left[ 2 Q + 3 (Q_- + Q_+) \right] \right. \nonumber \\
&& \left. + (3 Q_-^2 + 2 Q_- Q_+ + 3 Q_+^2) \mathrm{arcsin} \sqrt{\frac{Q - Q_-}{Q_+ - Q_-}}\right\}.
\end{eqnarray}
\end{subequations}
For $A < 0$ and real roots $Q_\pm$ satisfying $Q_- < Q_+$, the relevant range is $Q_+ < Q$ or $Q < Q_-$. The following forms can be used:
\begin{subequations}
\begin{eqnarray}
\int_{Q_+}^Q \frac{q dq}{\sqrt{R}} & = & \frac{1}{\sqrt{-A}} \int_{Q_+}^Q \frac{q dq}{\sqrt{(q - Q_-)(q - Q_+)}} \nonumber \\
& = & \frac{1}{\sqrt{-A}} \left[ \sqrt{(Q - Q_-)(Q - Q_+)} + (Q_- + Q_+) \mathrm{arsinh} \sqrt{\frac{Q - Q_+}{Q_+ - Q_-}} \right], \quad Q_+ < Q, \\
\int_{Q_+}^Q \frac{q^2 dq}{\sqrt{R}} & = & \frac{1}{\sqrt{-A}} \int_{Q_+}^Q \frac{q^2 dq}{\sqrt{(q - Q_-)(q - Q_+)}} \nonumber \\
& = & \frac{1}{4\sqrt{-A}} \left\{ \sqrt{(Q - Q_-)(Q - Q_+)} \left[2 Q + 3 (Q_- + Q_+) \right] \right. \nonumber \\
&& \left. + (3 Q_-^2 + 2 Q_- Q_+ + 3 Q_+^2) \mathrm{arsinh} \sqrt{\frac{Q - Q_+}{Q_+ - Q_-}} \right\}, \quad Q_+ < Q, \\
\int_Q^{Q_-} \frac{q dq}{\sqrt{R}}  & = & \frac{1}{\sqrt{-A}} \int_Q^{Q_-} \frac{q dq}{\sqrt{(Q_- - q)(Q_+ - q)}} \nonumber \\
& = & \frac{1}{\sqrt{-A}} \left[ - \sqrt{(Q_- - Q)(Q_+ - Q)} + (Q_- + Q_+) \mathrm{artanh} \sqrt{\frac{Q_- - Q}{Q_+ - Q}} \right], \quad Q < Q_-, \\
\int_Q^{Q_-} \frac{q^2 dq}{\sqrt{R}}  & = & \frac{1}{\sqrt{-A}} \int_Q^{Q_-} \frac{q^2 dq}{\sqrt{(Q_- - q)(Q_+ - q)}} \nonumber \\
& = & \frac{1}{4\sqrt{-A}} \left\{ - \sqrt{(Q_- - Q)(Q_+ - Q)} \left[ 2 Q + 3 (Q_- + Q_+) \right] \right. \nonumber \\
&& \left. + (3 Q_-^2 + 2 Q_- Q_+ + 3 Q_+^2) \mathrm{artanh} \sqrt{\frac{Q_- - Q}{Q_+ - Q}} \right\}, \quad Q < Q_-.
\end{eqnarray}
\end{subequations}

The roots $Q_\pm$ remain real, provided that
\[ E^2 \rho^2 + a^2 m^2 \sin^2 \vartheta \sin^2 \chi - m^2 \Delta \ge 0, \]
and $\Delta > 0$. As already argued, the first condition is always satisfied for $E > m$ and $r > r_-$. The second requires $\Delta > 0$.  For $r < r_+$, $A < 0$, and complex roots $Q_\pm$, one can use the form
\begin{eqnarray} \int \frac{qdq}{\sqrt{-A q^2 + B q + C}} & = & \frac{-A q^2 + B q+ C}{-A} - \frac{B}{2(-A)^{2/3}} \mathrm{artanh} \left( \frac{-2 A q + B}{2 \sqrt{-A} \sqrt{-A q^2 + B q + C}} \right), \\
\int \frac{q^2dq}{\sqrt{-A q^2 + B q + C}} & = & - \frac{(2 A q + 3B) \sqrt{-Aq^2 + Bq + C}}{4 A^2} \nonumber \\
&& - \frac{4 A C + 3 B^2}{8(-A)^{5/2}} \ln \left[ - A^2 \left(B - 2 A q - 2 \sqrt{-A} \sqrt{-Aq^2 + Bq + C} \right) \right].
\end{eqnarray}
It is easy to check that in this case
\begin{eqnarray}  
-1 < \frac{-2 A q + B}{2 \sqrt{-A} \sqrt{-A q^2 + B q + C}} < 1,
\end{eqnarray}
as required.

There is also the special case with $A = 0$. For $B>0$, we denote $Q_0 = -C/B$ and write
\begin{eqnarray}
    \int_{Q_0}^Q \frac{qdq}{\sqrt{Bq + C}} & = & \frac{1}{\sqrt{B}} \int_{Q_0}^Q \frac{qdq}{\sqrt{q - Q_0}} = \frac{2}{3\sqrt{B}}\sqrt{Q - Q_0}(Q + 2Q_0), \\
    \int_{Q_0}^Q \frac{q^2dq}{\sqrt{Bq + C}} & = & \frac{1}{\sqrt{B}} \int_{Q_0}^Q \frac{q^2dq}{\sqrt{q - Q_0}} = \frac{2}{15 \sqrt{B}} \sqrt{Q - Q_0} (3 Q^2 + 4 Q Q_0 + 8 Q_0^2).
\end{eqnarray}
For $B < 0$, we have
\begin{eqnarray}
    \int_Q^{Q_0} \frac{qdq}{\sqrt{Bq + C}} & = & \frac{1}{\sqrt{-B}} \int_Q^{Q_0} \frac{qdq}{\sqrt{Q_0 - q}} = \frac{2}{3\sqrt{-B}}\sqrt{Q_0 - Q}(Q + 2Q_0), \\
    \int_Q^{Q_0} \frac{q^2dq}{\sqrt{Bq + C}} & = & \frac{1}{\sqrt{-B}} \int_Q^{Q_0} \frac{q^2dq}{\sqrt{Q_0 - q}} = \frac{2}{15 \sqrt{-B}} \sqrt{Q_0 - Q} (3 Q^2 + 4 Q Q_0 + 8 Q_0^2),
\end{eqnarray}
where again $Q_0 = -C/B$.

\bibliographystyle{unsrt}
\bibliography{refs_kinetic}

\end{document}